\documentclass{jfm}

\usepackage{graphicx}
\usepackage{newtxtext}
\usepackage{newtxmath}
\usepackage{natbib}
\usepackage{color}
\usepackage{subfig}
\usepackage{tikz}
\usepackage{dashrule}
\usepackage{amssymb}
\usepackage{enumitem} 
\usepackage{rotating}
\usepackage{siunitx}
\usepackage{hyperref}
\usepackage[nameinlink,capitalise,noabbrev]{cleveref}
\hypersetup{
    colorlinks = true,
    urlcolor   = blue,
    citecolor  = black,
}
\usetikzlibrary{plotmarks}
\usetikzlibrary{mindmap}
\usetikzlibrary{spy}
\usetikzlibrary{arrows.meta,arrows}
\usetikzlibrary{positioning,shapes}

\usetikzlibrary{shapes.misc}

\tikzset{cross/.style={cross out, draw=black, minimum size=2*(#1-\pgflinewidth), inner sep=0pt, outer sep=0pt},
cross/.default={1pt}}

\tikzset{
  double -latex/.style args={#1 colored by #2 and #3}{    
    -latex,line width=#1,#2,
    postaction={draw,-latex,#3,line width=(#1)/3,shorten <=(#1)/4,shorten >=4.5*(#1)/3},
  },
  double round cap-latex/.style args={#1 colored by #2 and #3}{    
    round cap-latex,line width=#1,#2,
    postaction={draw,round cap-latex,#3,line width=(#1)/3,shorten <=(#1)/4,shorten >=4.5*(#1)/3},
  },
  double round cap-stealth/.style args={#1 colored by #2 and #3}{
    round cap-stealth,line width=#1,#2,
    postaction={round cap-stealth,draw,,#3,line width=(#1)/3,shorten <=(#1)/3,shorten >=2*(#1)/3},
  },
  double -stealth/.style args={#1 colored by #2 and #3}{
    -stealth,line width=#1,#2,
    postaction={-stealth,draw,,#3,line width=(#1)/3,shorten <=(#1)/3,shorten >=2*(#1)/3},
  },
}

\graphicspath{{./}{./images/}}

\newcommand{\markerltria}{\raisebox{0pt}{\tikz{\node[draw,scale=0.3,regular polygon, regular polygon sides=3,rotate=90,fill=black](){};}}}
\newcommand{\markerrtria}{\raisebox{0.5pt}{\tikz{\node[draw,scale=0.3,regular polygon, regular polygon sides=3,rotate=270,fill=black](){};}}}
\newcommand{\markerdiam}{\raisebox{0.5pt}{\tikz{\node[draw,scale=0.4,regular polygon, regular polygon sides=4,rotate=45,fill=black](){};}}}
\newcommand{\markersqua}{\raisebox{0.5pt}{\tikz{\node[draw,scale=0.4,regular polygon, regular polygon sides=4,rotate=0,fill=black](){};}}}
\newcommand{\markerstar}{\raisebox{0.5pt}{\tikz{\node[draw,scale=0.3,star,star points=5,star point ratio=2.25,rotate=0,fill=black](){};}}}
\newcommand{\markercirc}{\raisebox{0.5pt}{\tikz{\node[draw,scale=0.4,circle,fill=black](){};}}}

\definecolor{r1}{RGB}{255,178,102}
\definecolor{r2}{RGB}{255,128,0}
\definecolor{r3}{RGB}{255,0,0}
\definecolor{r4}{RGB}{153,0,0}
\definecolor{b1}{RGB}{0,255,255}
\definecolor{b2}{RGB}{0,165,255}
\definecolor{b3}{RGB}{0,0,255}
\definecolor{b4}{RGB}{0,0,153}
\definecolor{g1}{RGB}{0,0,0}
\definecolor{g2}{RGB}{96,96,96}
\definecolor{g3}{RGB}{160,160,160}
\definecolor{g4}{RGB}{202,202,202}

\newcommand{\RomanNumeralCaps}[1]
\linenumbers

\title{On the solidity parameter in canopy flows}

\author{Alessandro Monti\aff{1}
  \corresp{\email{alessandro.monti@oist.jp}},
  Shane Nicholas \aff{2},
  Mohammad Omidyeganeh \aff{2},
  Alfredo Pinelli \aff{2},
  \and Marco E. Rosti\aff{1}
  \corresp{\email{marco.rosti@oist.jp}}
  }

\affiliation{\aff{1}Okinawa Institute of Science and Technology OIST, 1919-1 Tancha, Onna, Kunigami District, Okinawa 904-0495, Japan
\aff{2}City University of London, SMCSE, Northampton Square, London EC1V 0HB, UK}

\begin{document}
\maketitle

\begin{abstract}
    We have performed high-fidelity simulations of turbulent open-channel
    flows over submerged rigid canopies made of cylindrical filaments
    of fixed length $l=0.25H$ ($H$ being the domain depth) mounted on 
    the wall with an angle of inclination $\theta$.
    The inclination is the free parameter that sets the density of 
    the canopy by varying its frontal area. The density of the canopy,
    based on the solidity parameter $\lambda$, is a
    widely accepted criterion defining the ongoing canopy flow regime, 
    with low values ($\lambda \ll0.1$) indicating the 
    \textit{sparse} regime, and higher values ($\lambda > 0.1$)
    the \textit{dense} regime.    
    All the numerical predictions have been obtained considering the same 
    nominal bulk Reynolds number (i.e. $Re_b=U_b H/\nu = 6000$). 
    We consider nine configurations of canopies, with $\theta$ 
    varying symmetrically around $\ang{0}$ in the range
    $\theta\in [\pm \ang{78.5}$], where positive angles define
    canopies inclined in the flow direction (\textit{with the grain}) 
    and $\theta=\ang{0}$ corresponds to the wall-normally mounted canopy.  
    The study compares canopies with identical solidity obtained 
    inclining the filaments in opposite angles and assesses the efficacy 
    of the solidity as a representative parameter.
    It is found that when the canopy is inclined, the actual flow regime
    differs substantially from the one of a straight canopy that shares 
    the same solidity indicating that criteria solely based on this 
    parameter are not robust.
    Finally, a new phenomenological model
    describing the interaction between the coherent structures populating 
    the canopy region and the outer flow is given.
\end{abstract}
    

\section{Introduction}
Simple structural elements such as beams or elastic filaments 
interacting with fluid flows have been largely studied because of 
their massive use in many technological applications and importance 
in environmental and biological flows.
For instance, suspension of fibers are often employed in low-Reynolds number
flows for studying biological transport processes such 
as microorganisms swimming \citep{LAUGA2009}, while in turbulent
flows fibers have been adopted, usually mixed with polymers 
\citep{LEE1974}, for drag reduction purposes \citep{PASCHKEWITZ2004}.
Recently, suspended fibers have been studied in turbulent flows
to exploit their usage as a proxy of turbulence statistics, in 
particular using an end-to-end length of the fiber as reference
length-scale for quantifying two-point statistics 
\citep{ROSTI2018b,ROSTI2020,olivieri_mazzino_rosti_2021d}, with the development of the novel 
technique of Fiber Tracking Velocimetry \citep{BRIZZOLARA2020}.

Surfaces of anchored filamentous layers exposed to fluid flows are
commonly found in nature, paving the way for novel bio-inspired 
technologies \citep{ALVARADO2017}. 
At microscales, ciliated walls and flagella are commonly 
found in living organs (e.g. microvilli, cilia in the bronchial 
epithelium, papillae of tongues, cilia of kidney cells) participating 
to a number of physiological processes like locomotion, digestion, 
circulation, respiration and reproduction \citep[][]{LODISH2007}. 
Enlarging the range of scales considered, the interaction of surfaces 
covered by complex texture with surrounding fluid flows are adopted 
in nature for a wide variety of tasks: decrease skin 
friction drag \citep[e.g. seal fur, see][]{ITOH2006}, control of flight 
aerodynamics \citep[e.g. birds feathers, see][]{BRUECKER2014}.
An active branch of research concerns the interaction of vegetative
plants immersed in the atmospheric environment (terrestrial
canopies) and water (aquatic canopies) \citep{RAUPACH1981,FINNIGAN2000,NEPF2012}. 
In terrestrial canopies, the exchange of mass, heat and momentum between 
the canopy layer and the environmental surrounding regulate the micro-climate 
providing, for instance, plants with carbon dioxide for the photosynthesis 
\citep{RAUPACH1981}; in aquatic environment, instead, vegetation significantly 
contributes in creating habitats for microorganisms by influencing 
the nutrient transport and deposition, by improving water quality (especially 
useful in the grey-waters treating) and by regulating the solar light uptake 
\citep{MARS1999,GHISALBERTI2002,LUHAR2008,WILCOCK1999}.

\begin{figure}
    \centering
    \includegraphics[width=0.65\linewidth]{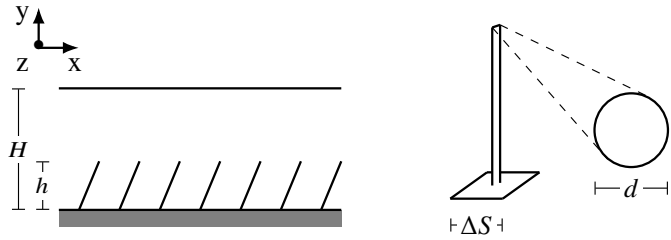}
    \caption{Geometrical parameters governing a canopy flow \citep{NEPF2012}.}
    \label{fig:geom}
  \end{figure}

The mentioned examples largely differ among them, with mechanical
properties highly depending on the tasks the filamentous layer has to
address. Therefore, a large variety of parameters must be accounted 
to correctly characterize the specific behaviour of 
each canopy configuration immersed in a fluid flow. 
These parameters span from purely geometrical properties (e.g. aspect 
ratio, size and shape of the stems, level of submersion, the angle of
inclination of the root of the stems) to mechanical 
aspects (e.g. flexibility, density ratio, active or passive motions);
including them all in a parametric study makes the analysis of canopy flows 
a very challenging topic.  In previous investigations,  researchers focused on finding a 
reduced set of parameters to characterize common behaviours that
helped to identify a standard classification of the flows.
The geometric argument has been thoroughly debated 
and, in the bulk of literature, the level of submersion, defined as the ratio between the flow depth 
$H$ and the canopy height $h$, and the solidity, 
a parameter that associates the frontal area of the canopy layer to the area 
of the canopy bed, have been extensively adopted to 
classify canopy flows \citep[see the reviews][]{NEPF2012,BRUNET2020}.
In particular, the former is used to distinguish emergent canopies
($H/h\le1$), where the resulting flow is dominated by the balance between the 
drag offered by the canopy elements and the driving pressure gradient, with
turbulence dominated by the vortices shed by the stems of canopy 
\citep{NEPF2000}, from submerged canopies ($H/h > 1$), 
where the flow is more complex due to the several
scales involved, e.g. the diameter of the 
stems $d$, the height of the canopy $h$, the average distance between the 
stems $\Delta S$, the size of the domain $H$, to mention a few of them 
\citep[][see also \cref{fig:geom}]{NEPF2012}. In the literature,  some of the parameters mentioned above
have been merged to define the so-called solidity,
\begin{equation}\label{eq:lambda1}
\lambda=\int_0^h \! d(y)/\Delta S^2\, \mathrm{d}y,
\end{equation}
an indicator of the density of the canopy that has been used to classify 
the submerged canopy flows into regimes that range from sparse to dense 
based on a threshold value (i.e. $\lambda_{t} \approx 0.1$) defined by means of
experimental evidence \citep{POGGI2004a,NEPF2012}.
In particular, it has been widely accepted that for values much smaller
than $\lambda_{t}$, the flow velocity above and within the canopy shows a
behaviour comparable to flows bounded by a solid wall covered by 
roughness elements (sparse regime); conversely,  the form drag offered by the stems becomes dominant (dense
regime) for large values of 
$\lambda$, and the mean velocity profile shows the typical two inflection
points caused by the drag discontinuity at the tip of the canopy (upper 
inflection point, located at the edge of the canopy layer)
and by the merging of the inflected profile below the canopy tip and the 
boundary layer developed in the proximity of the wall (lower inflection
point). In this regime, a stratified model with three separated layers
of the flow has been proposed \citep{BELCHER2003,POGGI2004a,NEZU2008}. 
The layers can be identified as follows: 
above the upper inflection point, $y/h > 1$,  an outer region with a behaviour typical of a boundary 
layer over a rough wall can be identified \citep{RAUPACH1996,FINNIGAN2000};
in the canopy region, instead, $y/h \ll 1$,  the flow is assumed to be characterized by the wakes shed by
the canopy elements, similarly to an emergent canopy
configuration. Finally, in the region in between, the flow is assumed to be dominated by a mixing-layer of constant 
thickness \citep{GHISALBERTI2004}.

Most of the past studies were of experimental nature that are of difficult 
realization and can hardly display, especially within the canopy,
a complete portrait of the mechanisms that characterize the mutual 
interaction between the stratified layers, due 
to the natural impedance enforced by the presence of the canopy layer itself.
With the increasing computational power and the introduction of advanced
techniques that enabled a full implementation of the canopy layer, 
new high-fidelity numerical studies have surged in the last few years
\citep{SHARMA2018, MONTI2019,MONTI2020,TSCHISGALE2021}.
In particular, \citet{MONTI2020} carried out a set of high-fidelity simulations
(wall-resolved LES) of open-channel flows bounded by a rigid, wall-normally 
mounted canopy simulated via a state-of-the-art immersed-boundary method. 
In their work, a parametric study has been investigated, choosing as free parameter 
the height of the canopy layer (thus implicitly setting the solidity $\lambda$).
The different heights analysed have been selected to span canopy flows from a 
marginally sparse regime to a dense one. With this study, a detailed 
characterisation of the canopy regimes has been given, providing the literature
with new insights on the structures populating the inner and the outer regions.
To better identify the transition from the
sparse to a dense regime, the authors presented a new criterion built on a 
simple physical model that establishes if the largest vortex of the outer
region could reach the bed, based on the geometrical
properties of the filamentous layer; with this model,
a new threshold value of solidity was found as lower bound for the dense regime,
i.e. $\lambda_{t}\approx 0.15$. 
The model was built upon the geometrical parameters characterising
the solidity,  but the latter may be a questionable parameter for
classifying canopy flows. For instance, simply considering rigid, cylindrical
stems with uniform length and diameter, inclined with opposite angles
$\theta$ in the direction of the flow \citep[$\theta>0$
flow {\em with the grain}, $\theta<0$ flow {\em against the grain},][]{ALVARADO2017}, 
the solidity value remains unchanged, while a very different
behaviour of the flow can be expected. This consideration may be extended to a more general
flexible canopy. Therefore, using $\lambda$ alone to classify
the flow may not be the adequate choice.

To address this uncertainty, in this work we analyse a set of rigid canopies
assembled with cylindrical stems, inclined at a certain angle $\theta$ in the 
streamwise direction. The angle of inclination is systematically varied from 
a wall-normal condition $\theta=\ang{0}$, to a condition that matches the lowest 
solidity value analysed in \citet{MONTI2020}, i.e. $\lambda=0.07$,
resulting in a inclination $\theta=\pm \ang{78.5}$. 
The whole analysis will be carried out by means of highly-resolved 
Large-Eddy Simulations, with the canopy layer simulated with a stem-by-stem 
approach implemented via an extensively validated Immersed-Boundary method.

The manuscript is organised as follows. \Cref{sec:method} 
describes the numerical method used to perform the simulations.
\Cref{sec:results} describes the obtained results that combines
statistical results with instantaneous realizations.
Finally, \cref{sec:conclusions} outlines the most 
important conclusions of the present work.

\section{The numerical method}\label{sec:method}
The turbulent flows over rigid canopies have been simulated by means of
a numerical solver \citep[SUSA,][]{OMID2013a} that solves the 
incompressible Navier-Stokes equations. In particular, we adopted a 
Large-Eddy Simulation (LES) approach, where the velocity and pressure field
obtained are a result of a high-pass filtering operation.
In a Cartesian frame of reference, where $x_1$, $x_2$ and $x_3$ 
(sometimes also referred to as $x$, $y$ and $z$) are adopted to identify
the streamwise, wall-normal and spanwise directions, with $u_1$,
$u_2$ and $u_3$ the corresponding velocity components ($u$, $v$ and $w$), the dimensionless incompressible 
LES equations for the resolved fields $\overline{u}$ and $\overline{p}$ 
read as 
\begin{equation}
  \frac{\partial \overline{u}_i}{\partial t} + \overline{u}_j 
  \frac{\partial \overline{u}_i}{\partial x_j} = 
- \frac{\partial \overline{P}}{\partial x_i} + \frac{1}{Re_b} 
  \frac{\partial^2 \overline{u}_i}{\partial x_j \partial x_j} 
+ \frac{\partial \tau_{ij}}{\partial x_j} + f_i, \hspace*{2em}
  \frac{\partial \overline{u}_i}{\partial x_i} = 0.
  \label{eq:NS-LES}
\end{equation}
In \cref{eq:NS-LES}, $Re_b=U_b H/\nu$ is the Reynolds 
number based on the bulk velocity $U_b$, the open channel height 
$H$ and the kinematic viscosity $\nu$, while
$\tau_{ij}=\overline{u_i u_j}-\overline{u}_i \overline{u}_j$ 
is the subgrid Reynolds stress tensor \citep{LEONARD1975} (from now
on, the overbar will be dropped to simplify the notation). To close
the equations, an eddy viscosity approach used to model the unresolved
subgrid stress tensor was adopted. In particular, we employ the Integral Length-Scale Approximation (ILSA) proposed 
by \citet{PIOMELLI2015} \citep[see also][]{ROUHI2016}.
The incompressible LES equations \cref{eq:NS-LES} are spatially
discretised with a second-order accurate, cell-centred finite 
volume method. Pressure and velocity are evaluated at the centres 
of the cells in a colocated grid fashion and, to avoid the appearance
of spurious pressure oscillations, the corrective approach proposed
by \citet{RHIE1983} has been adopted.
To advance the equations in time, we adopted a second-order, 
semi-implicit fractional-step method \citep{KIM1985},
where the implicit Crank-Nicolson scheme is implemented for 
the wall-normal diffusive terms and an explicit Adams-Bashforth 
scheme is applied to all other terms.
The Poisson equation for the pressure, required to enforce the solenoidal 
condition of the velocity field, is decoupled into a series of two-dimensional 
Helmholtz equations in the wavenumber space applying a fast Fourier transform
along the spanwise direction and solved through the iterative 
biconjugate gradient stabilized method with an algebraic 
multigrid preconditioner \citep[\textit{boomerAMG}, see][]{YANG2002}. 
The code is parallelised using the domain decomposition 
technique. 

The canopy is implemented as a set of stems
represented as rigid, solid, slender cylindrical rods of 
finite cross-sectional area, parallely mounted onto the impermeable 
bottom wall with an angle of inclination that constitute a free
parameter in this work.
The enforcement of the boundary conditions on the surface of the
rigid cylinders (zero-velocity) is obtained by means of an 
immersed boundary method (IBM) that deals with the presence of the 
rods by using a set of nodes ({\em Lagrangian nodes}) distributed 
along the length of each canopy element that do not necessarily
conform with the fluid grid. More specifically, at every time-step 
the employed IBM \citep{PINELLI2010} associates to every Lagrangian node a set of 
distributed body forces whose intensity can be computed by enforcing 
the no-slip condition on the nodes. The distributed set of body 
forces is defined on a compact support centred on each node of the 
Lagrangian mesh used to define the stems. 
The size of the support is related to the local grid size 
and defines the hydrodynamic thickness of the filament. An appropriate
study that investigates the adequate number of Langrangian nodes to be 
used to satisfactorily replicate the flow around a set of filaments has 
been done previously by \citet{MONTI2019}, who compared the 
outcomes obtained with the current methodology to the results from a 
simulation with an immersed boundary method that directly imposes the correct
boundary conditions on the surface of the filaments \citep{FADLUN2000}. 
From that study, we
concluded that a Lagrangian lattice with four points per cross-section were
enough to adequately reproduce the physics of the problem. Therefore, 
we indirectly set the diameter of the filaments to be around 
$2.2 ~\Delta x$ \citep{MONTI2019}, or $2.2~\Delta z$, since the mesh spacing 
is the same in the $x$ and $z$ directions.

\begin{figure}
\centering
  \subfloat[]{\includegraphics[width=0.49\linewidth]{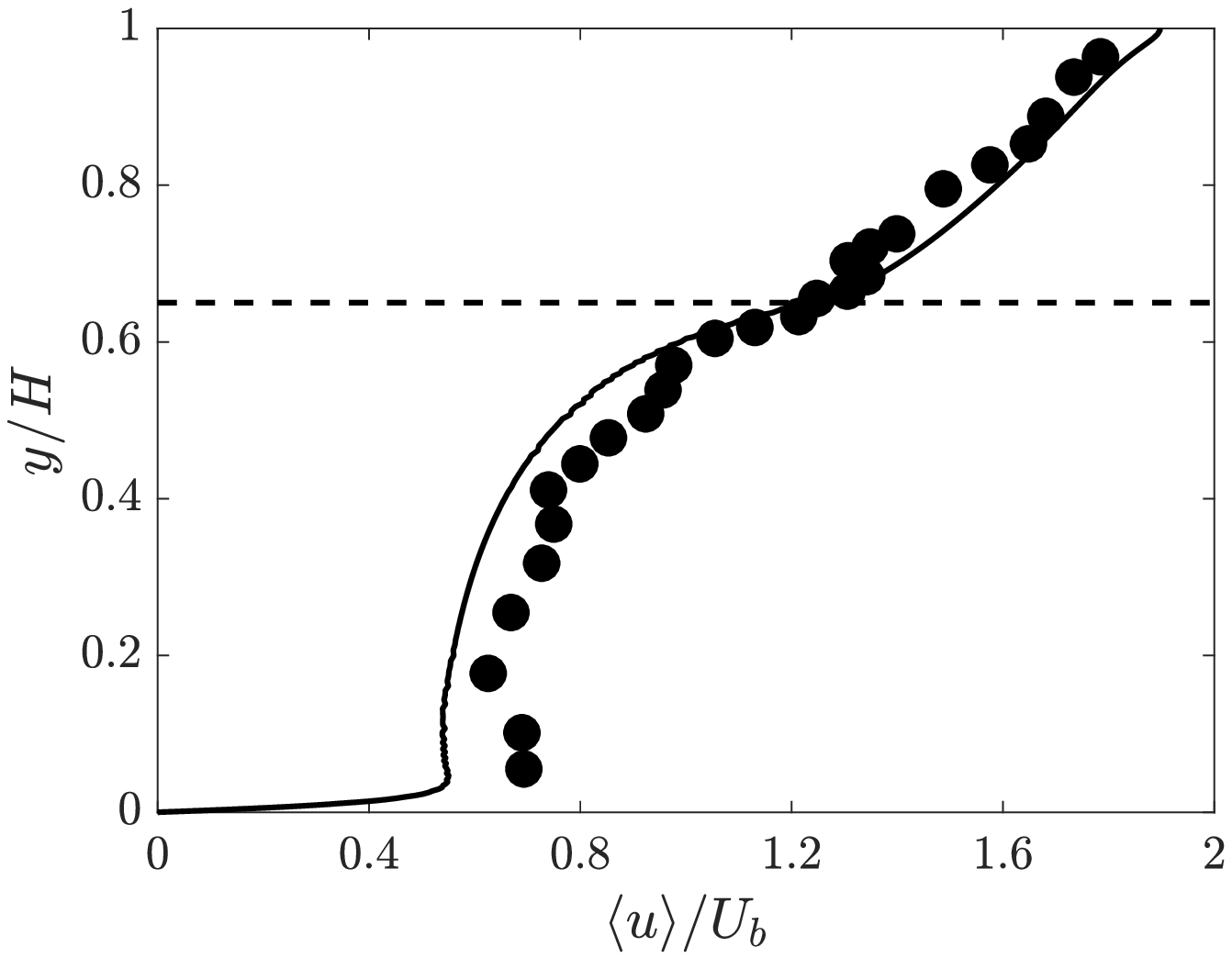}}
  \subfloat[]{\includegraphics[width=0.49\linewidth]{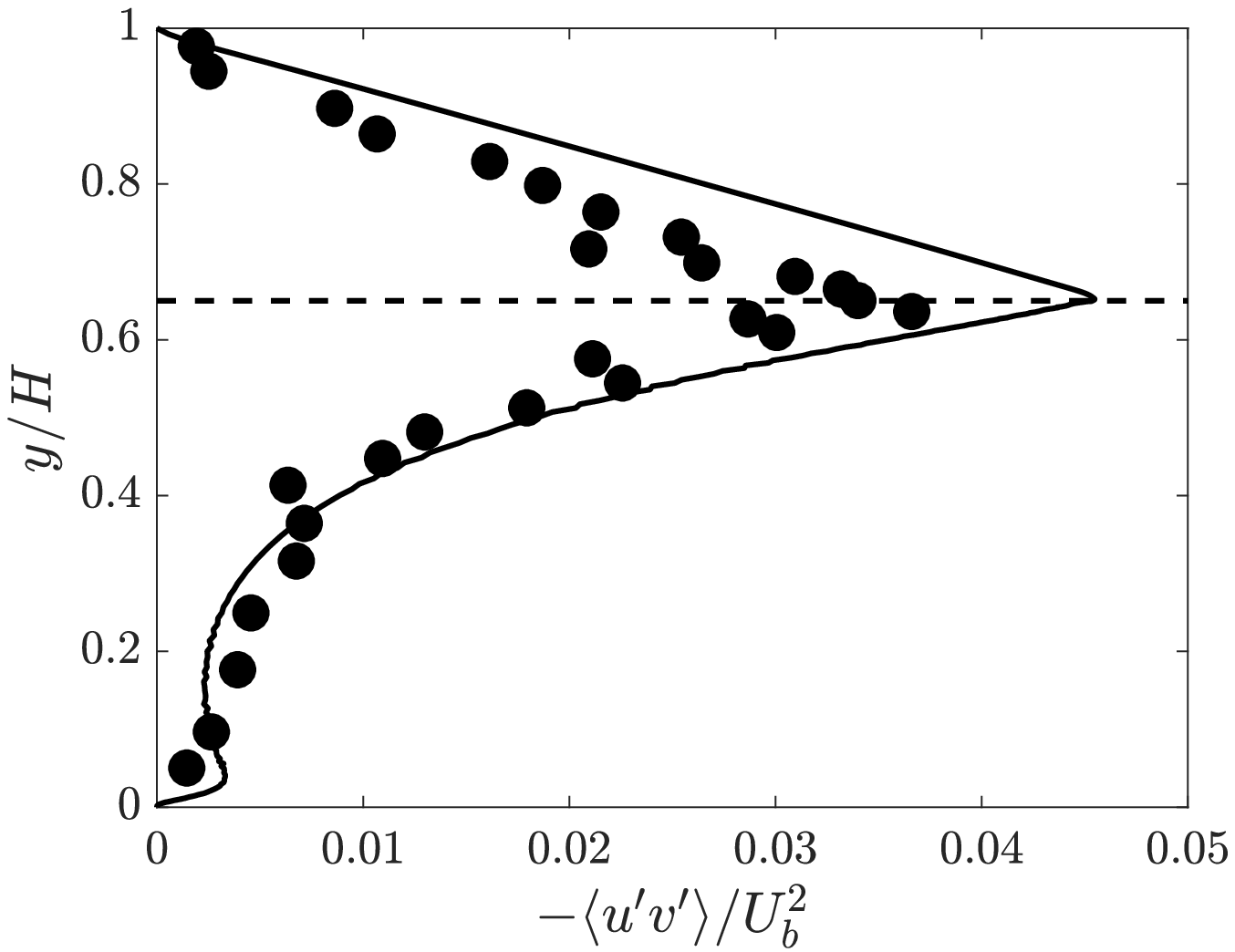}}
  \caption{Validation results \citep[see for more details][]{MONTI2019}. 
   (a) Mean velocity profile and 
   (b) Reynolds shear-stress distribution from our simulations
           (solid line) compared with the experimental values R31 by 
           \citep{SHIMIZU1991} (dotted curve). 
           The dashed line shows the location 
           of the canopy tip at $y=h$.}
  \label{fig:validation}
\end{figure}

\begin{table}
  \centering
  \setlength{\tabcolsep}{0.7em}
  \begin{tabular}{l|c|c}\hline
                  &Current    &Ref. \citep{SHIMIZU1991}            \\ \hline
  $Re_b$          &$7070$     &$7070$                              \\
  $Re_{\tau,in}$  &$535$      &--                                  \\
  $Re_{\tau,out}$ &$1310$     &--                                  \\
  $L_x/H\times L_y/H\times L_z/H$  &$2\pi \times 1 \times 1.5\pi$   &--  \\
  $h/H$           &$0.65$     &$0.65$                              \\ 
  $N_x\times N_y \times N_z$  &$480 \times 350 \times 360$    &--  \\
  $\lambda$       &$0.41$     &$0.41$                    \\ \hline
  Resolution &                &                                    \\
  $\Delta x^{+}_{in}\times \Delta y^{+}_{w,in}\times                  
   \Delta z^{+}_{in}$
  &$6 \times 0.15 \times 6$    &--                                  \\
  $\Delta x^{+}_{out}\times \Delta y^{+}_{\mathrm{h},out}\times \Delta z^{+}_{out}$ 
  &$20 \times 0.5 \times 20$   &--                                  \\ \hline
  \end{tabular}
  \caption{Validation case parameters.}
  \label{tab:domain_val}
\end{table}
Finally, to prove the appropriateness of the method, 
we report here the results of the validation campaign \citep{MONTI2019}, 
where we directly compare an appositely set-up simulation
with the experimental results (R31) by \citet{SHIMIZU1991}, with
wall-normally mounted rigid filaments of height $h/H=0.65$, 
solidity $\lambda=0.41$ and bulk Reynolds number $Re_b=7070$.
The comparison between the velocity profile and the Reynolds shear
stress obtained is shown in \cref{fig:validation}, with pretty good
agreement of the results.
The parameters of the simulation used for the validation case 
are provided in \cref{tab:domain_val} together with the 
corresponding experimental values \citep{SHIMIZU1991}.
Note that the viscous units used to compute the friction Reynolds numbers 
listed in \cref{tab:domain_val} (and therefore the resolution parameters) 
are based on the total shear stress at the solid wall ($in$ subscript) and 
at the canopy tip ($out$ subscript), with $\Delta y^{+}_{w,in}$ and 
$\Delta y^{+}_{h,out}$ indicating the resolution of the first computational 
cell at the wall and at the canopy tip, respectively.

\begin{figure}
  \centering
  \subfloat[]{\includegraphics[width=0.5\linewidth]{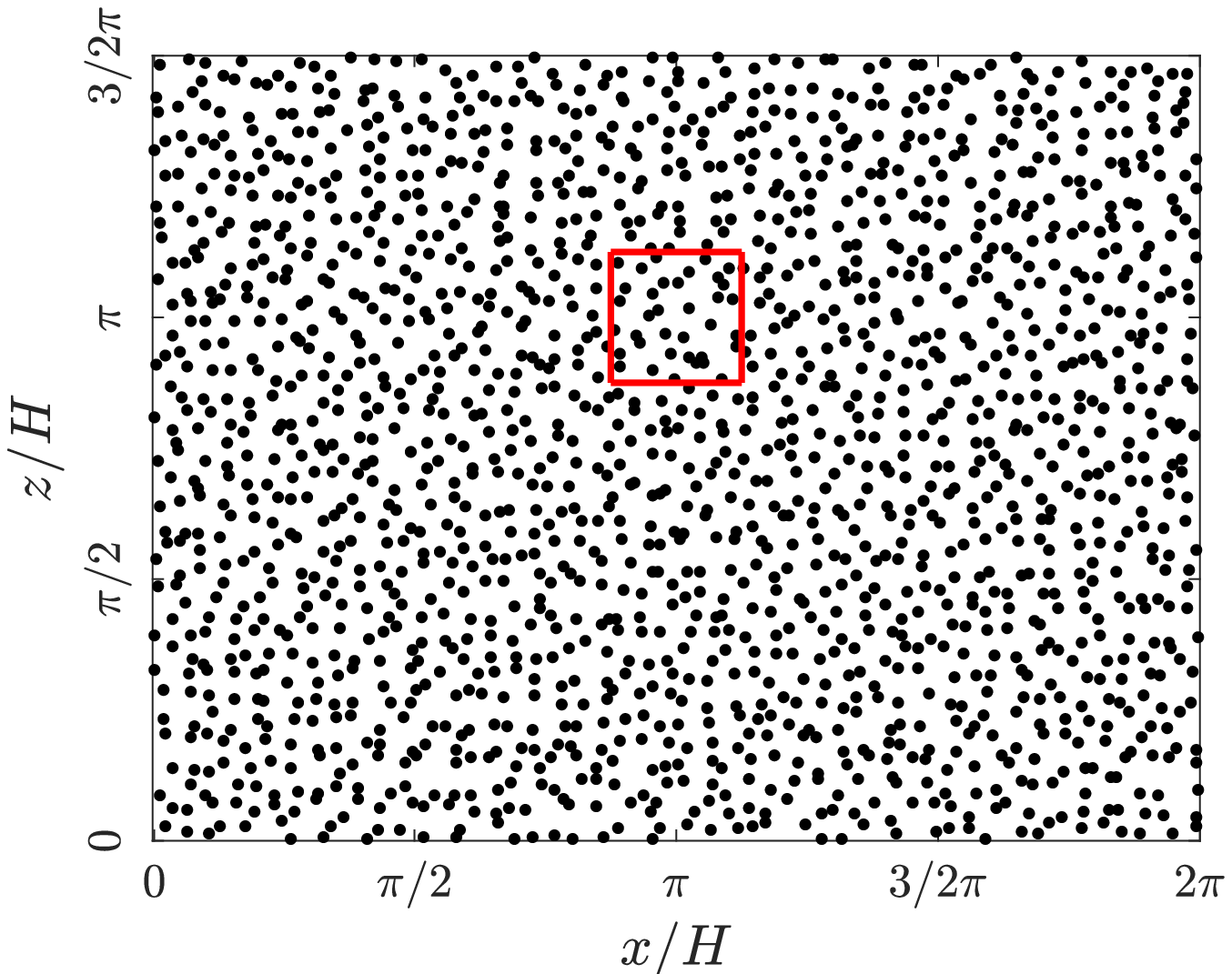}} \hspace{2em}
  \subfloat[]{\includegraphics[width=0.4\linewidth]{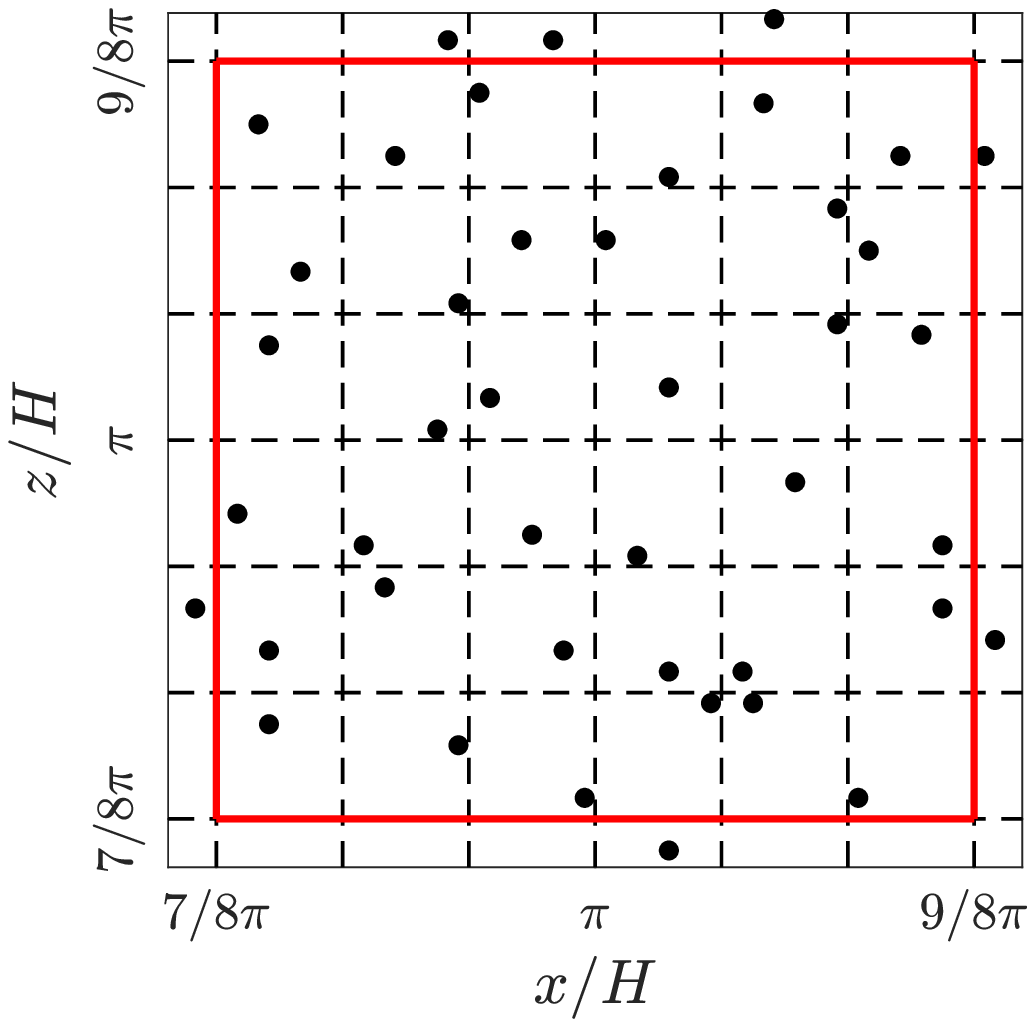}}
  \caption{(a) Filaments distribution on the bottom of the computational domain.
           The red box is zoomed out in panel (b), where the random allocation
           of each filament within a $\Delta S \times \Delta S$ tile is highlighted.}
  \label{fig:wall}
  \end{figure}

As we mentioned above, the stems are distributed on the
bottom wall. In particular, we have subdivided the latter in a 
Cartesian lattice of uniform squares of area $\Delta S^2$, with the 
filaments within each tile positioned randomly. The use of a random 
assignment on each tile prevents preferential flow channeling effects. 
A sketch of the distribution of the stems on the channel bottom wall 
is shown in \cref{fig:wall}.

\begin{figure}
\centering
%
%
%
%
%
%
%
%
  \includegraphics[width=0.65\linewidth]{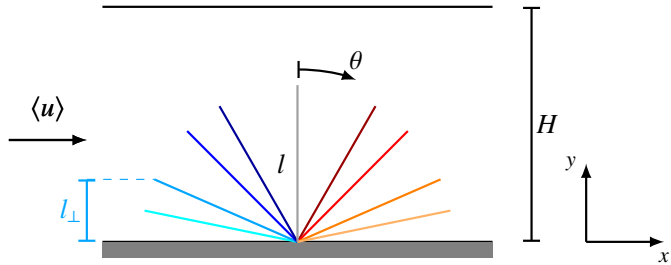}
  \caption{Sketch of the inclined canopy cases considered. 
           The colour scheme refers to the angle of inclination
           selected and will be kept in the whole manuscript.
           From left to right, in clockwise order: 
           $\theta=[-\ang{78.5}$, 
           $-\ang{66.5}$, $-\ang{45}$, $-\ang{30}$, $\ang{0}$,
           $\ang{30}, \ang{45}$, $\ang{66.5}$, $\ang{78.5}]$.}
  \label{fig:inclined}
\end{figure}

In order to assess the general usefulness of the solidity (defined in \cref{eq:lambda1}) 
as critical parameter in the framework of canopy flows, we adjust the size of the tile and the
angle of inclination of the stems (given the length of the filaments) 
to match solidity values that span from the quasi-sparse regime to the
dense one \citep{NEPF2012,BRUNET2020}. In particular, for stems with a uniform 
cross-sectional circular area of diameter $d$, the solidity simply reads as
\begin{equation}
\lambda=\frac{d\,l_\perp}{\Delta S^2},
\label{eq:lambda}
\end{equation}
where $l_\perp = l\cos(\theta)$ is the projection of the length of the 
filament $l$ along the wall-normal direction, defining the height of the
canopy layer $l_\perp=h$, with $\theta$ being the angle
of inclination positive in the clockwise direction counted from
the wall-normal direction (see \cref{fig:inclined}; note that the 
colour-scheme used to indicate the different inclinations will be kept
for the remaining part of the manuscript).
In this work, to vary the value of the solidity,
we fix the size of the tile $\Delta S$ (fixing the number of filaments in 
the streamwise and spanwise direction to $nf_x\times nf_z$), the diameter 
and the length of the filaments, and we vary the value of the inclination 
$\theta$.
In particular, we chose eight angles, symmetrically ranging from 
$\theta=\pm\ang{78.5}$ around $\theta=0$ (i.e 
$\lambda=[0.07,0.14,0.25,0.30,0.35]$).
\begin{table}
  \begin{center}
    \setlength{\tabcolsep}{0.7em}
    \begin{tabular}{c|c|c|c|c|c|c|c|c}
    \hline
    \rule{0pt}{4ex}
    $\theta$ & $l/H$ & $l_\perp/H$ & $\Delta S/H$ 
        & $\lambda$ & $nf_{x}\times nf_{z}$ & $N_y$ & $Re_\tau$
        & $\Delta x^+\times \Delta y^+_{h}\times \Delta z^+$\\ [2mm]\hline
    \rule{0pt}{4ex}
    {\color{b1} $\ang{-78.5}$} & $0.25$ & $0.05$  
            & $\pi/24$ & $0.07$ & $48\times 36$ & $230$ & $478.3$
            & $5.22 \times 0.24 \times 5.22$ \\[2mm]
    {\color{b2} $\ang{-66.5}$} & $0.25$ & $0.10$  
            & $\pi/24$ & $0.14$ & $48\times 36$ & $290$ & $656.7$
            & $7.16 \times 0.26 \times 7.16$ \\[2mm]
    {\color{b3} $\ang{-45}$}   & $0.25$ & $0.175$ 
            & $\pi/24$ & $0.25$ & $48\times 36$ & $290$ & $905.8$
            & $9.88 \times 0.36 \times 9.88$ \\[2mm]
    {\color{b4} $\ang{-30}$}   & $0.25$ & $0.215$ 
            & $\pi/24$ & $0.30$ & $48\times 36$ & $300$ & $1099.2$
            & $11.99 \times 0.44 \times 11.99$ \\[2mm]
    {\color{g3} $\ang{0}$}   & $0.25$ & $0.25$ 
            & $\pi/24$ & $0.35$ & $48\times 36$ & $300$ & $1157.5$
            & $12.63 \times 0.35 \times 12.63$ \\[2mm]
    {\color{r4} $\ang{30}$}    & $0.25$ & $0.215$ 
            & $\pi/24$ & $0.30$ & $48\times 36$ & $300$ & $966.9$
            & $10.55 \times 0.39 \times 10.55$ \\[2mm]
    {\color{r3} $\ang{45}$}    & $0.25$ & $0.175$ 
            & $\pi/24$ & $0.25$ & $48\times 36$ & $290$ & $831.4$
            & $9.07 \times 0.33 \times 9.07$ \\[2mm]
    {\color{r2} $\ang{66.5}$}  & $0.25$ & $0.10$  
            & $\pi/24$ & $0.14$ & $48\times 36$ & $290$ & $612.6$ 
            & $6.68 \times 0.25 \times 6.68$ \\[2mm]
    {\color{r1} $\ang{78.5}$}  & $0.25$ & $0.05$  
            & $\pi/24$ & $0.07$ & $48\times 36$ & $230$ & $472.7$
            & $5.16 \times 0.24 \times 5.16$ \\[2mm]
    \hline
    \end{tabular}
    \caption{Set of the parameters for the inclined canopies. From left to right:
             the angle of inclination; the length of the filaments; the wall-normal
             projection of the filaments (height of the canopy layer $l_\perp=h$); 
             the average spacing between the filaments; the solidity; the number 
             of filaments in the streamwise and spanwise directions; 
             the number of nodes of the computational mesh in the
             wall-normal direction; the friction Reynolds number, $Re_\tau=u_\tau H/\nu$, 
             where $u_\tau$ is computed evaluating the value of the total shear stress at the 
             canopy tip; the resolution of the computational domain in wall units, where 
             $\Delta y_h^+$ is evaluated in the region of maximum shear, i.e. at the edge of
             the canopy.}
    \label{tab:par}
  \end{center}
\end{table}
The nine cases share the same computational box of size
$L_x/H = 2\pi $, $L_y/H = 1$ and $L_z/H = 3/2 \pi$, similar
to other works \citep{BAILEY2013,MONTI2020}.

The numerical domain is
set to be periodic in both the streamwise (i.e. $x$) and the
spanwise (i.e. $z$) directions; 
at the bottom wall, a no-slip boundary condition is imposed 
while, at the top surface, a free-slip condition is set to 
mimic an open-channel free surface.
The Cartesian computational lattice is uniformly distributed
in the horizontal directions, while a stretched distribution (with
ratio between neighbouring cells kept below $4\%$) is adopted
in the wall-normal direction. The latter, in particular, is built using
two tangent-hyperbolic functions that concentrate the nodes in the regions
where higher shears are expected, i.e. at the edge of the canopy layer
and close to the solid wall.
The total number of nodes is equal to $N_x=576$ and $N_z=432$, while $N_y$ ranges
between $N_y\in[230,300]$, with the lower and 
upper cases set for the most 
inclined and the wall-normally mounted cases, respectively.
The number of nodes has been selected such that 
the spacings in wall units satisfy the standard 
values suggested for wall-bounded flows \citep{KIM1987}; the wall-units are
estimated using the maximum value of the viscous length scale based on
the local shear stress \citep[further explanations on the evaluation
of the local viscous scales are provided 
in the next section and in][]{MONTI2019}. Note that, the maximum value
of the local shear stress is obtained at the canopy edge; therefore,
the wall-normal grid spacing in wall-units considered is evaluated 
considering the value at the tip of the canopy.
Finally, to drive the flow, a uniform pressure gradient is applied 
in the streamwise direction. 
In particular, at each time step, the mean streamwise pressure 
gradient is adjusted to fix the flow rate to a constant 
value corresponding to a bulk Reynolds number of $Re_b=U_b~H/\nu=6000$,
a value close to the ones already available in the literature
\citep{BAILEY2013,SHIMIZU1991}. The detailed parameters of the simulations
are listed in \cref{tab:par}.

\begin{figure}
  \centering
  \subfloat[]{\includegraphics[width=0.49\linewidth]{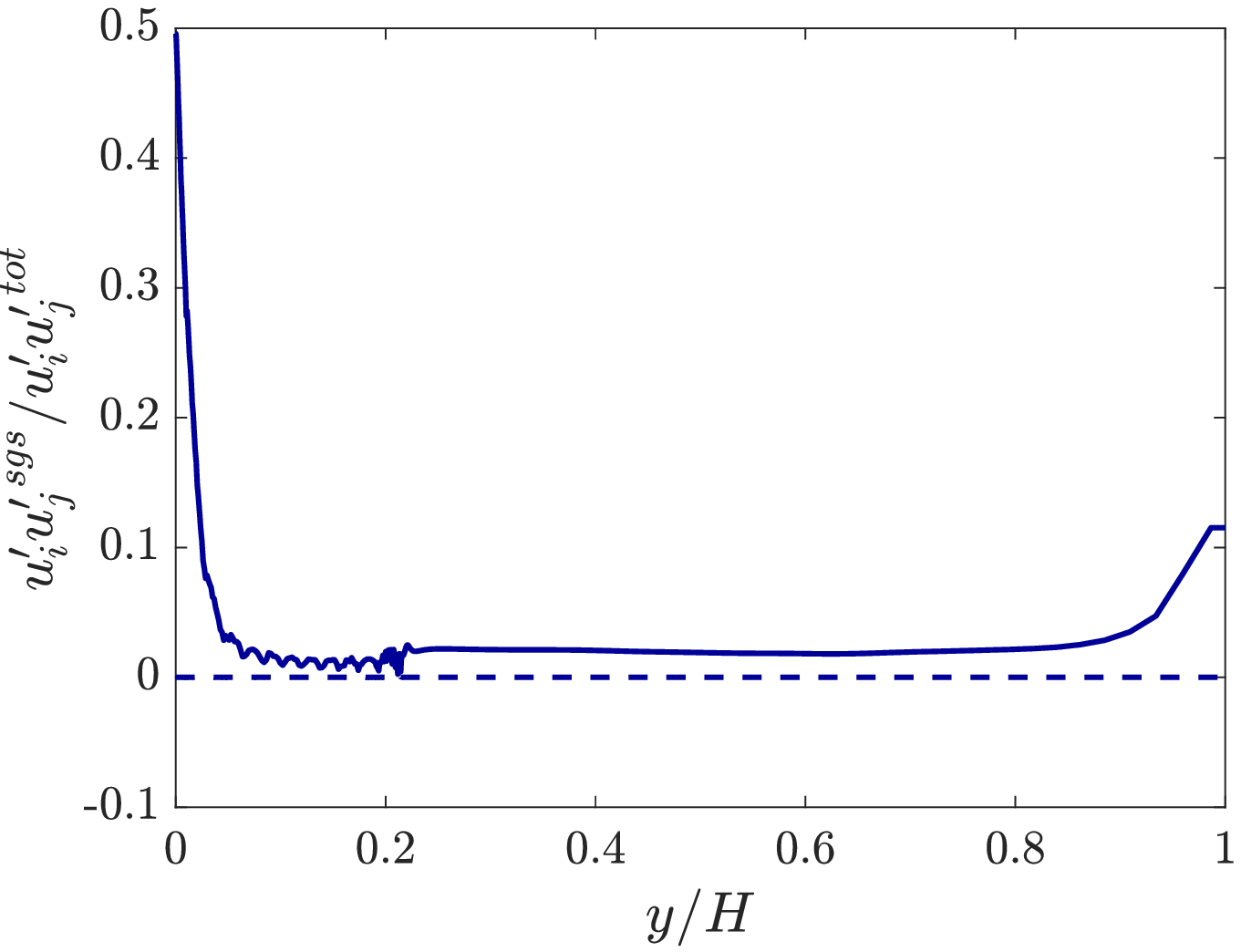}}
  \subfloat[]{\includegraphics[width=0.49\linewidth]{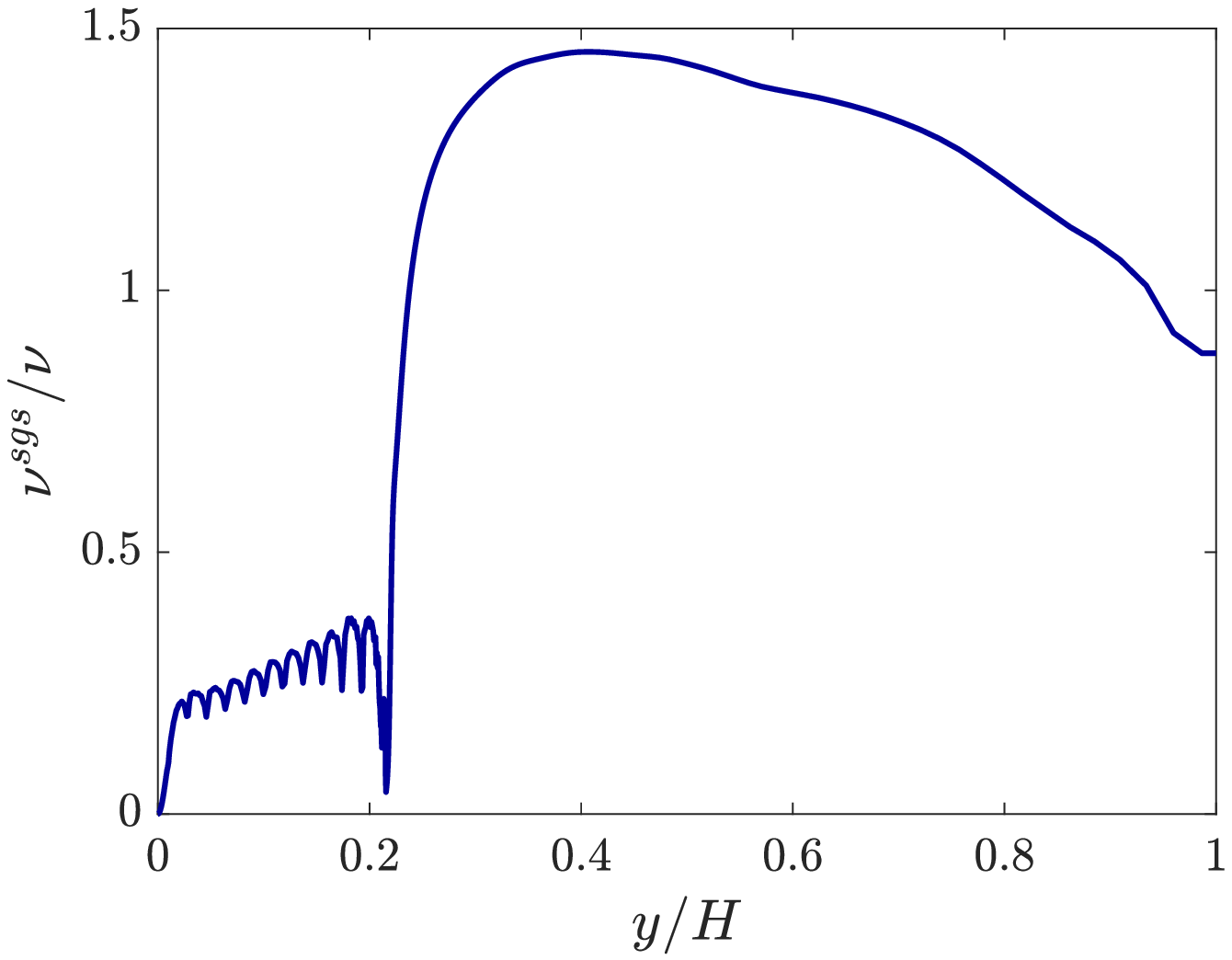}}
  \caption{(a) Dashed line: ratio between the sub-grid energy and the total 
               fluctuating energy along the wall-normal direction.
               Solid line: ratio between the sub-grid shear stress and the total 
               fluctuating shear stress along the wall-normal direction.
           (b) Ratio between the eddy viscosity and the physical viscosity 
               along the wall-normal direction.
           In both panels, the superscript $sgs$ indicates the subgrid stress
           tensor (eddy viscosity in panel (b)) component, while in panel (a) 
           the superscript $tot$ refers to the total part of the stress tensor, 
           i.e.\ the summation of the resolved and subgrid parts.
           The reference case chosen is $\theta=-\ang{30}$, consistent with
           the colormap in \cref{fig:inclined}.} 
  \label{fig:LESsgs}
\end{figure}

For the sake of completeness, we point out that the formulation of the problem
used in this work can be considered as a coarse Direct Numerical Simulation (DNS) in 
the outer portion of the flow that progressively becomes highly resolved as the canopy 
is approached. In the outer flow region, the subgrid stress contribution plays only 
the role of a very mild and stabilising numerical dissipation. Indeed, the ratio between 
the total and the subgrid energies averaged in time and in the two homogeneous directions, 
shown in the left panel of \cref{fig:LESsgs} (dashed line) along the channel height, 
is always below $10^{-5}$. Concerning the subgrid stress activity along the streamwise 
direction (dominant in a shear driven flow), the LES model always contributes with 
a value far below $0.1$, excluding the region in the proximity of the wall (not so 
relevant for canopy flows), as shown by the ratio between the subgrid shear stress and the 
total one \citep{ROUHI2016} averaged in time and in the two homogeneous directions 
in the left panel of \cref{fig:LESsgs} (solid line). These a-posteriori checks allowed us
to avoid introducing any particular treatment for the coupling between the LES and the 
IBM.
A further indication that the LES filter operates at the end of the turbulence cascade is
provided in the right panel of \cref{fig:LESsgs} showing that the ratio between the time 
and space averaged eddy viscosity and the physical one is always of order unity or less
throughout the whole channel.
Note that the curves shown in \cref{fig:LESsgs} refer to the case $\theta=-\ang{30}$;
the other scenarios show similar trends and, as such, are not reported for the sake of
clarity.

\section{Results}\label{sec:results}
The results collected in this section present the flow statistics 
characterising the flow and illustrate the emergence of various 
coherent structures related with the various regimes encountered 
(see \cref{tab:par}).
In particular, the section is structured as follows: the first part 
deals with the mean velocity profiles, analysing the location of the 
inflection points and the location of the (virtual) origin
of the boundary-layer developed in the outer part of the flow;
in the second part, we introduce the higher-order statistics and 
we extensively describe the turbulent coherent structures that 
characterise the flow.

Note that all the quantities shown in this section will be reported
in a non-dimensional form.

\subsection{Canopy properties}
\begin{figure}
  \centering
  \setbox1=\hbox{%
  \includegraphics[width=0.42\linewidth,height=4cm]{example-image-b}}
  \subfloat[]{\includegraphics[width=0.49\linewidth]{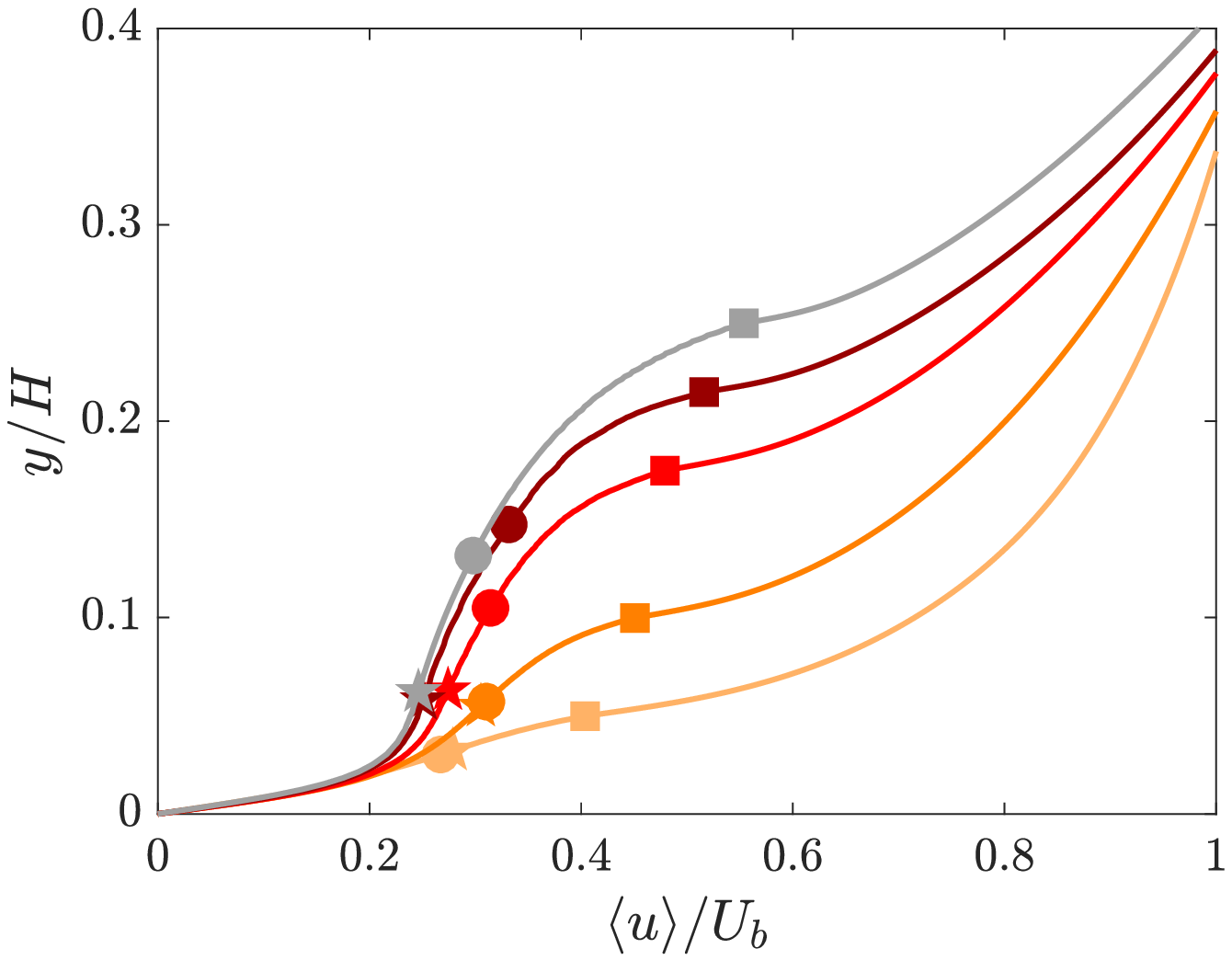}
              \llap{\makebox[\wd1][l]{\raisebox{3.3cm}{%
              \includegraphics[width=0.15\linewidth,height=1.5cm]
              {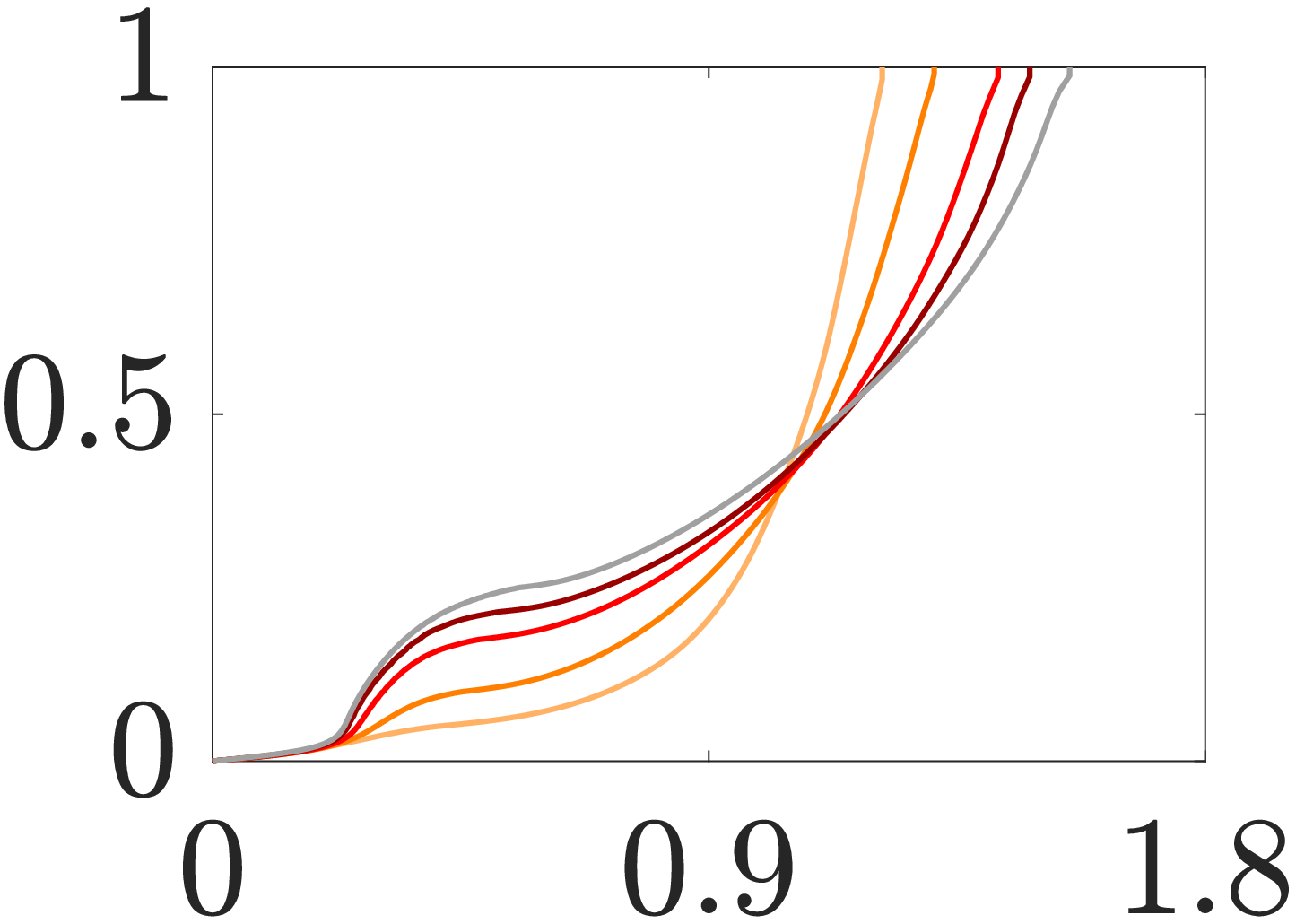}}}}
             }
  \subfloat[]{\includegraphics[width=0.49\linewidth]{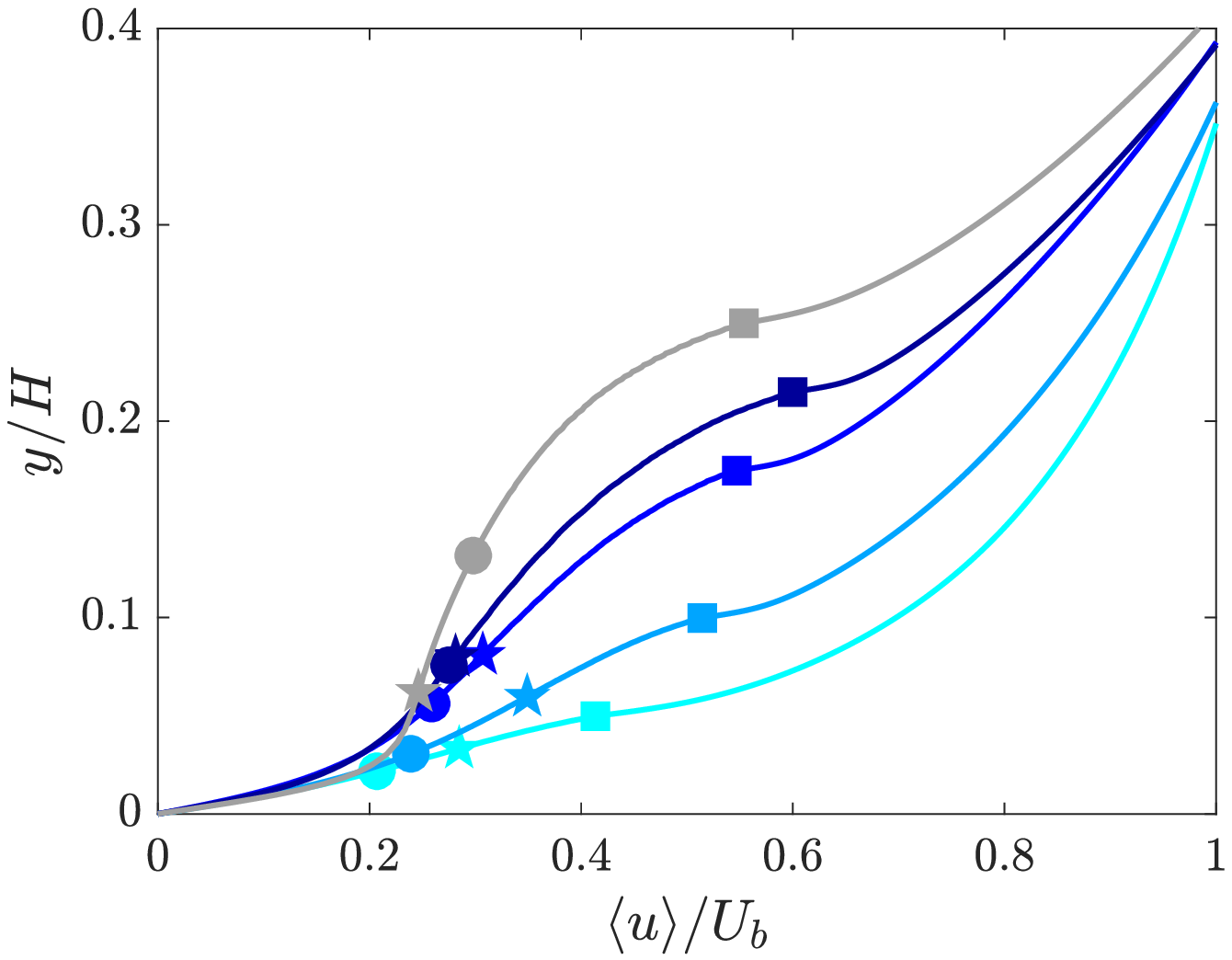}
              \llap{\makebox[\wd1][l]{\raisebox{3.3cm}{%
              \includegraphics[width=0.15\linewidth,height=1.5cm]
              {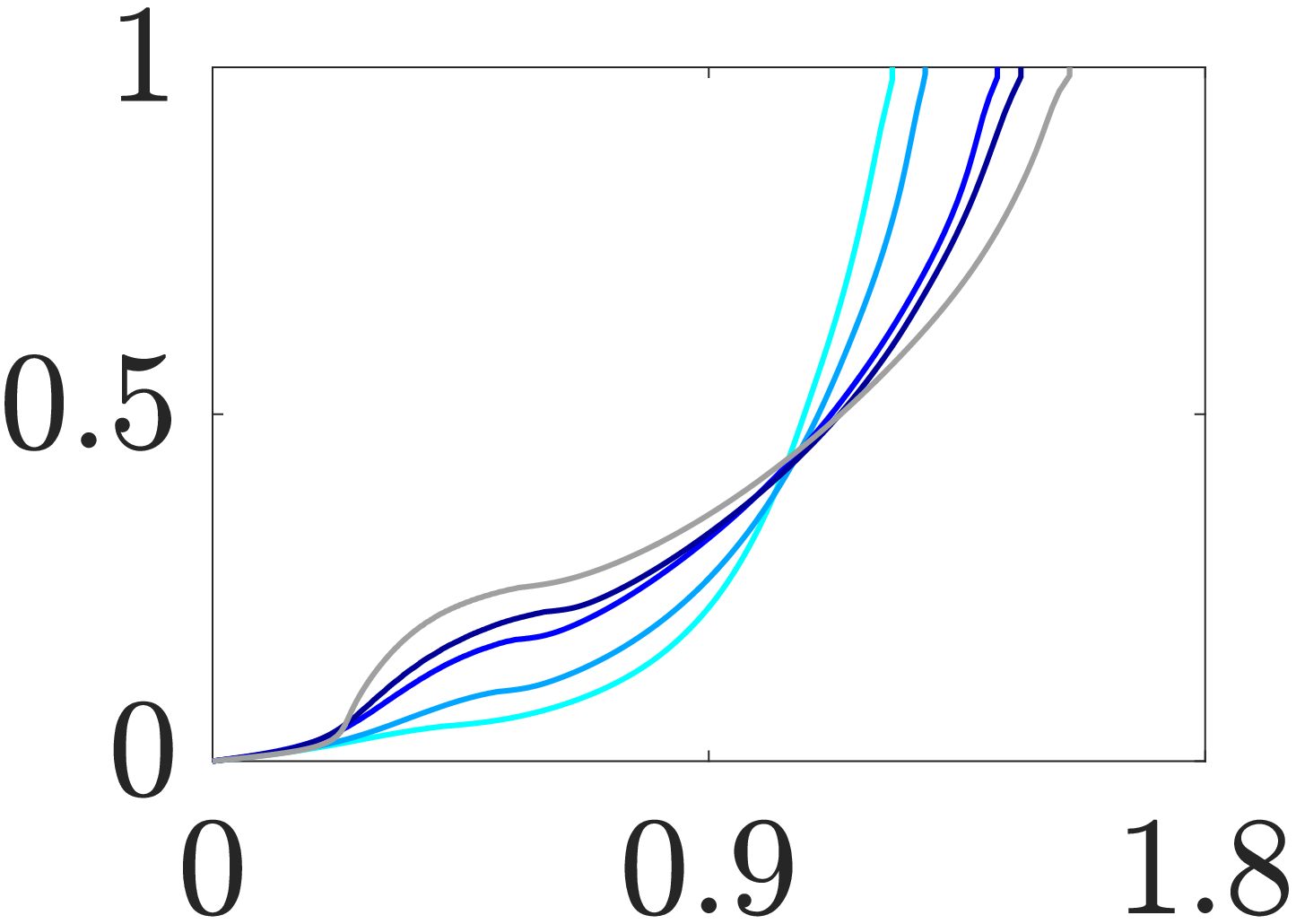}}}}
             } \\
  \caption{Mean velocity profiles for the canopies inclined with the grain
           (a) and against the grain (b). The small 
           inset on the left, top corner of each plot shows an enlarged
           view that visualizes the mean velocity profiles along the whole
           channel depth. 
           The three symbols indicate: the location of the first 
           inflection point (\protect\markerstar), the location of the virtual 
           origin (\protect\markercirc) and the location of the canopy tip, 
           i.e. the second inflection point, (\protect\markersqua).}
  \label{fig:velprofs}
\end{figure}

First, we analyse the effect of the canopy inclination on the mean velocity
profile; the aim is to show that positive and negative inclination angles
$\theta$ have a very different impact on the bulk statistics of the flow, thus proving
that a simple prediction of the canopy flow regimes based on the solidity 
parameter $\lambda$ is not always meaningful. \Cref{fig:velprofs} shows the mean 
velocity profiles close to the canopy tips, marked by the symbol
(\protect\markersqua), with the left panel showing the canopies inclined with
the grain $\theta>0$, and the right panel showing the cases inclined against the 
grain $\theta<0$. All the cases considered show the typical convex region of the velocity 
profile within the canopy layer, confined by two inflection points that arise
from the discontinuity of the drag at the canopy edge as a consequence of the 
sudden end of the stems (upper inflection point), and as a result of the 
merging of the boundary-layer in the proximity of the wall and the convex region
(lower inflection point, $y_{lip}$).
A higher or lower curvature of the convex region depends on the penetration level of the 
flow above the canopy. Since \cref{fig:velprofs} shows qualitatively that 
the mean velocity profile within the canopy, i.e. below the marker (\protect\markersqua), 
has a milder convexity when the filaments are inclined against the grain 
($\theta<0$), we expect to find a higher penetration of the outer flow
in these scenarios. 
\begin{figure}
  \centering
  \subfloat[]{\includegraphics[width=0.49\linewidth]{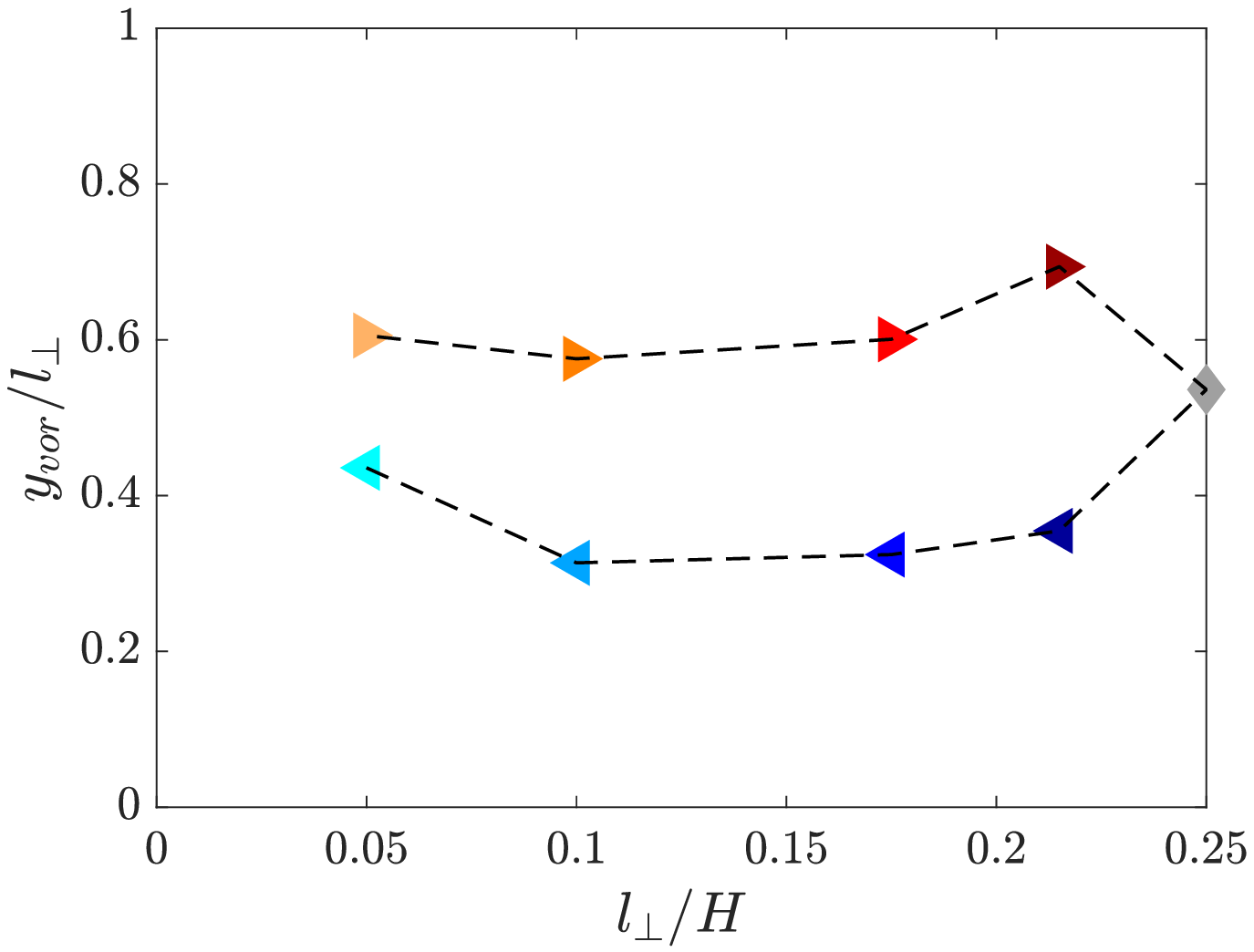}}
  \subfloat[]{\includegraphics[width=0.49\linewidth]{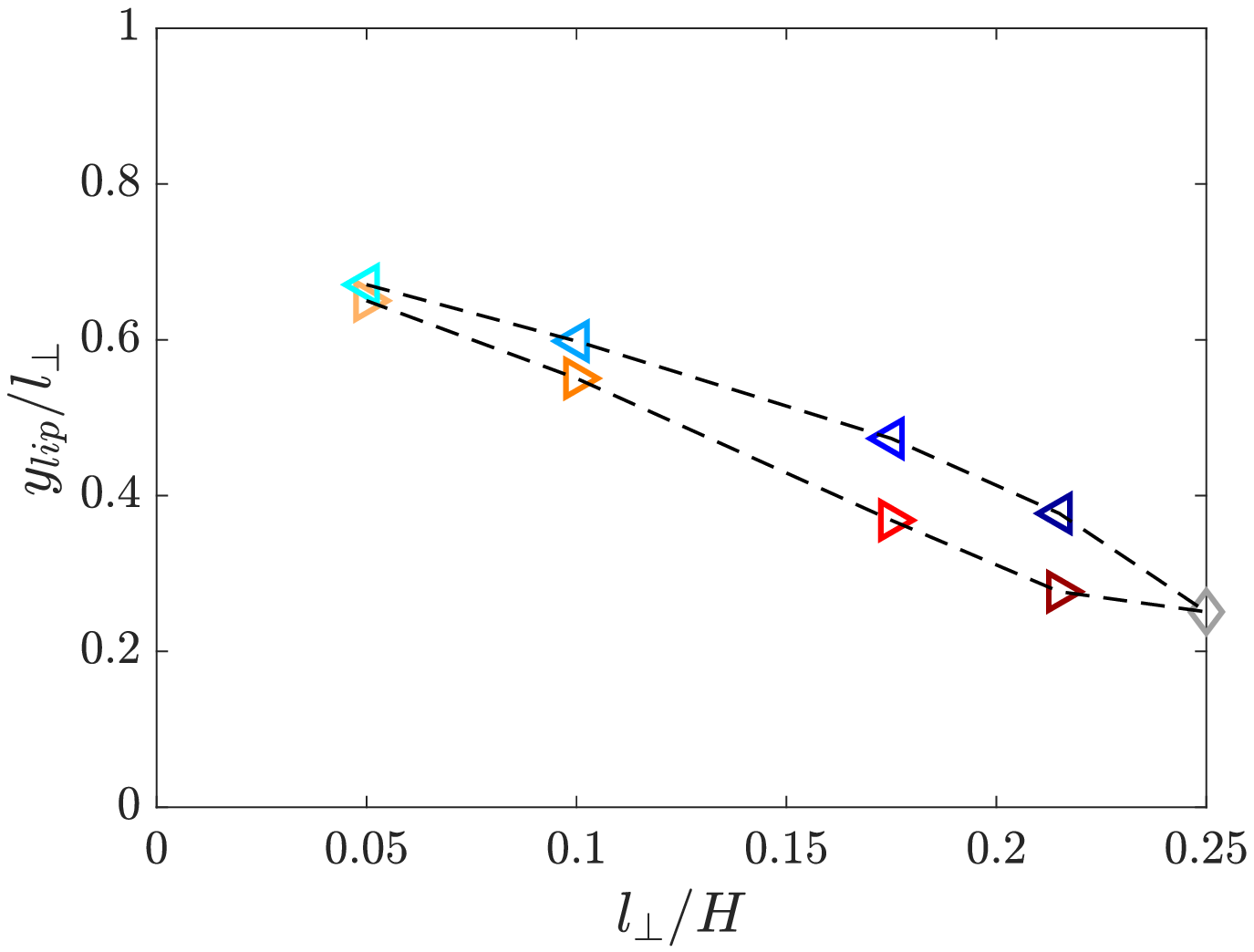}}
  \caption{(a) Hysteresis of the location of the virtual origin with the
               canopy inclination angle, represented using the wall-normal 
               projection of the canopy layer.
           (b) Hysteresis of the location of the inner inflection point with 
               the canopy inclination angle, represented using the wall-normal 
               projection of the canopy layer.
           The symbols (\protect\markerrtria) refer to the canopies inclined with
           the grain, while the symbols (\protect\markerltria) refer to the canopies
           inclined against the grain. 
           The symbol (\protect\markerdiam) refers to the wall-normally 
           mounted canopy. The colour scheme is the one used in \cref{tab:par}.}
  \label{fig:remarkable_points2}
  \end{figure}
To verify this, we analyse the locations of the most significant points of 
the mean velocity profile (the markers in \cref{fig:velprofs}), i.e. 
the already mentioned two inflection points enclosing
the convex region, and the virtual origin for the outer flow, $y_{vor}$, 
defined as the location of the effective wall for the outer boundary layer, 
that can be determined by enforcing the mean outer flow to take on a 
canonical logarithmic shape, as shown by \citet{MONTI2019}. 
In particular, we focus on the locations of the virtual origin and the 
inner inflection point in \cref{fig:remarkable_points2}, left 
and right panel respectively.
While the location of the inner inflection point only slightly differ between 
the positive and negative values of $\theta$, the location of the virtual
origin clearly reveals the large difference of the 
penetration of the outer flow in the two opposite configuration. 
The left panel of \cref{fig:remarkable_points2}
shows quantitatively the intuitive effect of the inclination: 
the negative angles promote the penetration of the outer flow structures within 
the canopy, while the positive ones shelter the layer from the large vortices.
\begin{figure}
  \centering
  \subfloat[]{\includegraphics[width=0.49\linewidth]{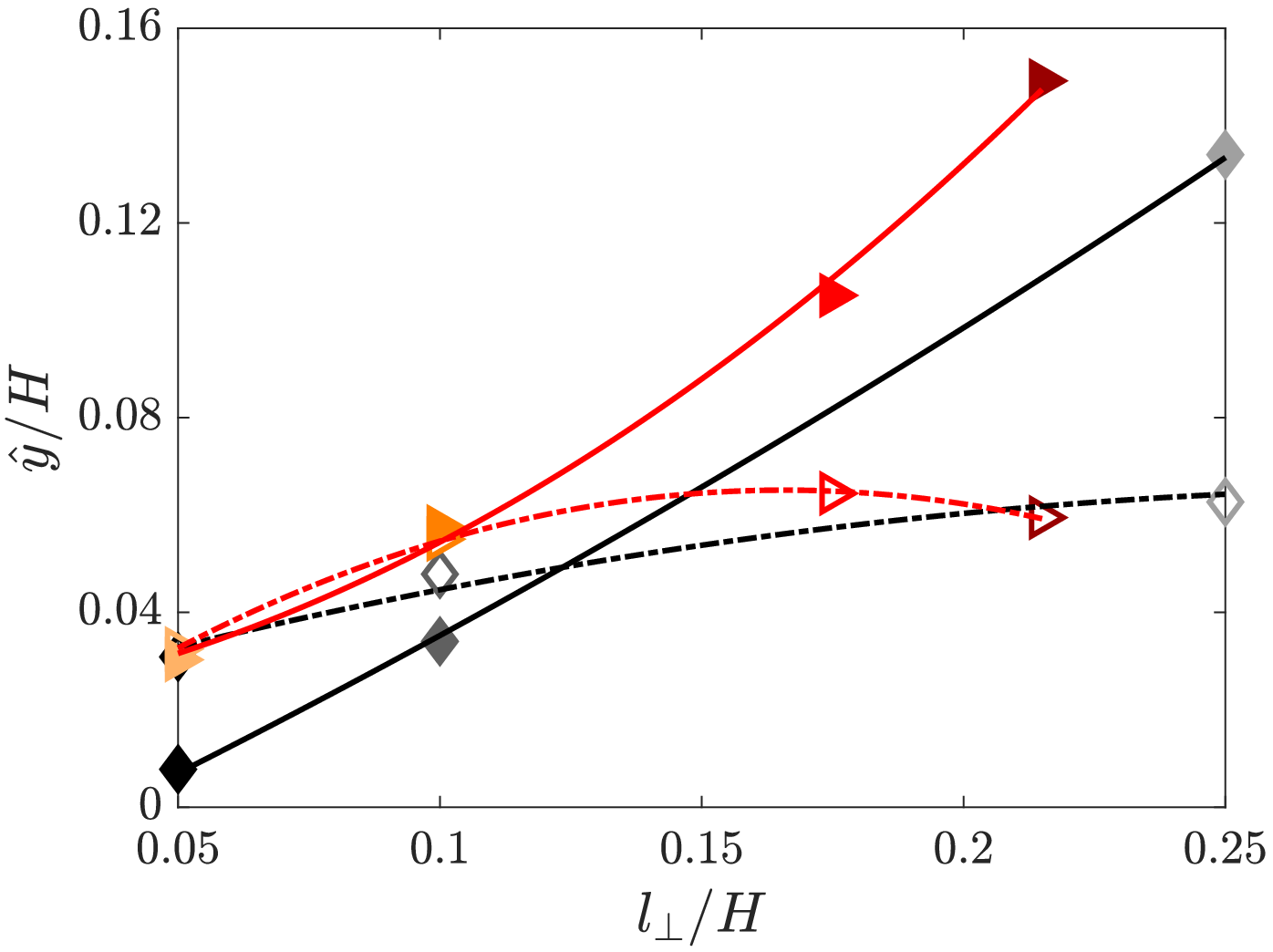}}
  \subfloat[]{\includegraphics[width=0.49\linewidth]{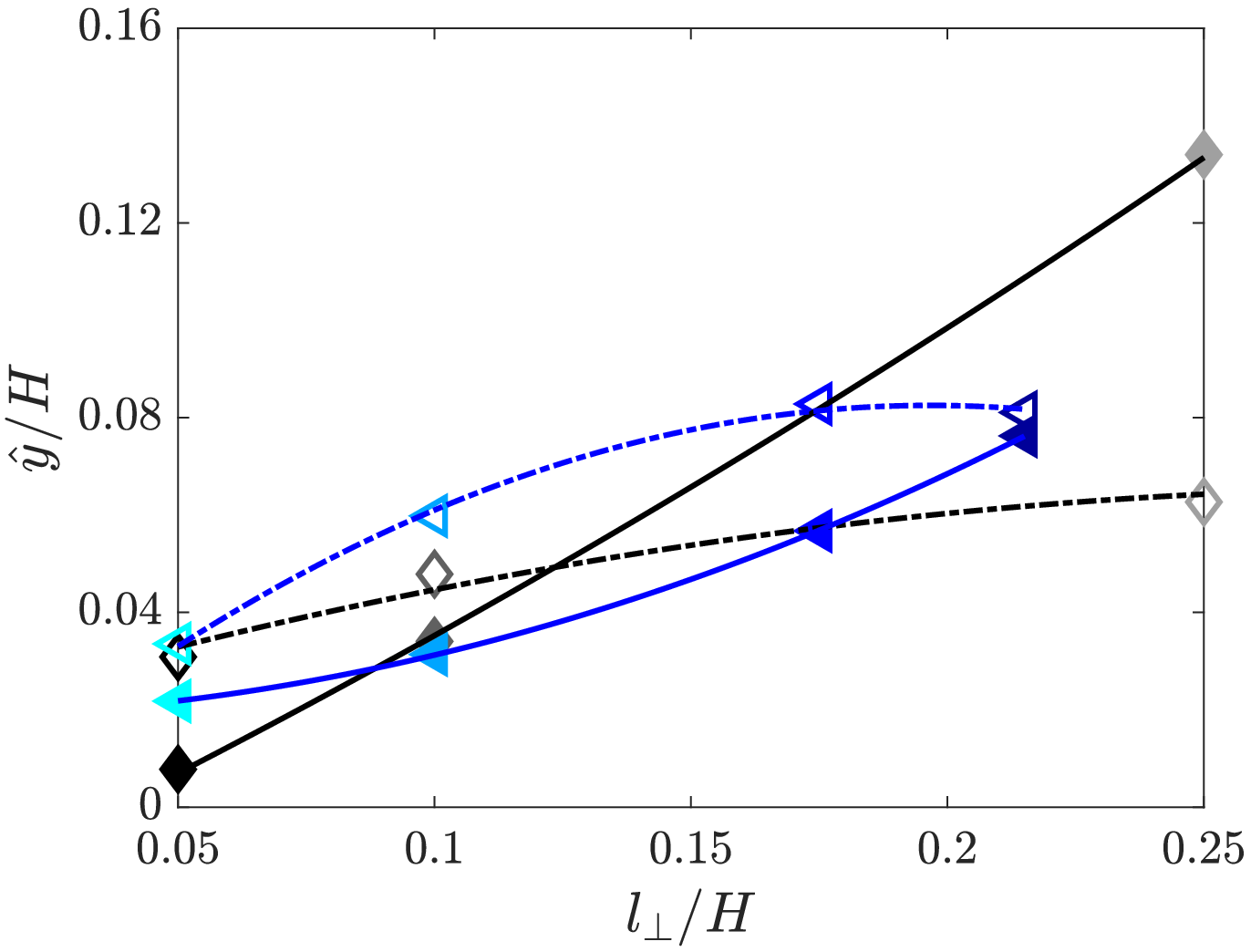}}
  \caption{Locations of the inner inflection points (empty symbols) and 
           of the virtual origins (filled symbols) along the wall-normal
           projection of the canopy. The abscissa indicates the canopy-case
           considered based on the wall-normal canopy projection $l_\bot$. 
           The lines represent the polynomial
           fits passing by the virtual origin (solid lines) and the inner
           inflection points (dashed lines). The red lines (a) refer 
           to the canopies inclined with the grain, while the blue lines (b) 
           refer to the ones inclined against the grain. 
           The black lines indicate the wall-normally mounted canopies data 
           from \citet{MONTI2020}. 
           The crossing point of the couple of lines of the same color indicates 
           the transition from a quasi-sparse to a dense regime.}
  \label{fig:remarkable_points}
  \end{figure}
To conclude the analysis on the features of the mean velocity profile,
we consider the relative location of the virtual origin and the inner
inflection point. According to \citet{MONTI2020}, these quantities
define the transition from a quasi-sparse canopy flow to a dense one, since
the location of the lower inflection point marks the end of the boundary-layer
close to the canopy bed and the virtual origin marks a lower limit for the 
outer flow; therefore, a crossing between these two points means that the two
regions overlap and the transitional to sparse scenario described by 
\citet{NEPF2012} appears.
\Cref{fig:remarkable_points} reports the trends of the locations of the virtual 
origins (solid lines) and the inner inflection points (dash-dotted lines)
compared to the data obtained for the wall-normally mounted canopies (black lines) 
taken from \citet{MONTI2020}.
In particular, the figure shows that, when the filaments are positively 
inclined (panel (a), red lines), the intersection between the two curves moves closer 
to the wall than when wall-normally mounted, meaning that a dense-like regime develops even for very low 
values of solidity. On the contrary, when the filaments have a negative inclination
(panel (b), blue lines), the intersection point is shifted towards the canopy tip, in
a sparse-like canopy fashion. Therefore, two canopies inclined by an opposite angle
$\theta$ but having the same solidity $\lambda$ may behave in a completely
different manner. This is a clear indication that a classification of the canopy
regimes based on the solidity alone, as it has been largely done in the literature, i.e.
$\lambda > 0.1$ \citep{NEPF2012,POGGI2004a,BRUNET2020}, can actually
be misleading.

\begin{figure}
  \centering
  \setbox1=\hbox{%
  \includegraphics[width=0.42\linewidth,height=4cm]{example-image-b}}
  \subfloat[]{\includegraphics[width=0.49\linewidth]{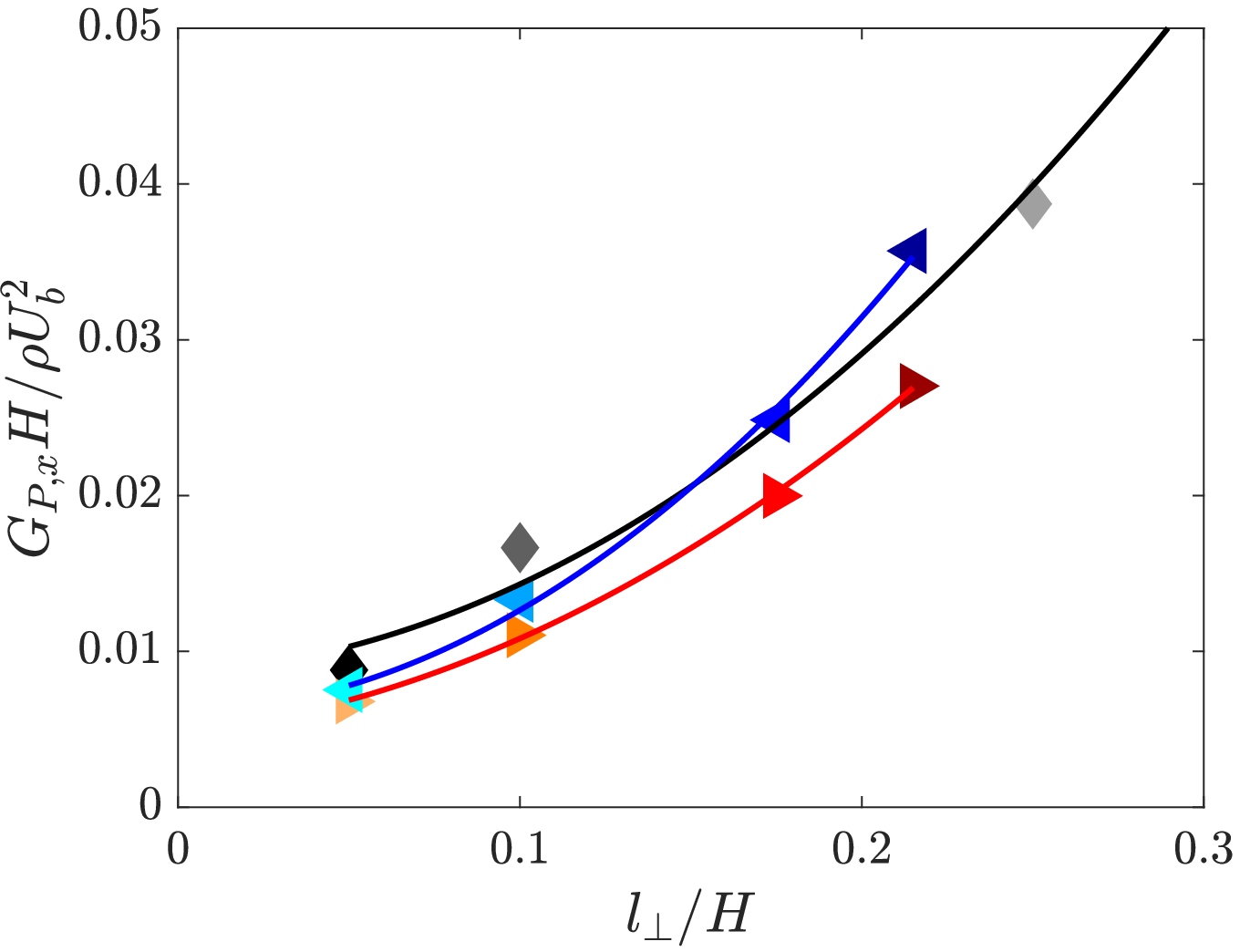}
              \llap{\makebox[\wd1][l]{\raisebox{2.7cm}{%
              \includegraphics[width=0.20\linewidth,height=2.0cm]
              {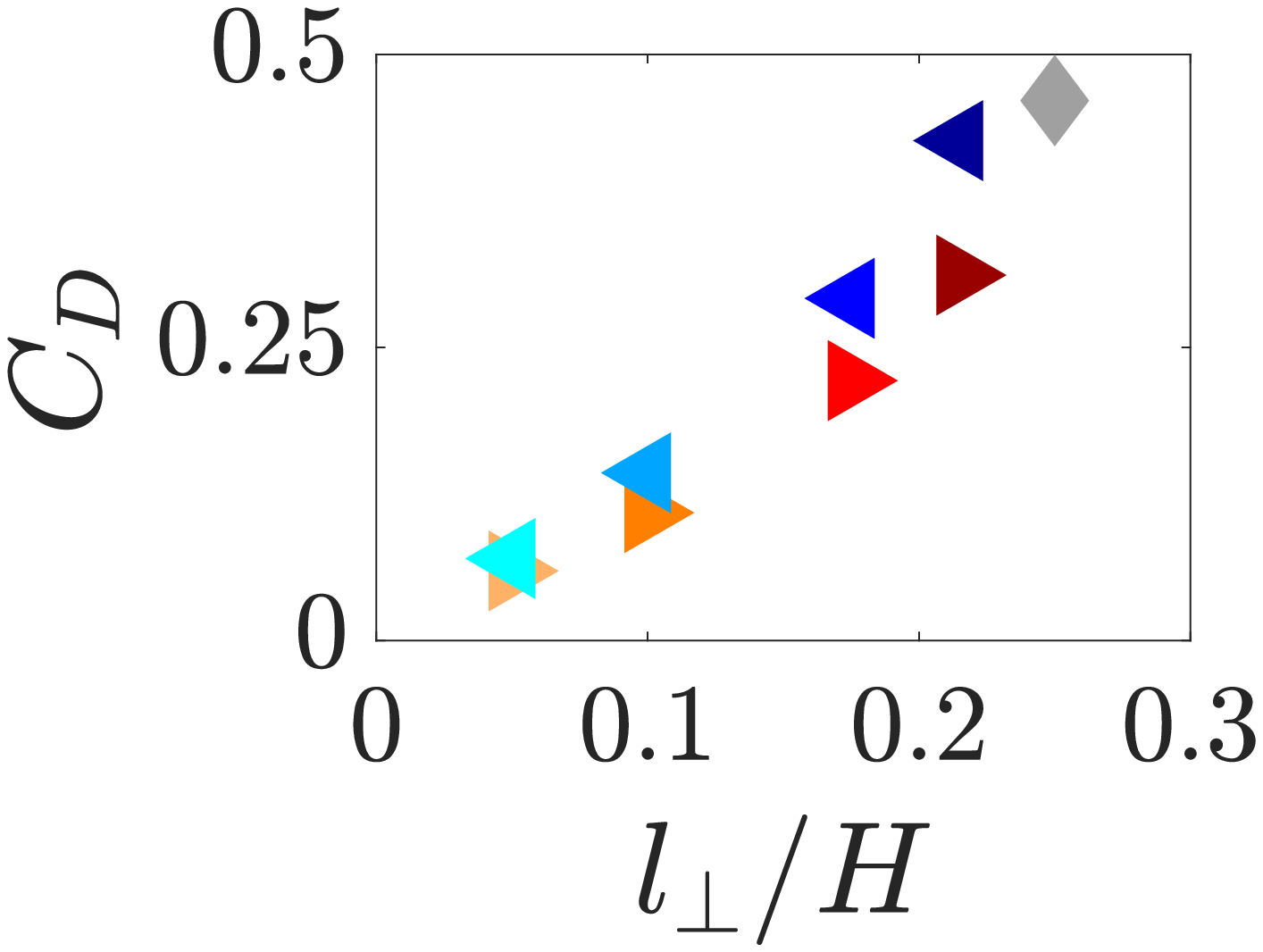}}}}
             }
  \subfloat[]{\includegraphics[width=0.49\linewidth]{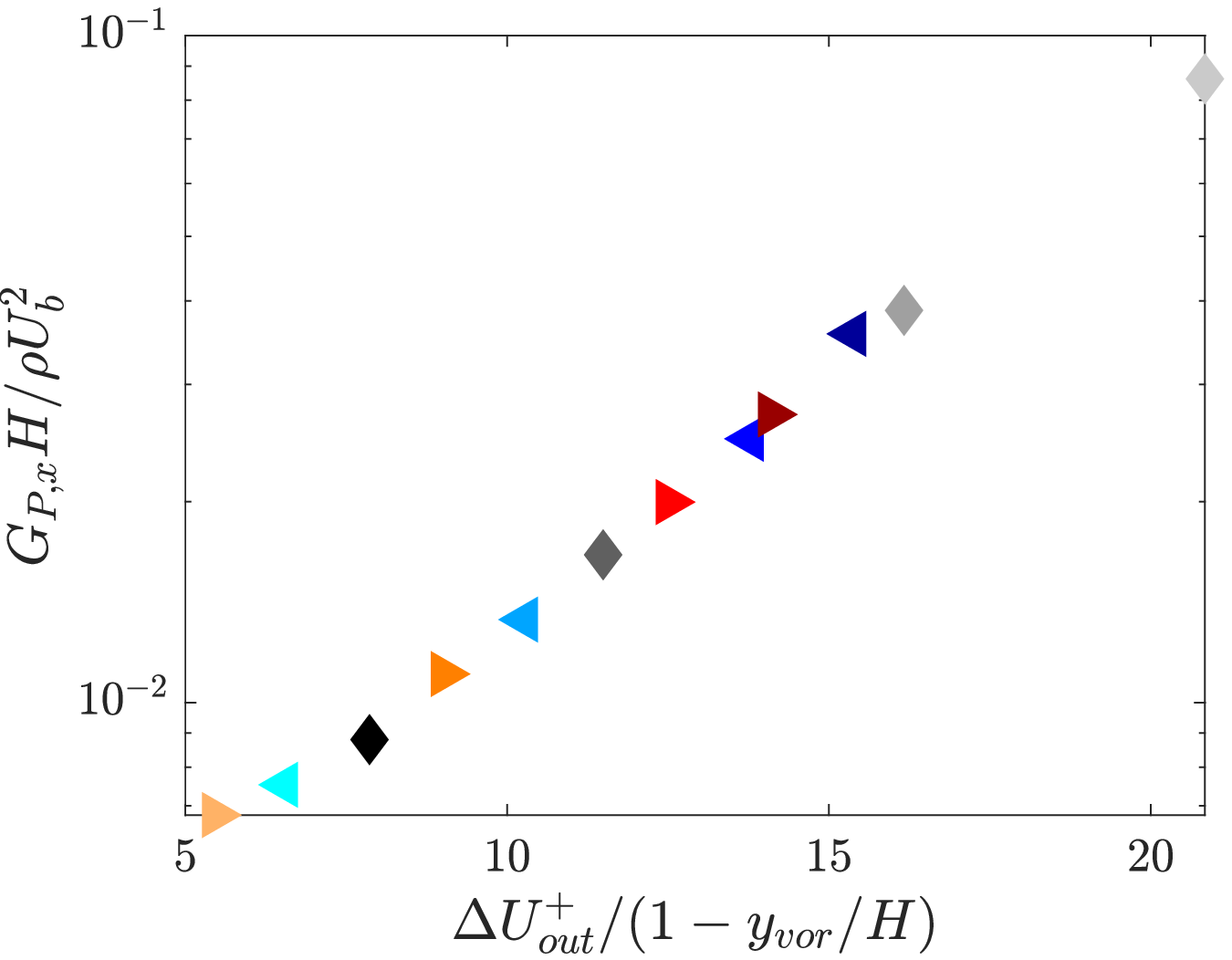}}
  \caption{Non-dimensional mean pressure gradient versus the height 
           of the canopy layer $l_\perp$ (a) and the
           roughness function $\Delta U^+_{out}$ related to the outer
           boundary-layer developed starting from the location of the 
           virtual origin, rescaled by the fraction of the domain 
           occupied by the latter (b).  
           The small inset on the left, top corner of panel (a) shows the
           drag coefficient $C_D$ provided by the canopies analysed in this
           work versus the height of the canopy layer $l_\perp$. 
           The shapes and colours of the symbols are the same adopted in
           \cref{fig:remarkable_points2}. 
           The grey-scale diamonds refer to the cases analysed in
           \citet{MONTI2020}.}
  \label{fig:dudpx}
  \end{figure}

The shape of the mean velocity profile described above is caused by the 
resistance that the flow encounters when flowing through the stems of the 
canopy layer. The drag exerted by the canopy can be quantified by the mean 
pressure gradient needed to move the flow. \Cref{fig:dudpx} (panel a) shows 
the mean pressure gradient as a function of the wall-normal projection of the canopy 
layer $l_\bot$.
As reference, the values of the wall-normally mounted canopies
studied in \citet{MONTI2020} have been added to the graph (diamonds, grey-scale).
The figure shows that a positive inclination always reduces the drag compared
to negatively inclined canopies, suggesting that the sheltering effect
is beneficial in these terms. The left panel of \cref{fig:dudpx} shows also
that for short canopy layers, i.e. $l_\bot/H < 0.15$, the inclination (negative and 
positive) is also beneficial compared to the 
wall-normally mounted canopies having the same frontal area (black solid line). 
However, when increasing $l_\bot/H$, only the cases with negative $\theta$ results to be 
drag increasing, (blue solid line) in \cref{fig:dudpx}.
To corroborate this result, the drag coefficient provided by the canopy layer,
\begin{equation}
    C_D = \dfrac{2 \mathbb{D}}{\rho U_b^2 H L_z},
    \label{eq:cd}
\end{equation}
is shown in the inset of panel (a) of \cref{fig:dudpx}. In \cref{eq:cd},
$\mathbb{D}$ is the integral drag force provided by the whole canopy,
$\rho$ is the density of the fluid, $U_b$ is the bulk velocity and $H\times L_z$
is the frontal area of the whole computational domain. Note that the trend of 
the drag coefficient reflects the pressure gradient's one since the former is the
main contribution to the latter in a canopy flow.
These behaviours are mainly due to the type of turbulent structures forming 
and living within the canopy layer, as it will be shown later in the manuscript.

\begin{figure}
  \centering
  \subfloat[]{\includegraphics[width=0.49\linewidth]{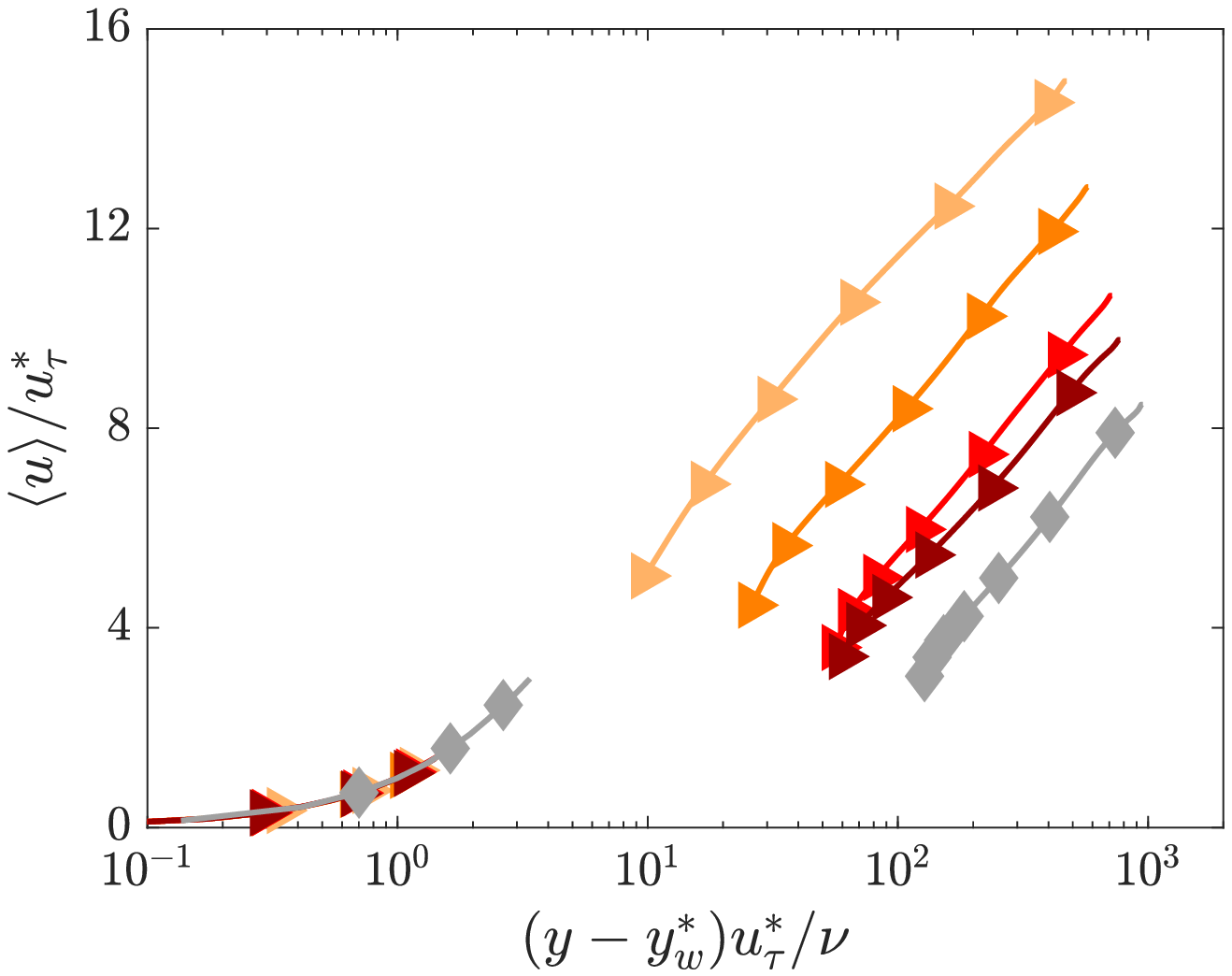}}
  \subfloat[]{\includegraphics[width=0.49\linewidth]{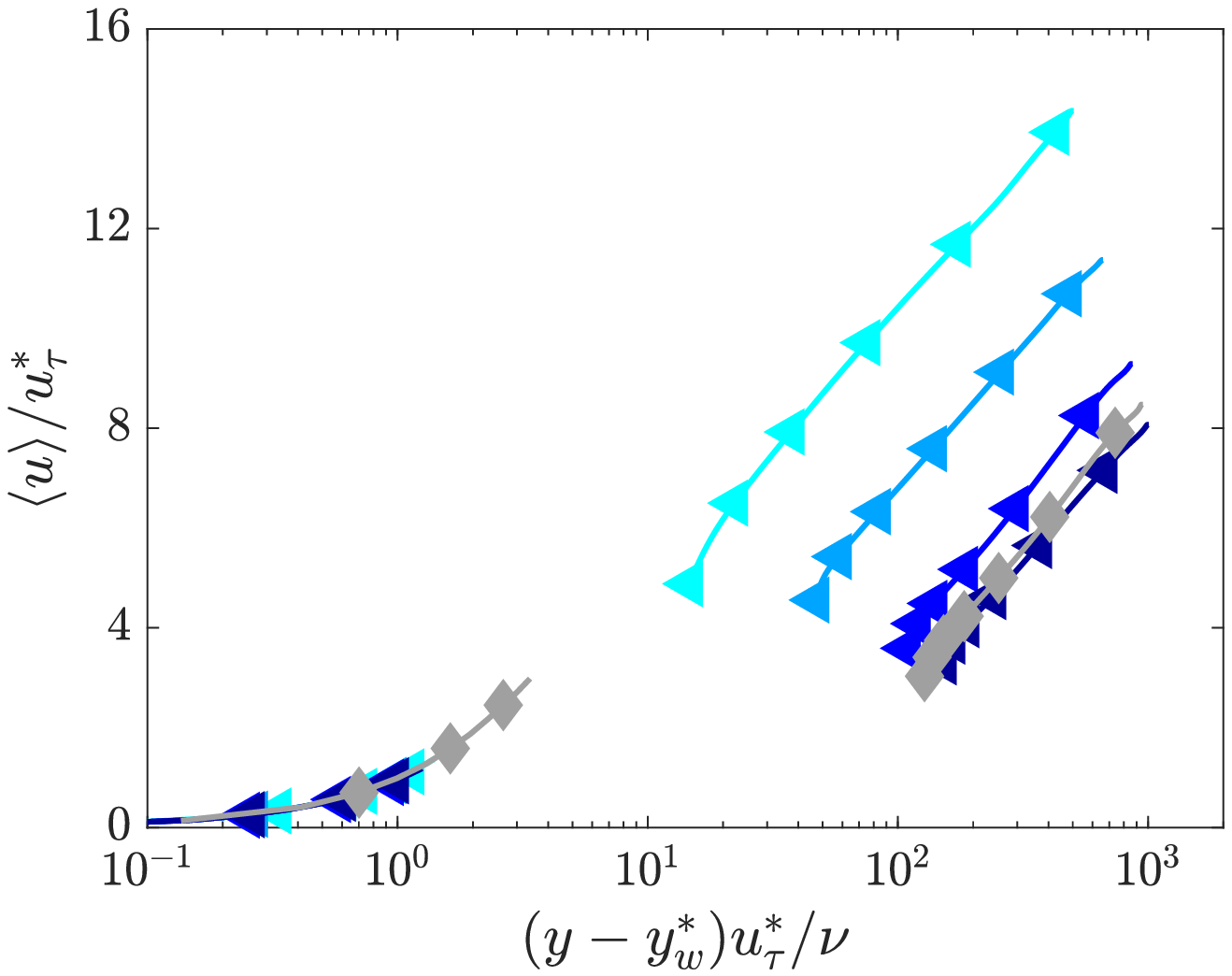}}
  \caption{Mean velocity profiles normalised using the viscous quantities
           defined in the inner layer, i.e.\ the friction velocity 
           $u_\tau^*=u_{\tau,in}$ computed at the bottom wall $y_w^*=0$,
           and the viscous quantities defined in the outer layer,
           i.e.\ the friction velocity $u_\tau^*=u_{\tau,out}$ computed 
           at the virtual origin $y_w^*=y_{vor}$.
           The abscissa represents the wall-normal coordinate rescaled with 
           the inner or outer wall units considering an origin located either 
           on the canopy bed or at the virtual origin $y_{vor}$. 
           The profiles in panel (a) refer to the canopies inclined with 
           the grain, while the profiles in panel (b) refer to the ones 
           inclined against the grain. The grey lines indicate the 
           wall-normally mounted canopies with $h/H=0.25$ from \citet{MONTI2020}.
           The shapes and colours of the symbols are the same adopted in
           \cref{fig:remarkable_points2}.} 
  \label{fig:dulog}
  \end{figure}

The right panel of \cref{fig:dudpx} 
shows that the pressure gradient is strictly related to the level of penetration 
of the outer flow within the canopy layer and the resistance felt by the outer 
flow caused by the tips of the canopy stems that can be thought of as elements 
of a distributed roughness. Indeed, 
the figure shows an almost exponential-law behaviour, independently of the canopy 
configuration, when the friction function
$\Delta U^+_{out}$ is normalised by the percentage of the open-channel domain 
occupied by the outer layer. 
The friction function is an offset
added to the logarithmic-law of the mean velocity in a turbulent boundary-layer
over a smooth wall to characterise the presence of roughness \citep{JIMENEZ2004} 
and, for the outer flow in a canopy, can be defined as in \citet{MONTI2019},
\begin{equation}
    \Delta U^+_{out} = \kappa^{-1} \log{\dfrac{(y-y_{vor})u_{\tau,out}}{\nu}} 
    + B - U^+_{out},
    \quad \quad \text{for} \,\,y > y_{vor}
    \label{eq:loglawout}
\end{equation}
where the first part of the right hand side is the logarithmic-law for a boundary-layer
over a smooth wall located at the virtual origin $y_{vor}$, and the second part is 
the mean velocity profile found in the outer flow in a canopy normalised by the friction
velocity $u_{\tau,out}$ computed at the virtual origin; in the equation, $\kappa=0.41$
is the K\'arm\'an constant, $\nu$ is the kinematic viscosity and $B=5.5$ is the 
additive constant for smooth walls. The mean velocity profiles obtained from
\cref{eq:loglawout} are shown in \cref{fig:dulog}, together with the inner part
of the profile normalised with the inner viscous units for the sake of completeness.
Relation \eqref{eq:loglawout} links together the
pressure gradient and the outer flow quantities only,  but the former 
is computed by considering the whole drag offered by the canopy, therefore 
the relation implicitly links the quantities of the outer and inner layers. 
The connection between the regions that characterize the canopy flows will be 
clarified in the following section, where a detailed description of the coherent 
structures will be given.


\subsection{Canopy structures}
Before describing the turbulent coherent structures that populate the 
canopy flows, we discuss the mean distributions of the velocity fluctuations 
since they contain important information concerning the turbulent properties 
of the flow.
\begin{figure}
  \centering
  \subfloat[]{\includegraphics[width=0.33\linewidth]{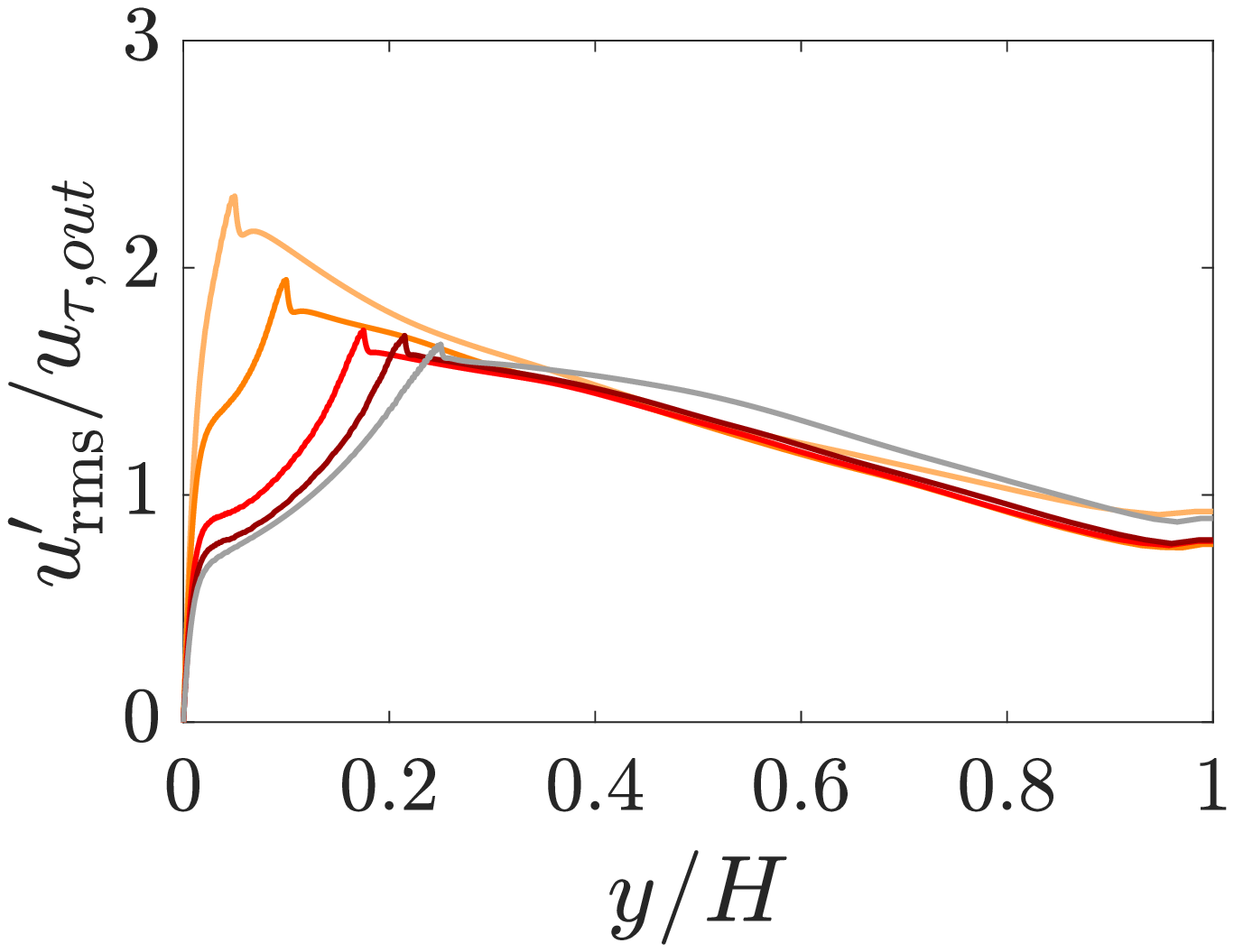}}
  \subfloat[]{\includegraphics[width=0.33\linewidth]{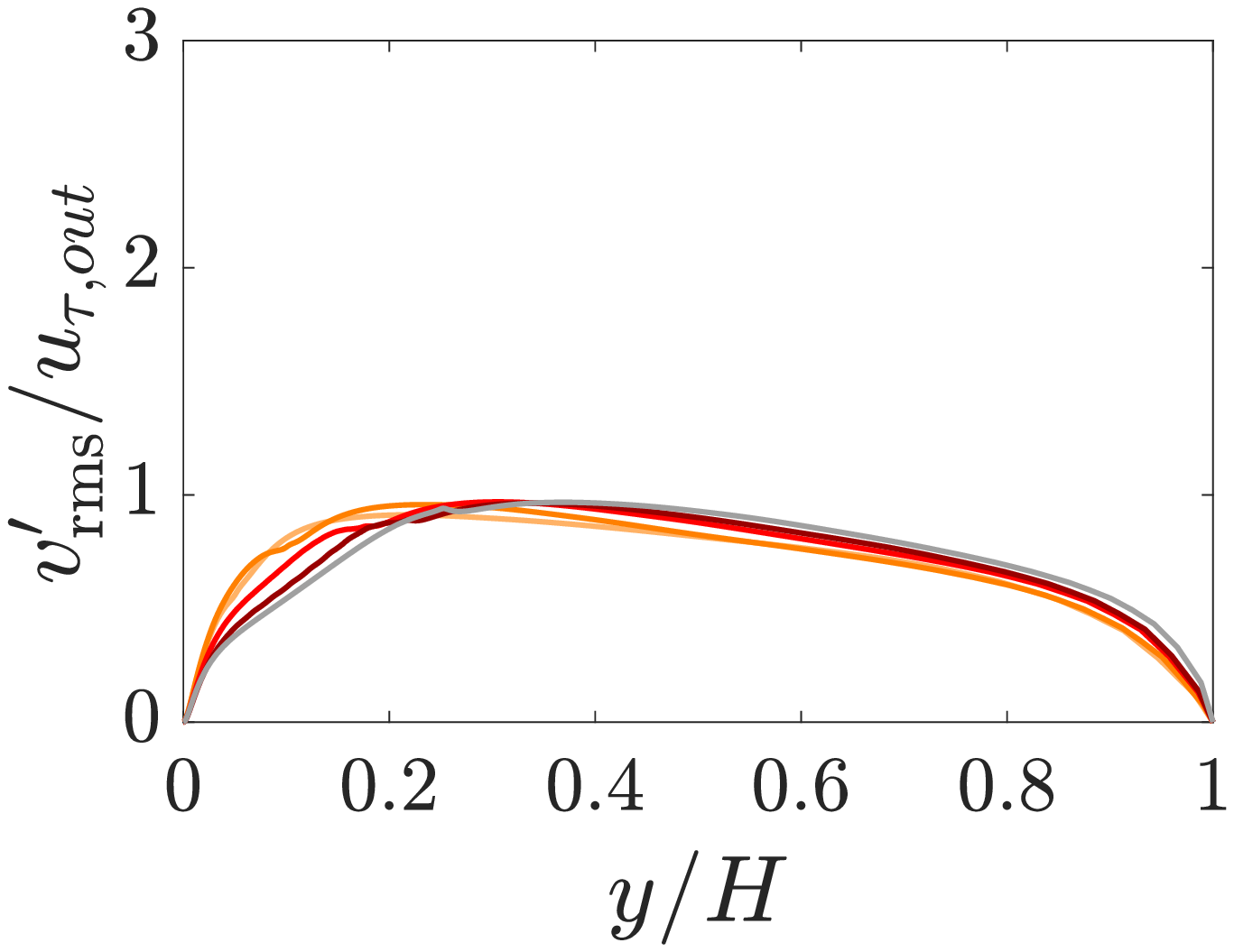}}
  \subfloat[]{\includegraphics[width=0.33\linewidth]{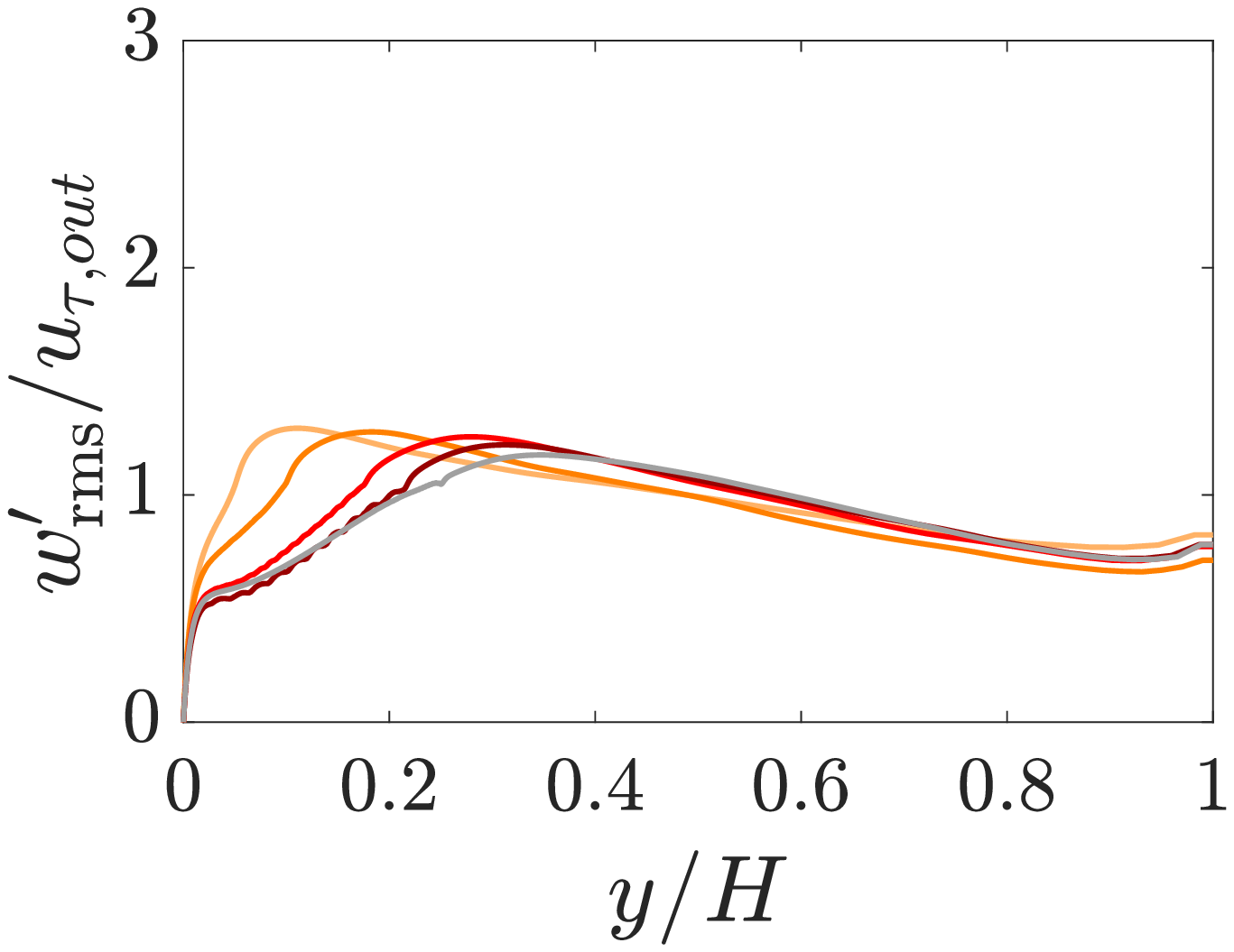}}\\
  \subfloat[]{\includegraphics[width=0.33\linewidth]{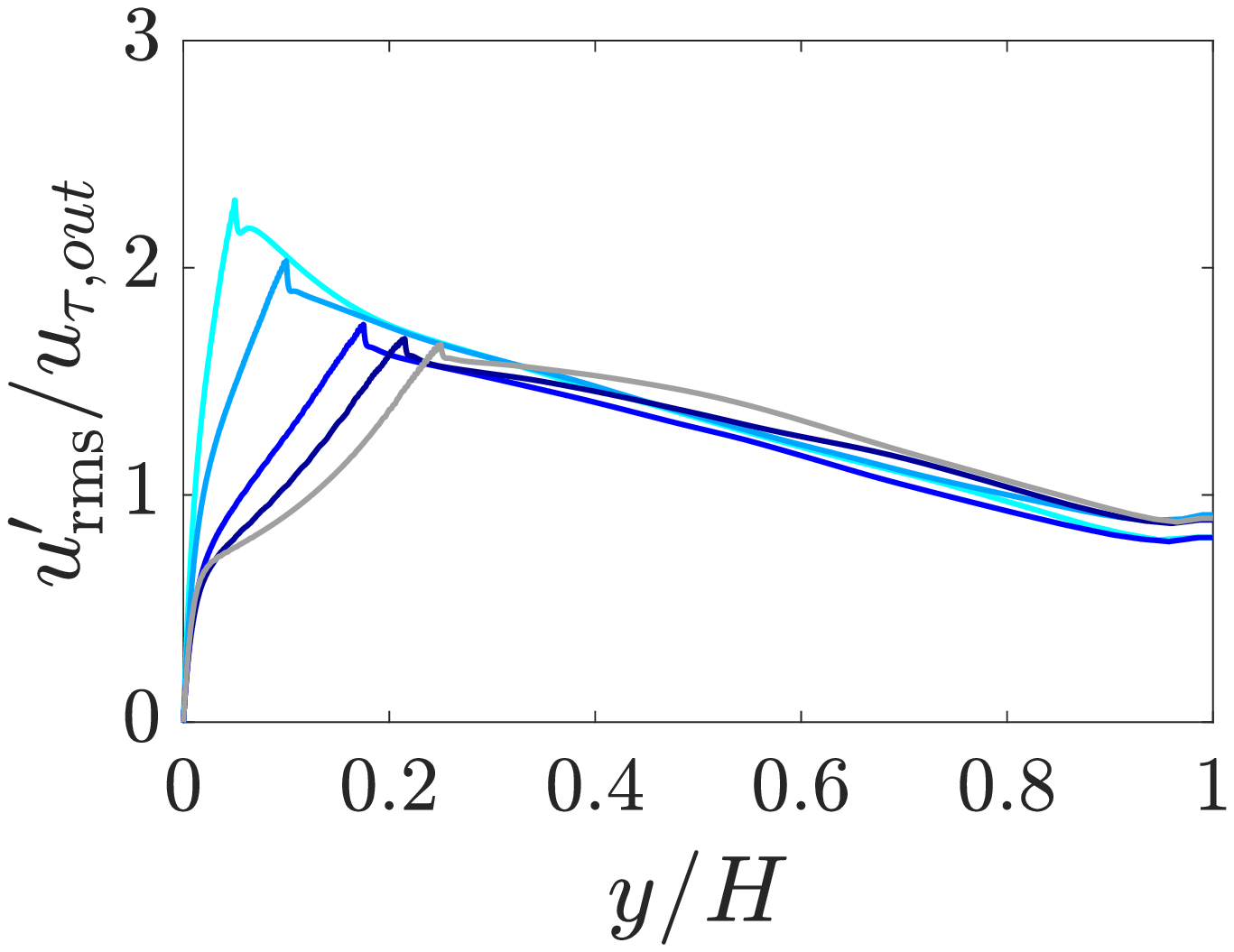}}
  \subfloat[]{\includegraphics[width=0.33\linewidth]{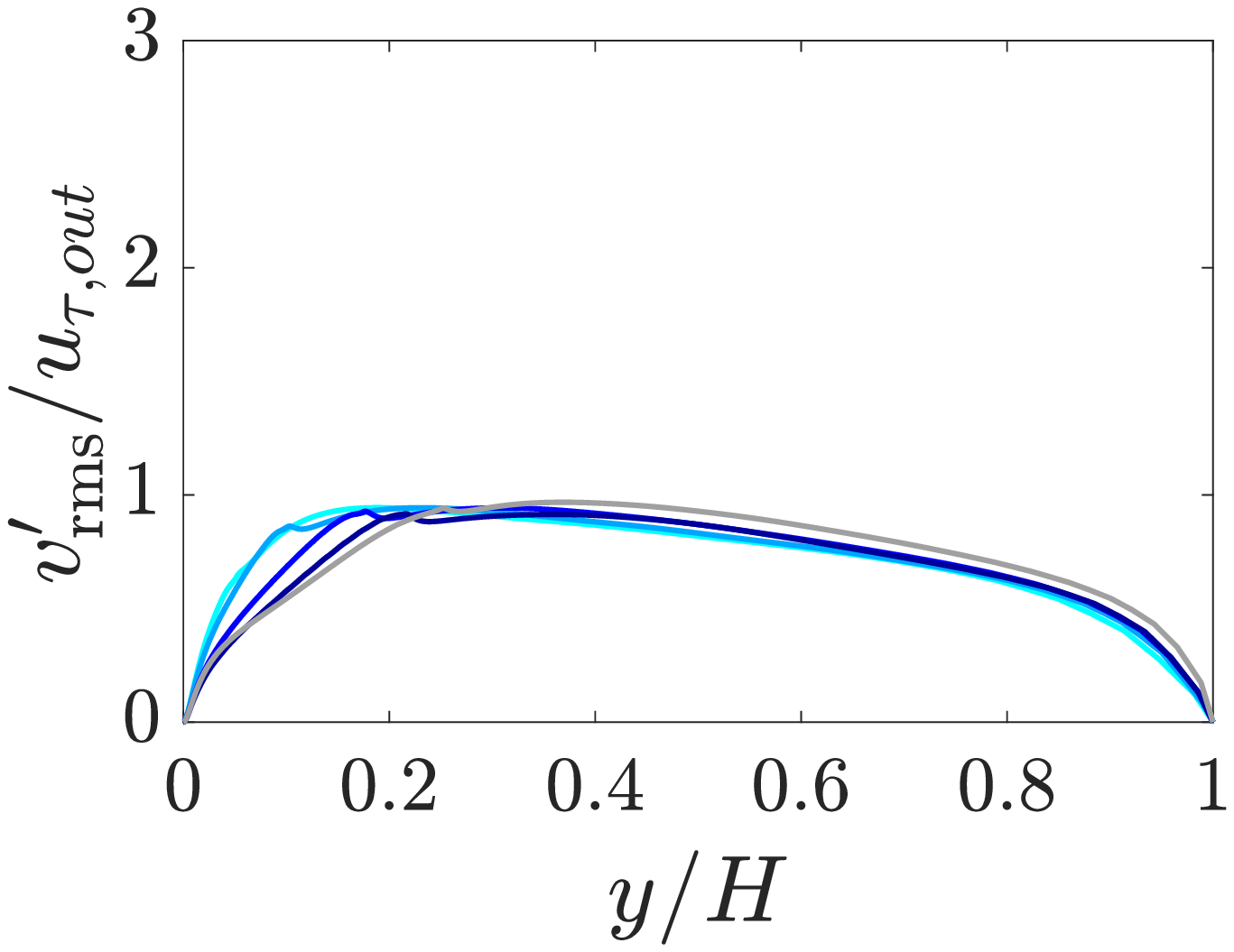}}
  \subfloat[]{\includegraphics[width=0.33\linewidth]{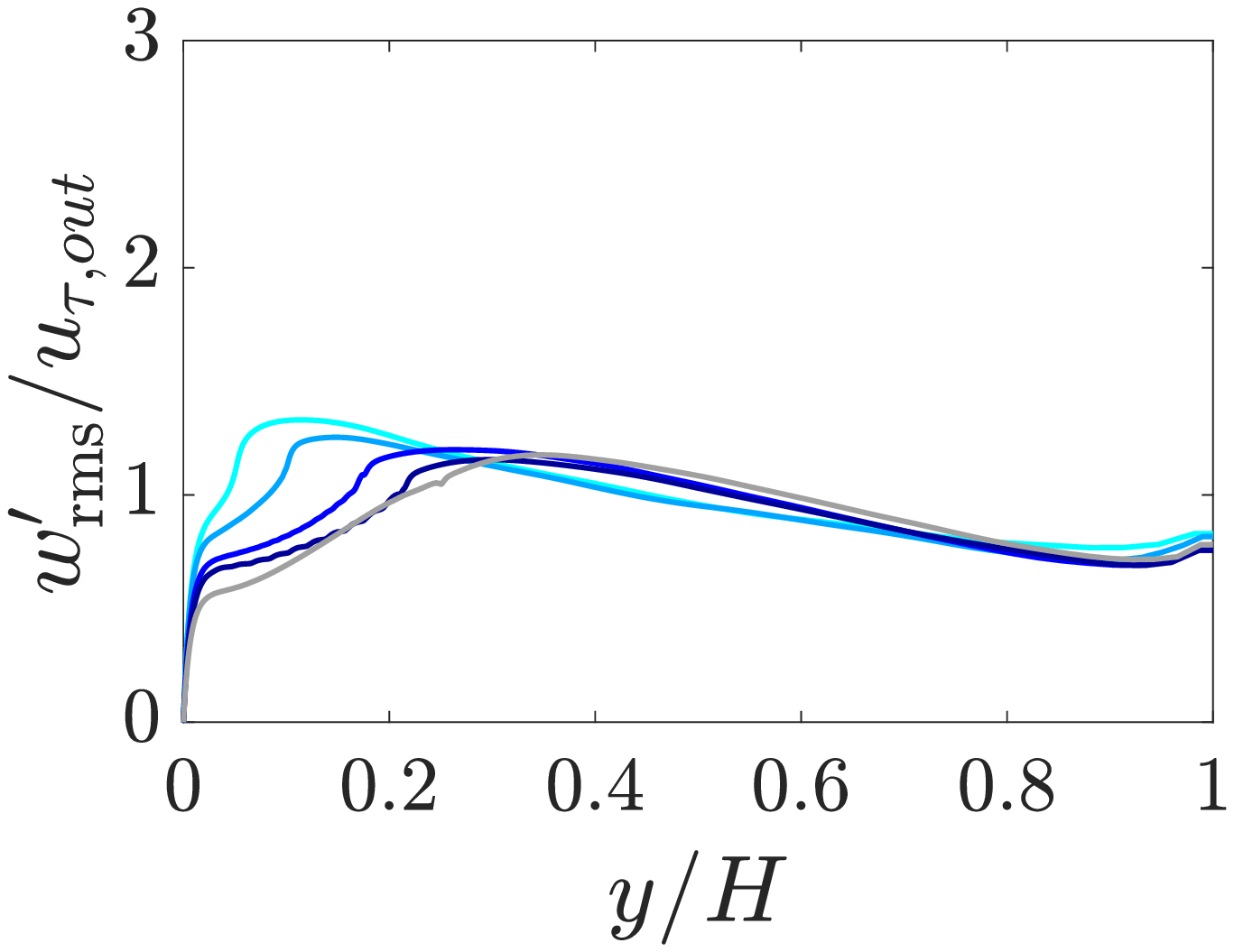}}\\
  \caption{Profiles of the RMS of the velocity fluctuations
           versus the wall-normal coordinate $y/H$. 
           The top row, panels (a)--(c), shows the distributions
           of the canopies inclined with the grain; the bottom row, panels (d)--(f),
           the distributions of the canopies inclined against the grain.
           From left to right, the columns of the figure show the streamwise,
           the wall-normal and the spanwise component.
           The distributions are normalized with the
           friction velocity computed at the virtual origin, $u_{\tau,out}$.
           Colours as in \cref{tab:par}.}
  \label{fig:secondOrderStatsNormOut}
  \end{figure}
\begin{figure}
  \centering
  \subfloat[]{\includegraphics[width=0.33\linewidth]{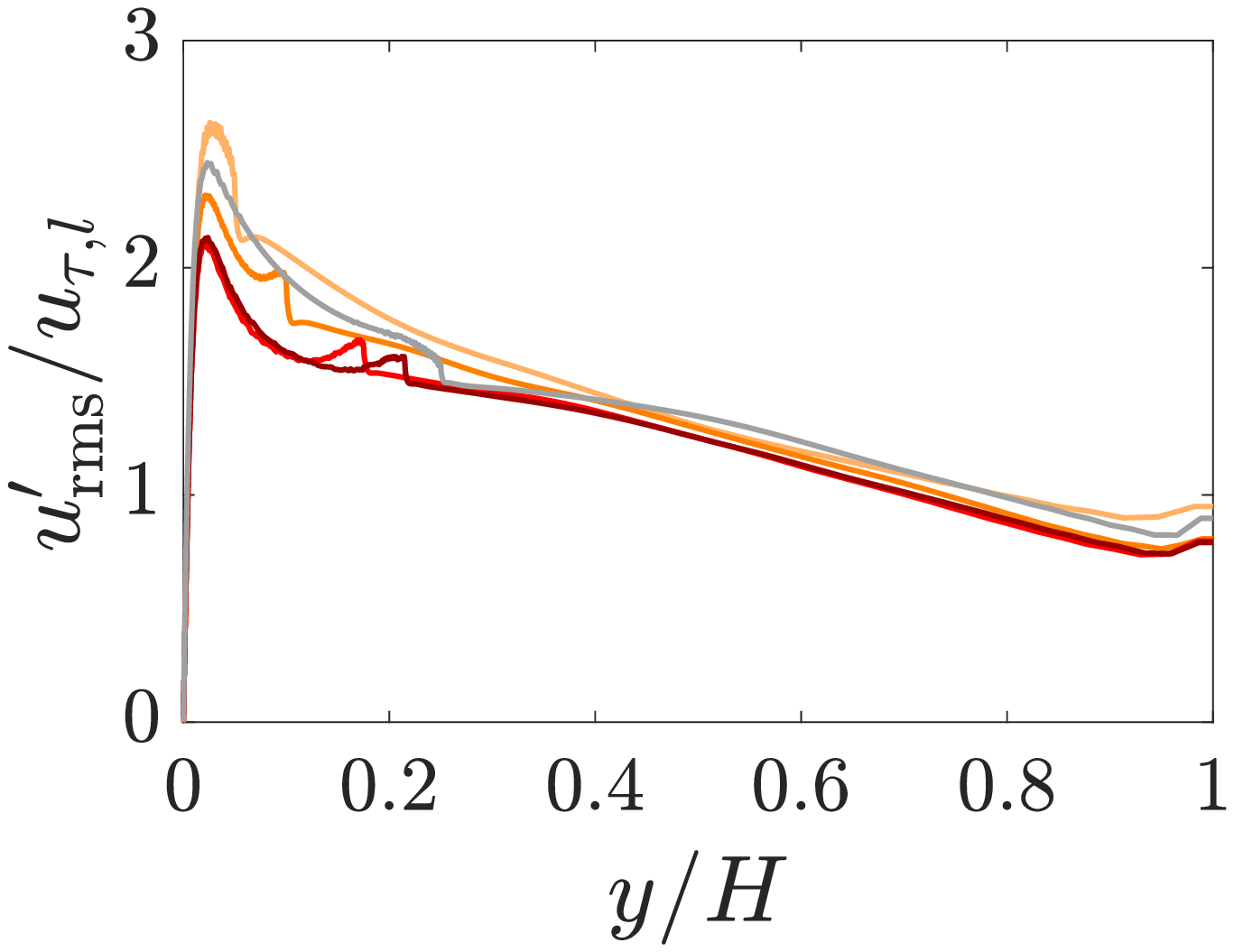}}
  \subfloat[]{\includegraphics[width=0.33\linewidth]{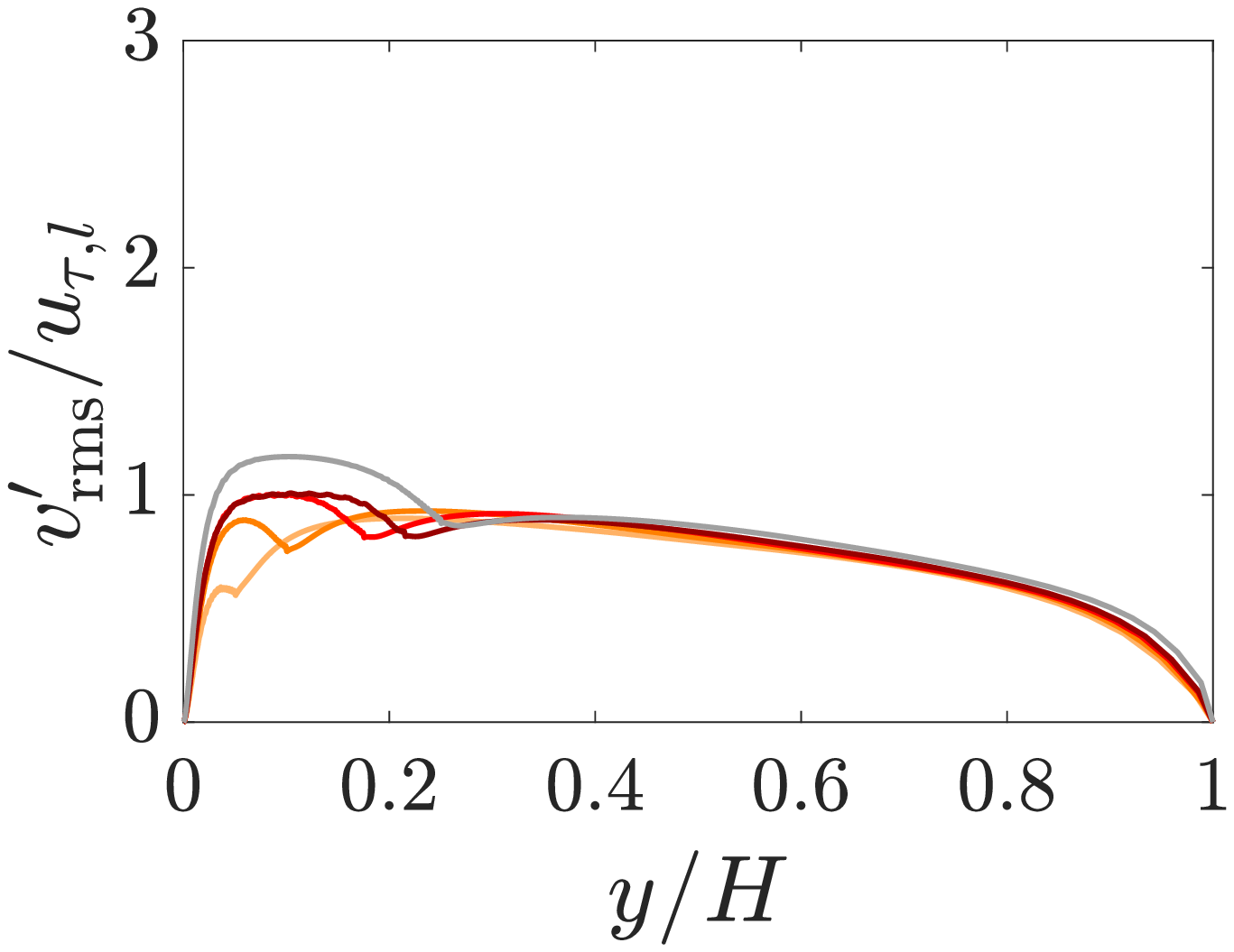}}
  \subfloat[]{\includegraphics[width=0.33\linewidth]{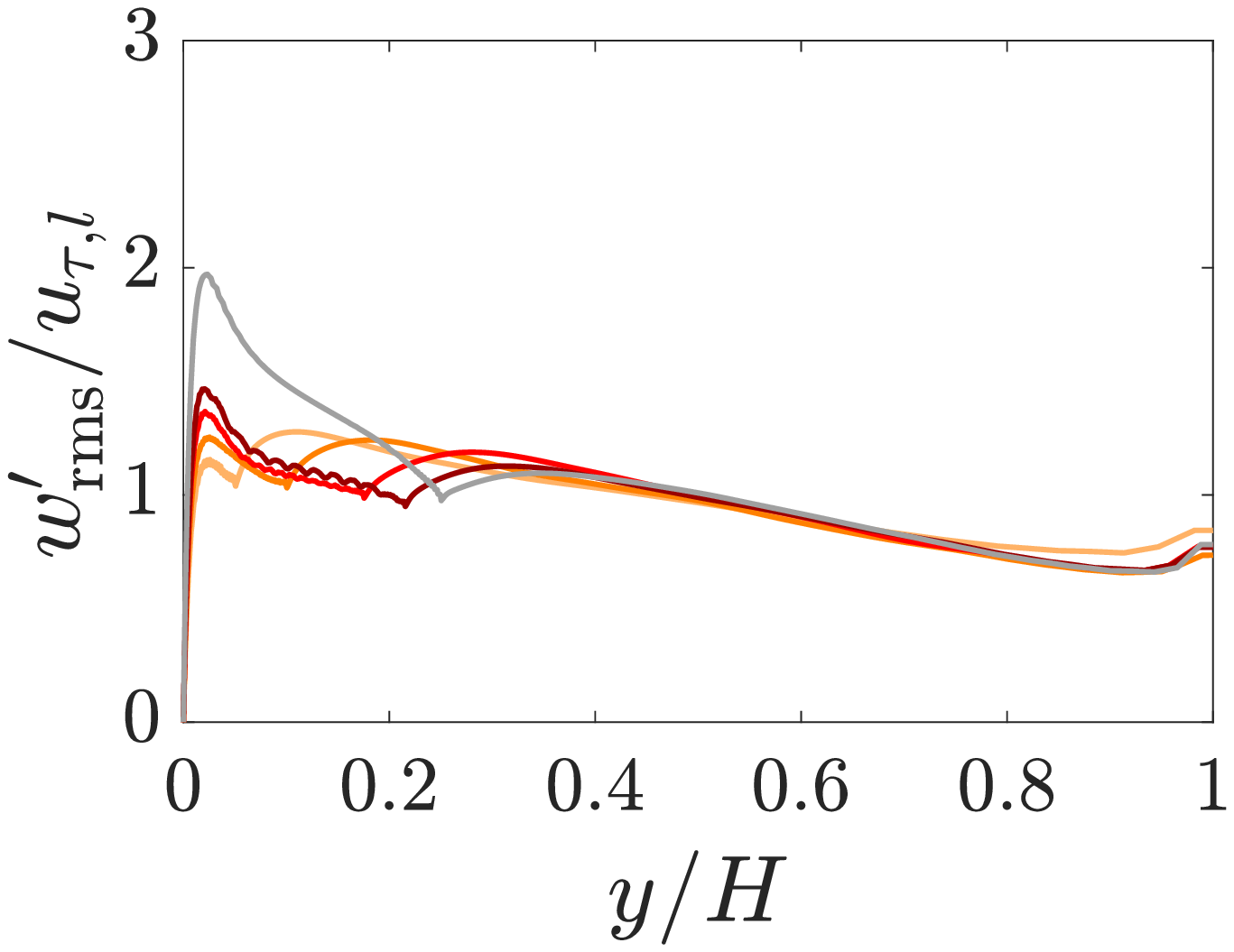}}\\
  \subfloat[]{\includegraphics[width=0.33\linewidth]{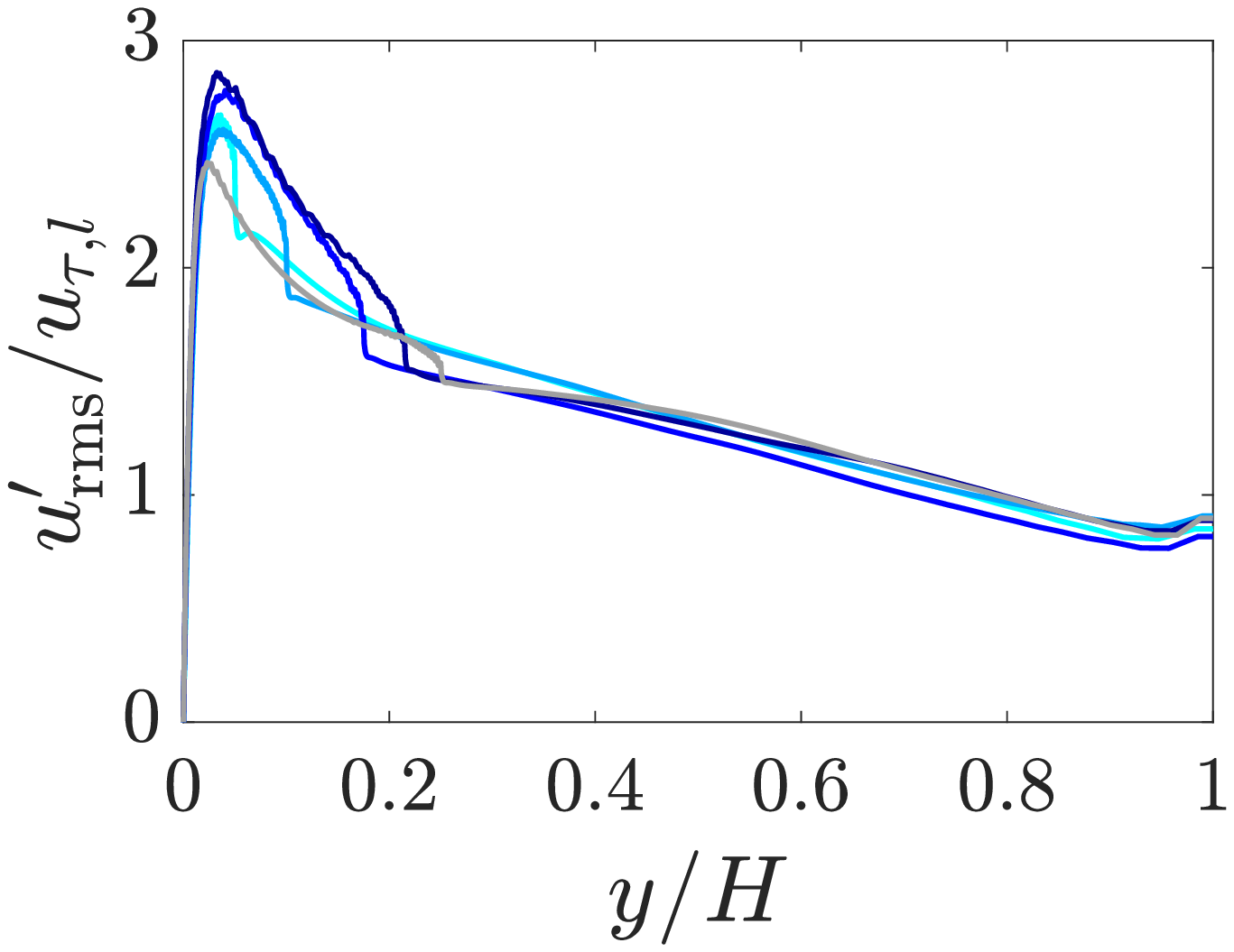}}
  \subfloat[]{\includegraphics[width=0.33\linewidth]{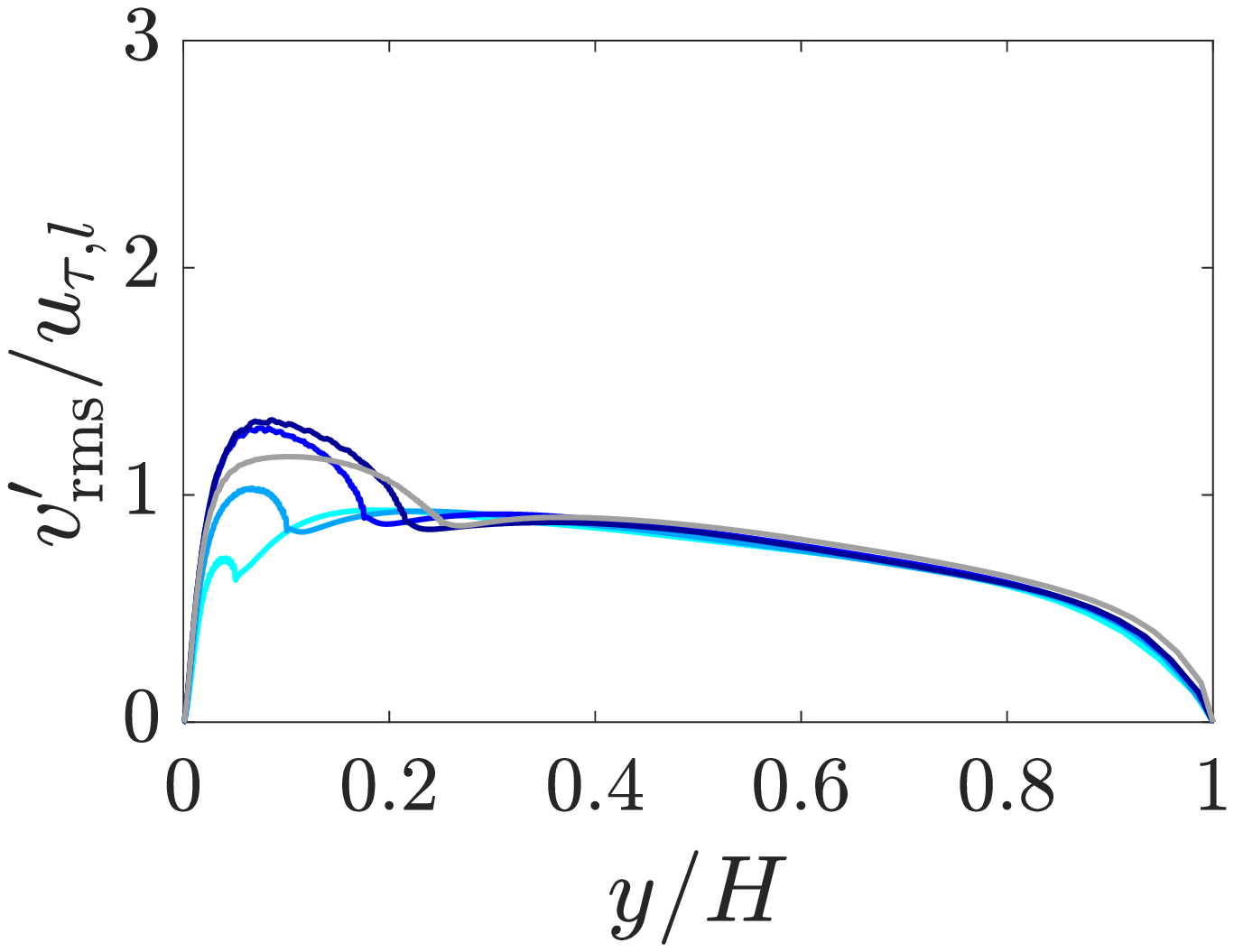}}
  \subfloat[]{\includegraphics[width=0.33\linewidth]{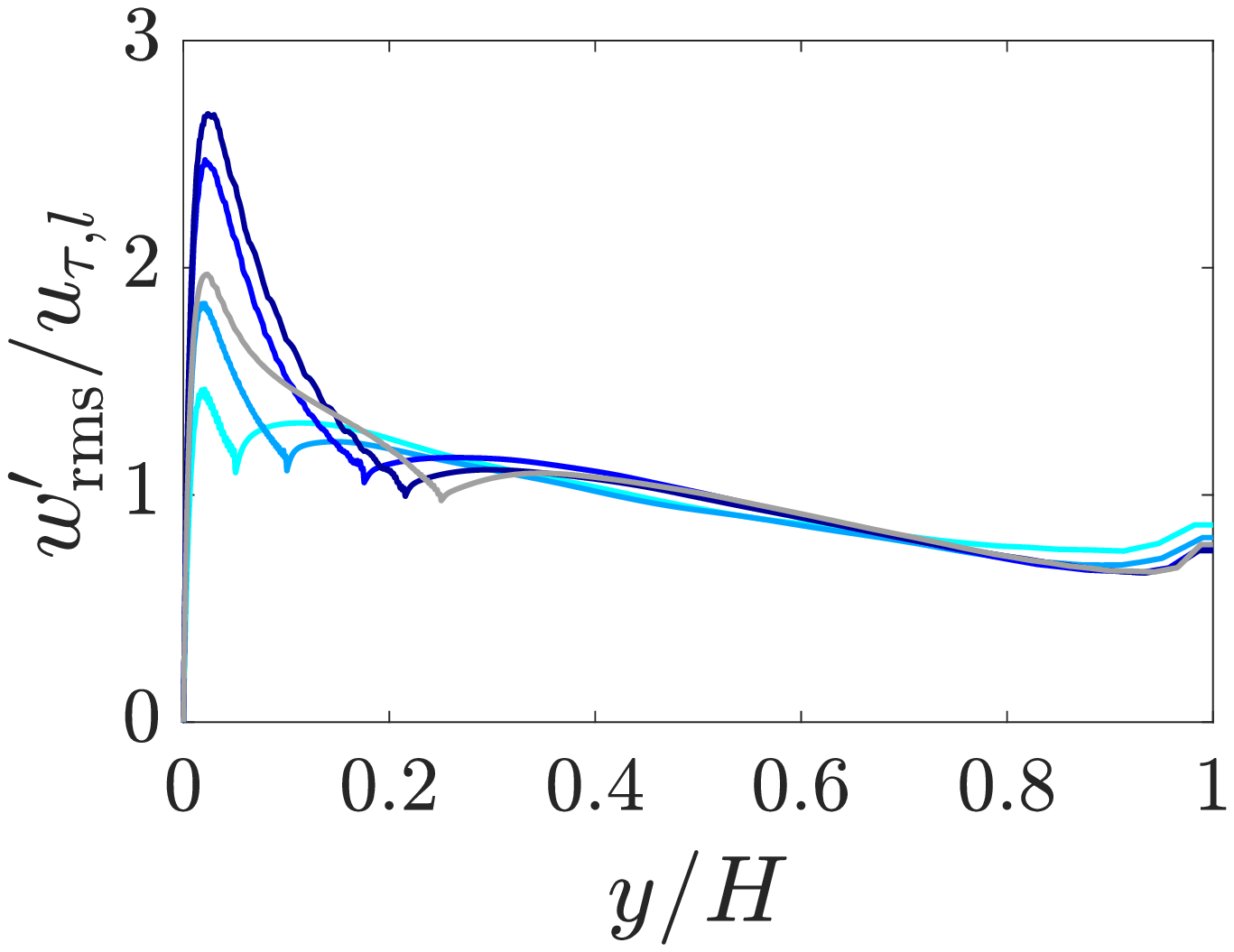}}\\
  \caption{Profile of the RMS of the velocity fluctuations
           versus the wall-normal coordinate $y/H$. 
           The top row, panels (a)--(c), shows the distributions
           of the canopies inclined with the grain; the bottom row, panels (d)--(f),
           the distributions of the canopies inclined against the grain.
           From left to right, the columns of the figure show the streamwise,
           the wall-normal and the spanwise component.
           The distributions are normalized with the local
           friction velocity defined in \cref{eq:friction_vel}.
           Colours as in \cref{tab:par}.}
  \label{fig:secondOrderStatsNormLoc}
  \end{figure}
\Cref{fig:secondOrderStatsNormOut} shows a comparison of the rms of the
velocity fluctuations
between the positively (top row) and negatively (bottom row) inclined
canopies, with the profiles of the wall-normally mounted canopy shown
as reference (grey lines). The velocity fluctuations are normalised with the 
external friction velocity $u_{\tau,out}$, obtained using the total 
stress at the virtual origin. The curves collapse in the region outside
the canopy, confirming once more the behaviour of the outer flow
as a boundary-layer over a virtual rough wall \citep{MONTI2019,MONTI2020}, 
while inside the canopy layer the comparison becomes more challenging
because of the change of the frontal area with the inclination. 
To overcome this problem, we introduce a velocity
scale based on the combination of the imposed pressure gradient and the 
drag exerted by the filaments,
\begin{equation}
  u_{\tau,l}(y) = 
                  \sqrt{\frac{\mu \, \mathrm{d}_y \langle u \rangle 
                - \rho \, \langle u'v' \rangle}{\rho \,(1-y/H)}}.
  \label{eq:friction_vel}
\end{equation}
In the above, $\mu\,\mathrm{d}_y
\langle u\rangle$ is the viscous shear stress and $\rho\,\langle u'v'
\rangle$ is the turbulent shear stress. For more details on the explicit
derivation of \cref{eq:friction_vel}, the reader is referred to \citet{MONTI2019}.
When normalising the rms of the velocity fluctuations with $u_{\tau,l}$, a better 
collapse is achieved within the canopy layer, as shown in \cref{fig:secondOrderStatsNormLoc}.
In particular, we obtain a set of profile comparable to the ones corresponding 
to an open channel flow over a smooth wall.
From \cref{fig:secondOrderStatsNormLoc}, we note that the maximum of the 
streamwise velocity fluctuations (leftmost panels) decreases 
as the angle of inclination for $\theta>0$ (top row) decreases, 
with the wall-normally mounted canopy inverting the trend (grey line). 
The opposite is true for the cases $\theta<0$ (bottom row), where the peak 
of $u'_{rms}$ increases by decreasing the inclination angle, dropping 
when $\theta=0$ (grey line).
The peak of the streamwise velocity fluctuations close to the wall suggests the 
presence of coherent structures in that region, similar to streaks. However,
the presence of the stems makes the development of classic streaks and the 
typical wall-cycle \citep{JIMENEZ1999} unlikely, so the peak of
$u'_{rms}$ is more likely related to the high-speed and low-speed regions evolving 
between the stems, similarly to a biperiodic flow past a set 
of cylinders.
Concerning the wall-normal fluctuations $v'_{rms}$ within the canopy, as 
already observed by \citet{MONTI2020}, the normalised distribution does not 
qualitatively differ from the one observed in an open-channel flow over a 
smooth wall, suggesting the filtering effect of the stems acts only along 
the homogeneous directions.
Although the observation is still valid in the context of inclined canopies,
we must point out that the intensity of $v'_{rms}$ within the canopy layer 
tends to decrease monotonically as the inclination $|\theta|$ increases,
with the negatively inclined canopies having a larger value, as expected,
since the penetration of the outer flow is facilitated.
Finally, the spanwise velocity fluctuations show the most interesting
behaviour, with a large peak related to the deviation of the streaky 
structures (peak of $u'_{rms}$) caused by the presence of the stems. 
The trend of its maximum is similar to that of $v'_{rms}$, suggesting that the
intensity of spanwise velocity structures close to the wall is dominated
by the penetration of the large outer structures for taller canopies 
(i.e. lower $|\theta|$), while for shorter canopies (i.e. higher $|\theta|$)
the outer and inner flows are mixed, with the former dictating the coherence 
near the bed.

\begin{figure}
  \centering
  \subfloat[]{\includegraphics[width=0.25\linewidth]{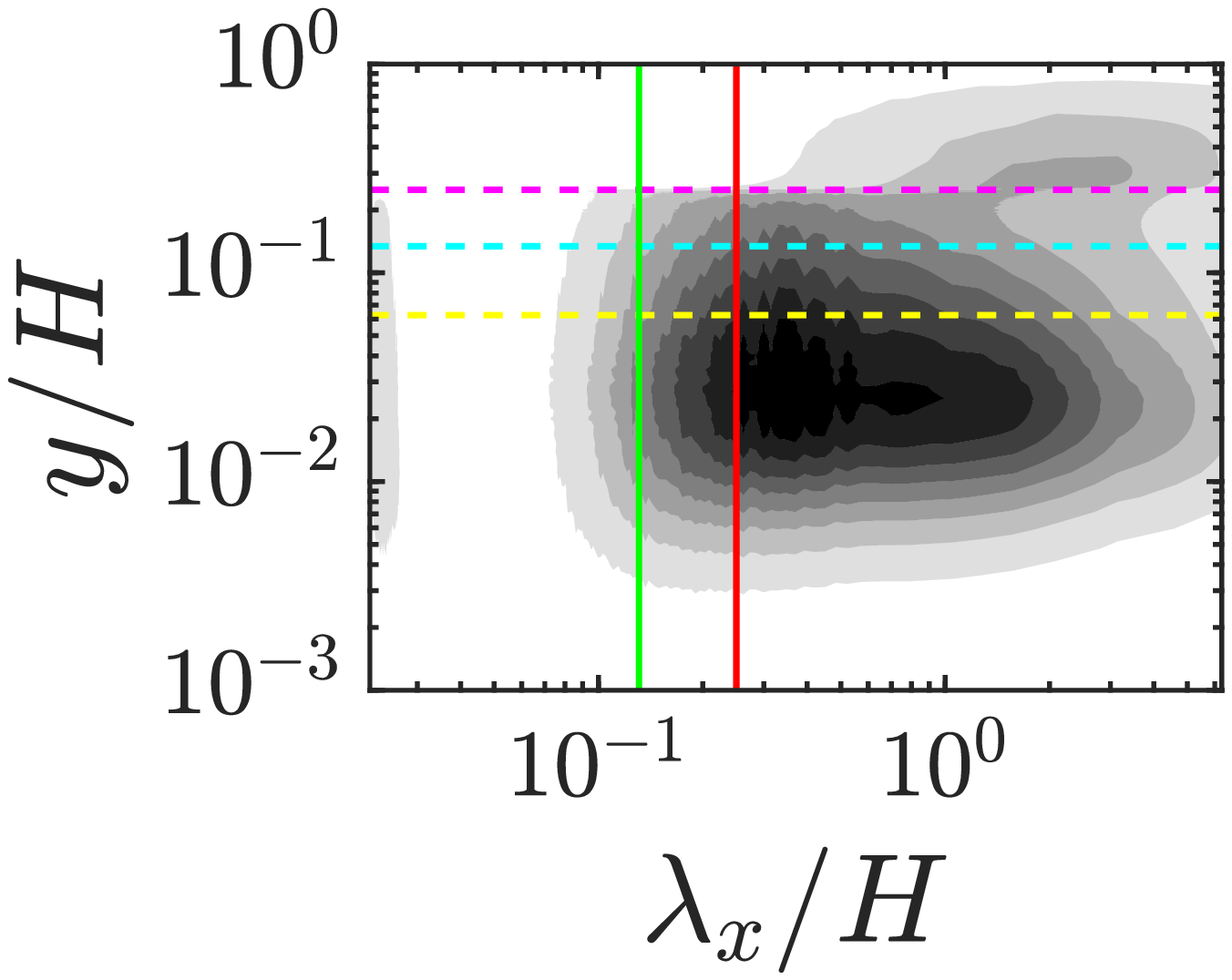}}
  \subfloat[]{\includegraphics[width=0.25\linewidth]{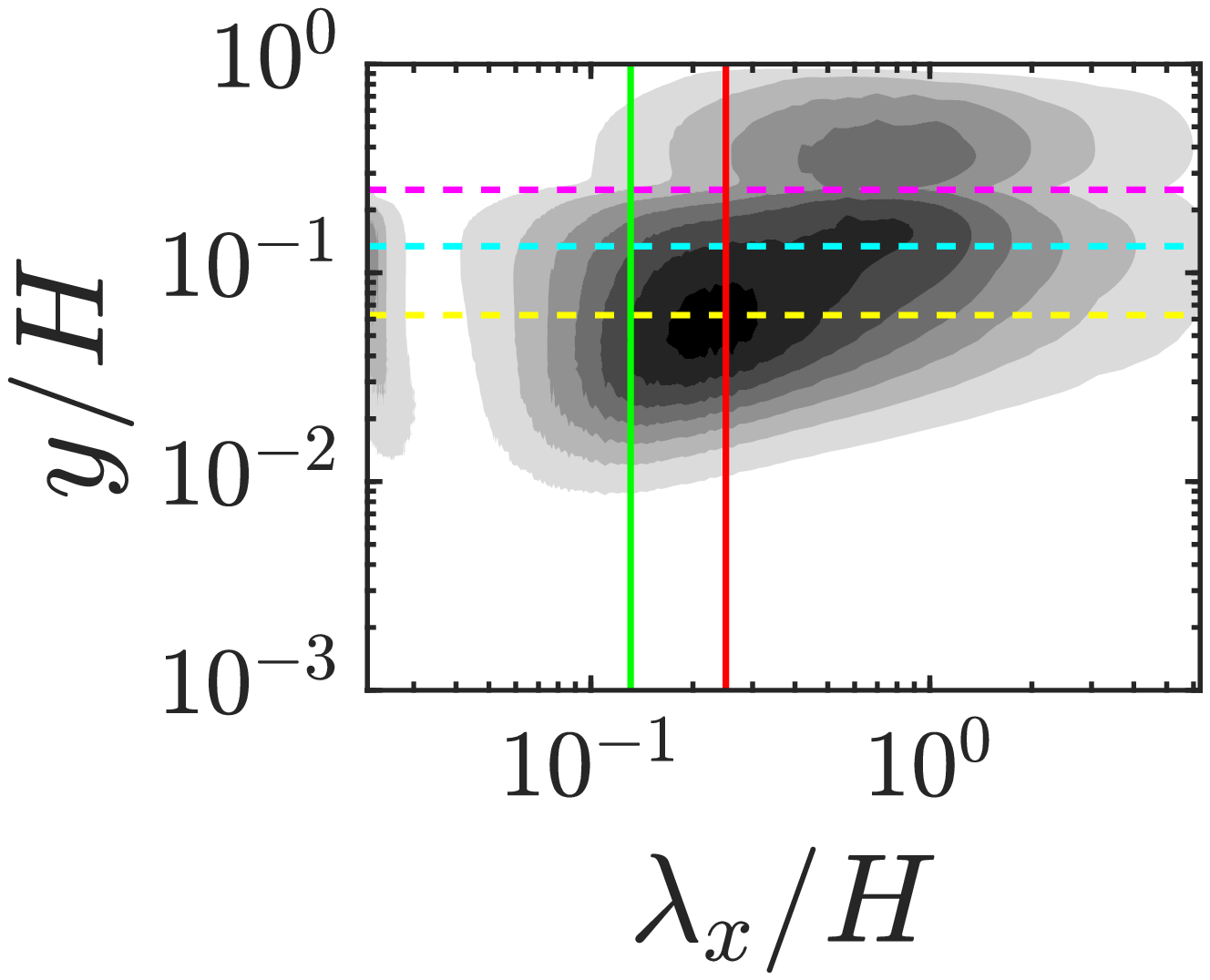}}
  \subfloat[]{\includegraphics[width=0.25\linewidth]{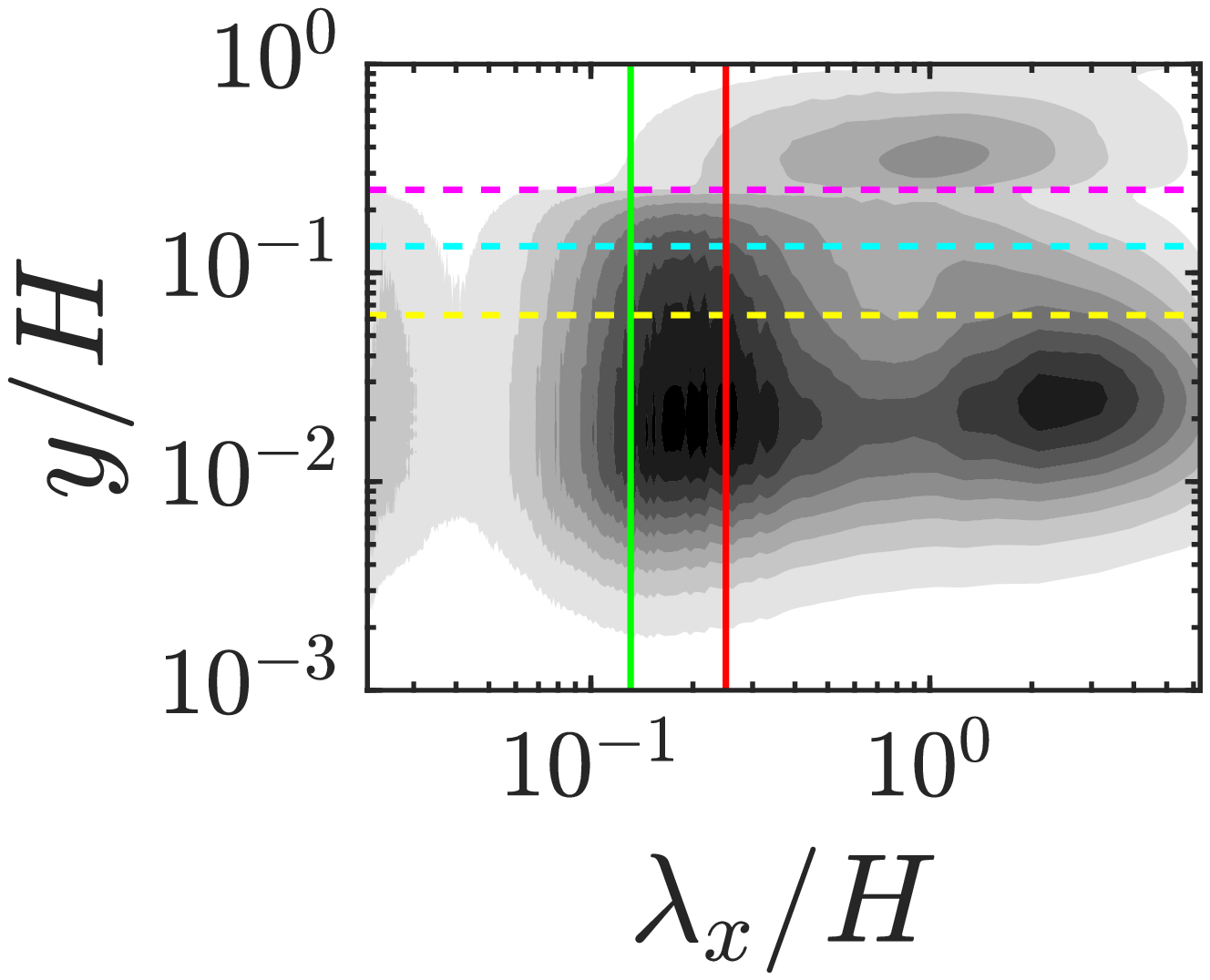}}
  \subfloat[]{\includegraphics[width=0.25\linewidth]{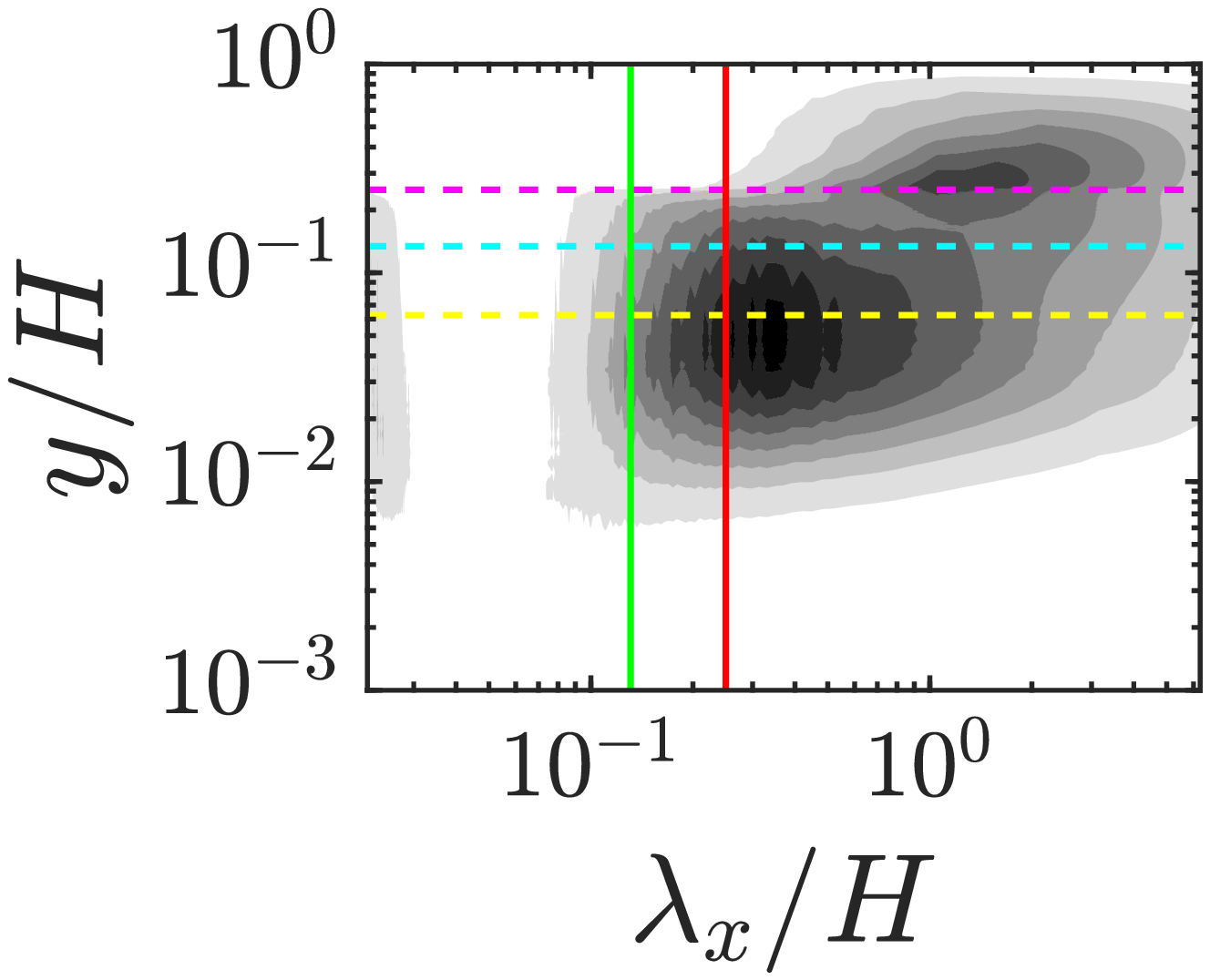}}\\
  \subfloat[]{\includegraphics[width=0.25\linewidth]{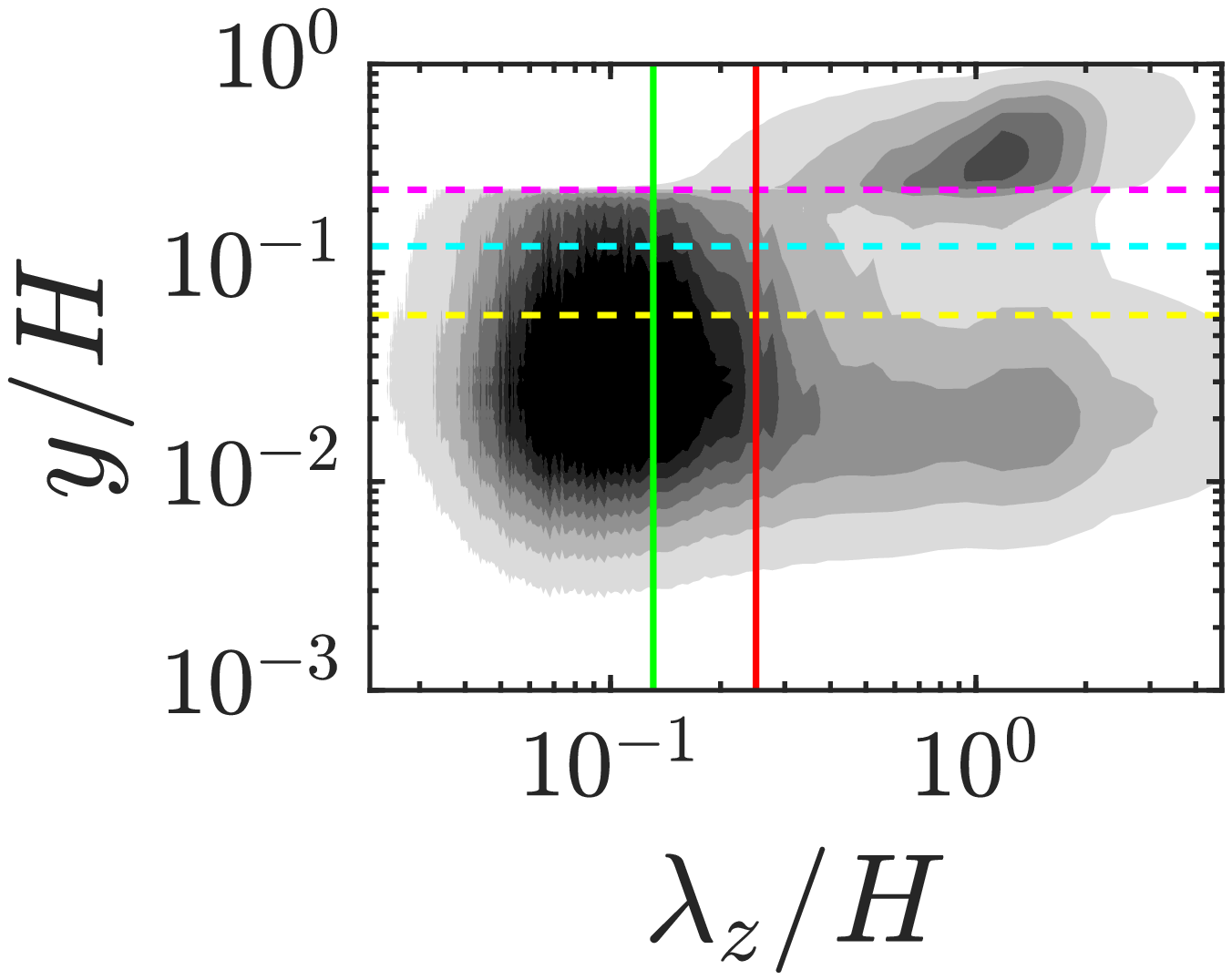}}
  \subfloat[]{\includegraphics[width=0.25\linewidth]{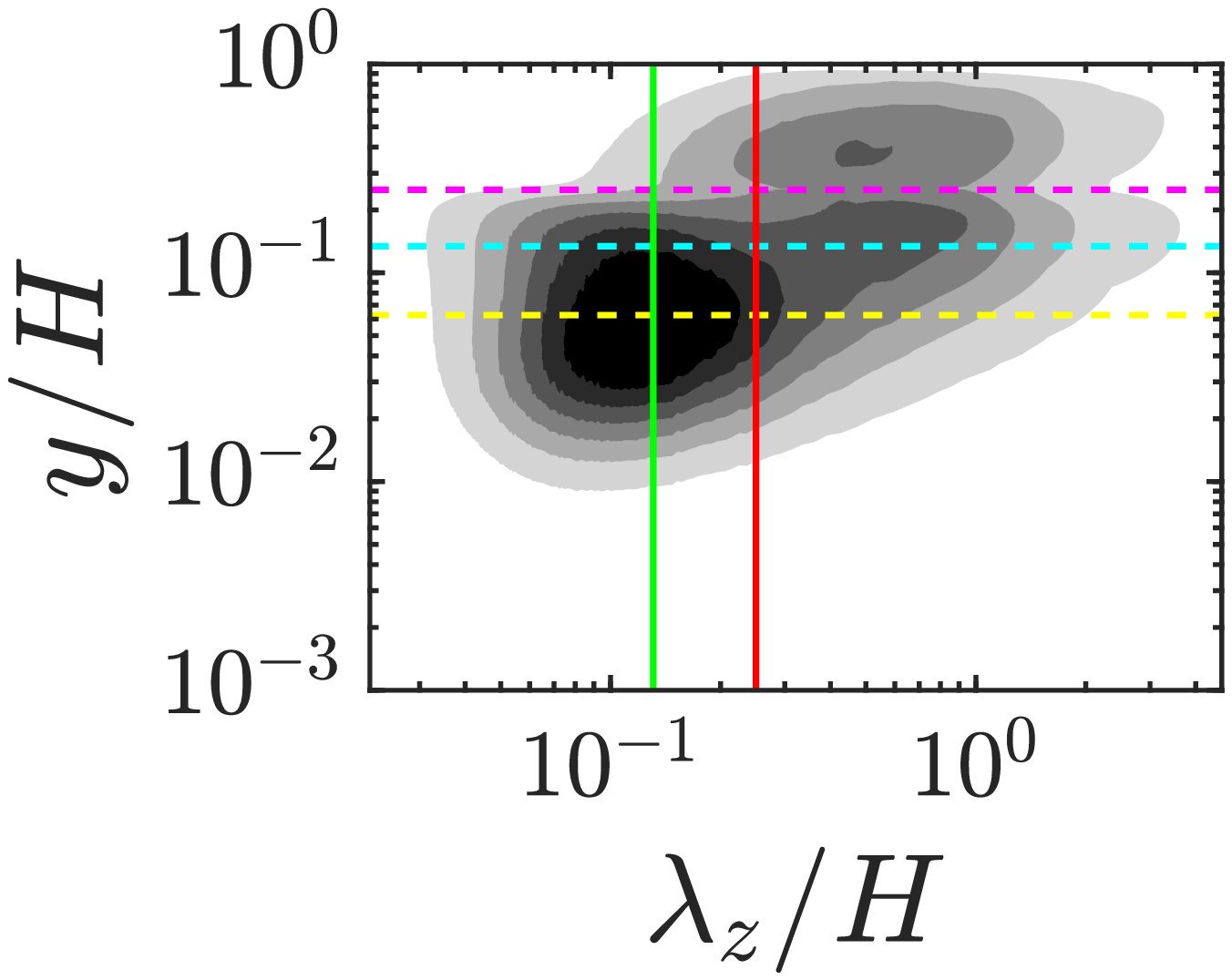}}
  \subfloat[]{\includegraphics[width=0.25\linewidth]{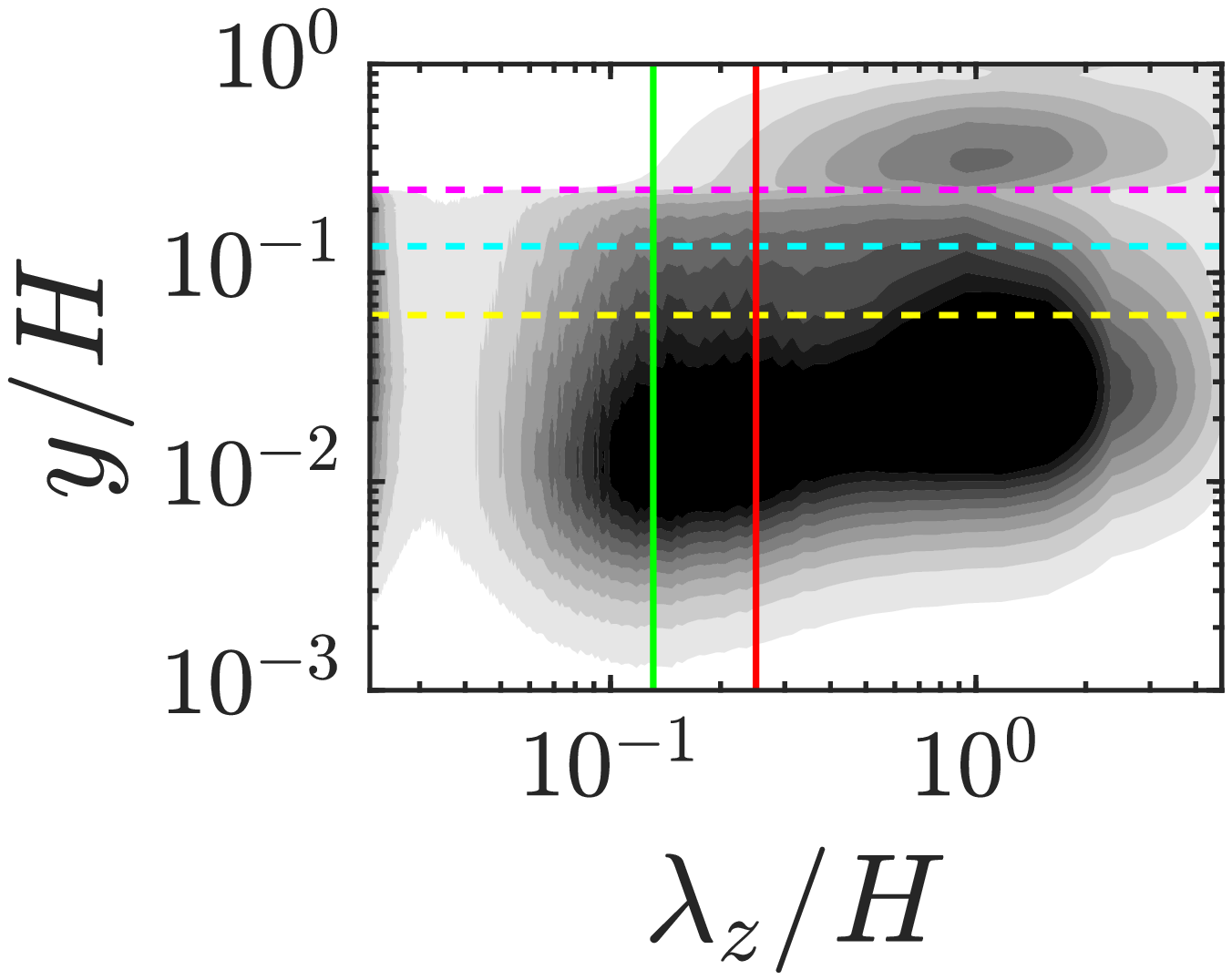}}
  \subfloat[]{\includegraphics[width=0.25\linewidth]{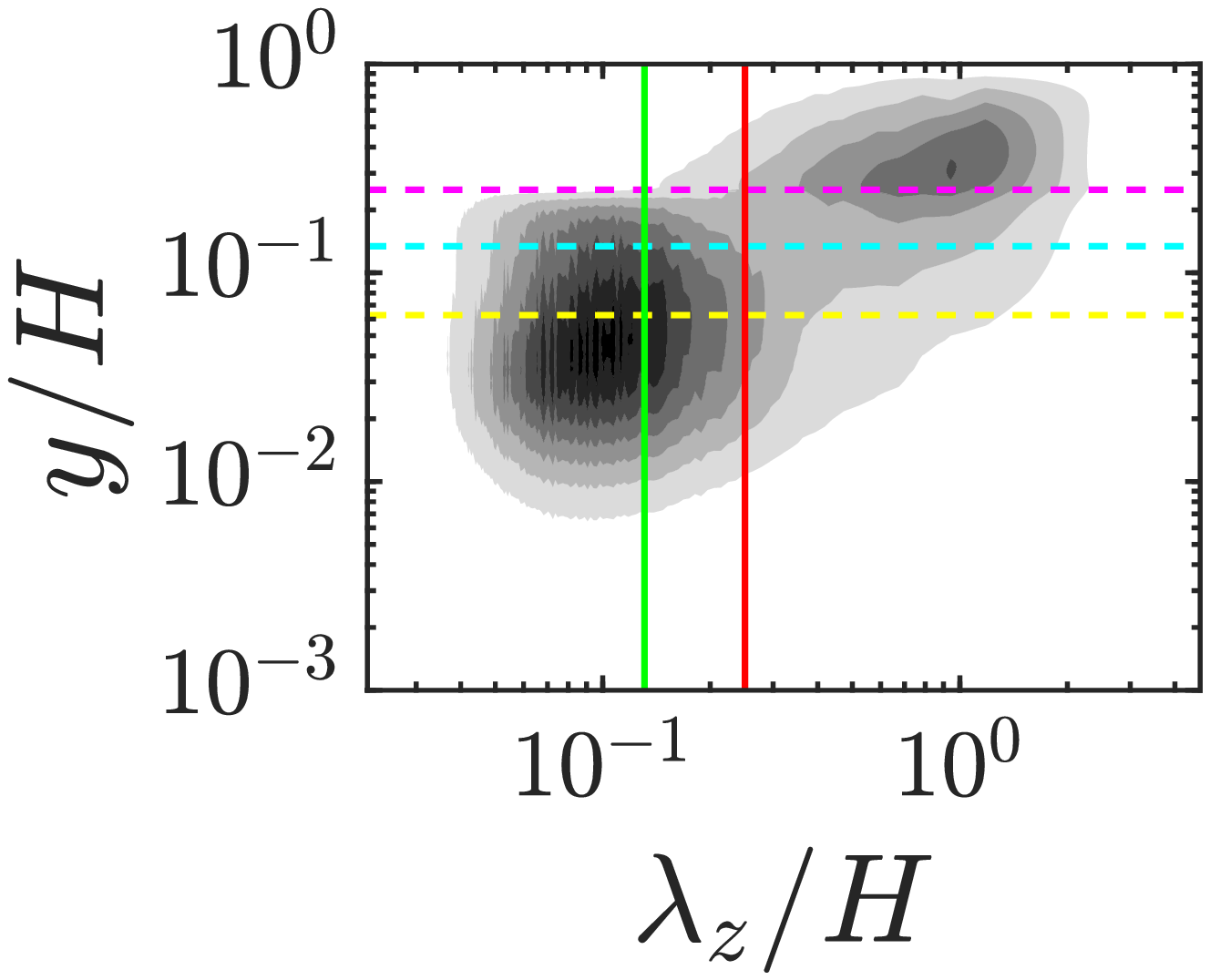}}\\
  \caption{Case $\theta=\ang{0}$. 
           Magnitude of the premultiplied spectra of the velocity components
           and cospectra of the Reynolds shear stress as a function of 
           the wall-normal coordinates $y/H$ and
           the streamwise wavelength $\lambda_x/H$ (top row), and the 
           spanwise wavelength $\lambda_z/H$ (bottom row). 
           From left to right, the columns are:
           $\kappa_x\Phi_{u'u'}/u_{\tau,l}^2$ with 
           grey levels range in $[0,0.8]$ with a $0.1$ increment; 
           $\kappa_x\Phi_{v'v'}/u_{\tau,l}^2$ with grey levels range in 
           $[0,0.3]$ with a $0.03$ increment; 
           $\kappa_x\Phi_{w'w'}/u_{\tau,l}^2$ 
           with grey levels range in $[0,0.5]$ with a $0.05$ increment;
           $\kappa_x|\Phi_{u'v'}|/u_{\tau,l}^2$ 
           with grey levels range in $[0,0.4]$ with a $0.02$ increment.
           The vertical solid lines are (red) $l_\perp/H$ and (green) $\Delta S/H$.
           Horizontal dashed lines: (yellow) location of the inner 
           inflection point, (magenta) canopy height (i.e.,~outer inflection point), 
           (cyan) location of the virtual origin.}
  \label{fig:premultipliedSpectraV0d250}
  \end{figure}

Further insight on the structures populating the flow can be obtained by
analysing the spectral energy content of the fluctuations of the velocity 
components.
We start by looking at the structures of the wall-normally-mounted canopy 
flow to describe the phenomenology that bonds together the outer and inner 
flows when the scale separation introduced by the tall stems is expected
(dense regimes).
In \cref{fig:premultipliedSpectraV0d250}, we present the one-dimensional 
premultiplied spectra of the velocity fluctuations and the magnitude of the
one-dimensional premultiplied cospectra of the Reynolds shear stress, as a
function of the distance from the wall, organised in a $2\times 4$ matrix of 
panels, where each column of the matrix shows, in order, the streamwise, 
wall-normal and spanwise component of the velocity fluctuations, and the
fourth column the magnitude of the cospectra of the Reynolds shear stress. 
In the top row, the quantities are plotted as a function of the streamwise
wavelength, while in the bottom row as a function of the spanwise wavelength.

\begin{figure}
  \centering
  \subfloat[]{\includegraphics[width=0.33\linewidth]{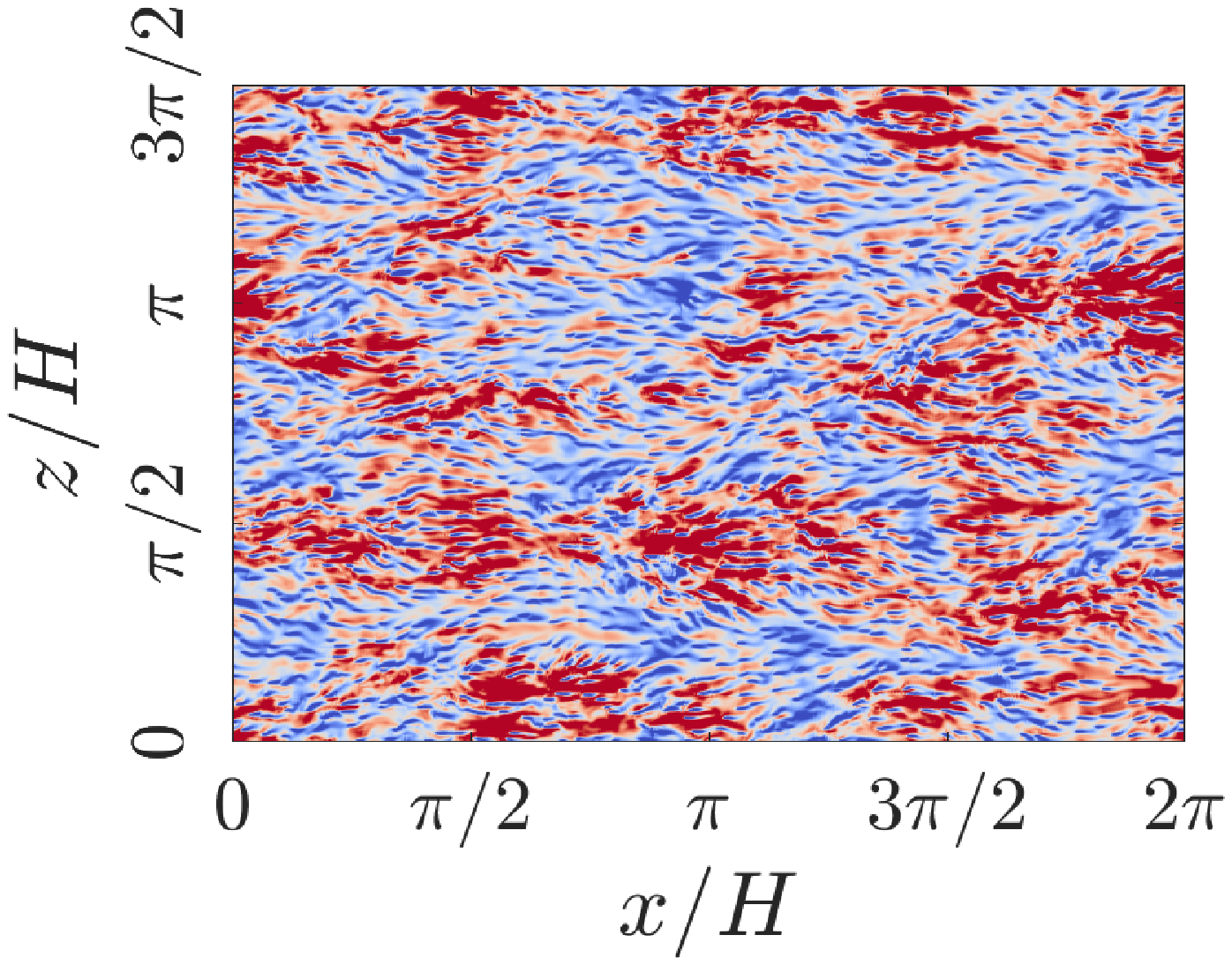}}
  \subfloat[]{\includegraphics[width=0.33\linewidth]{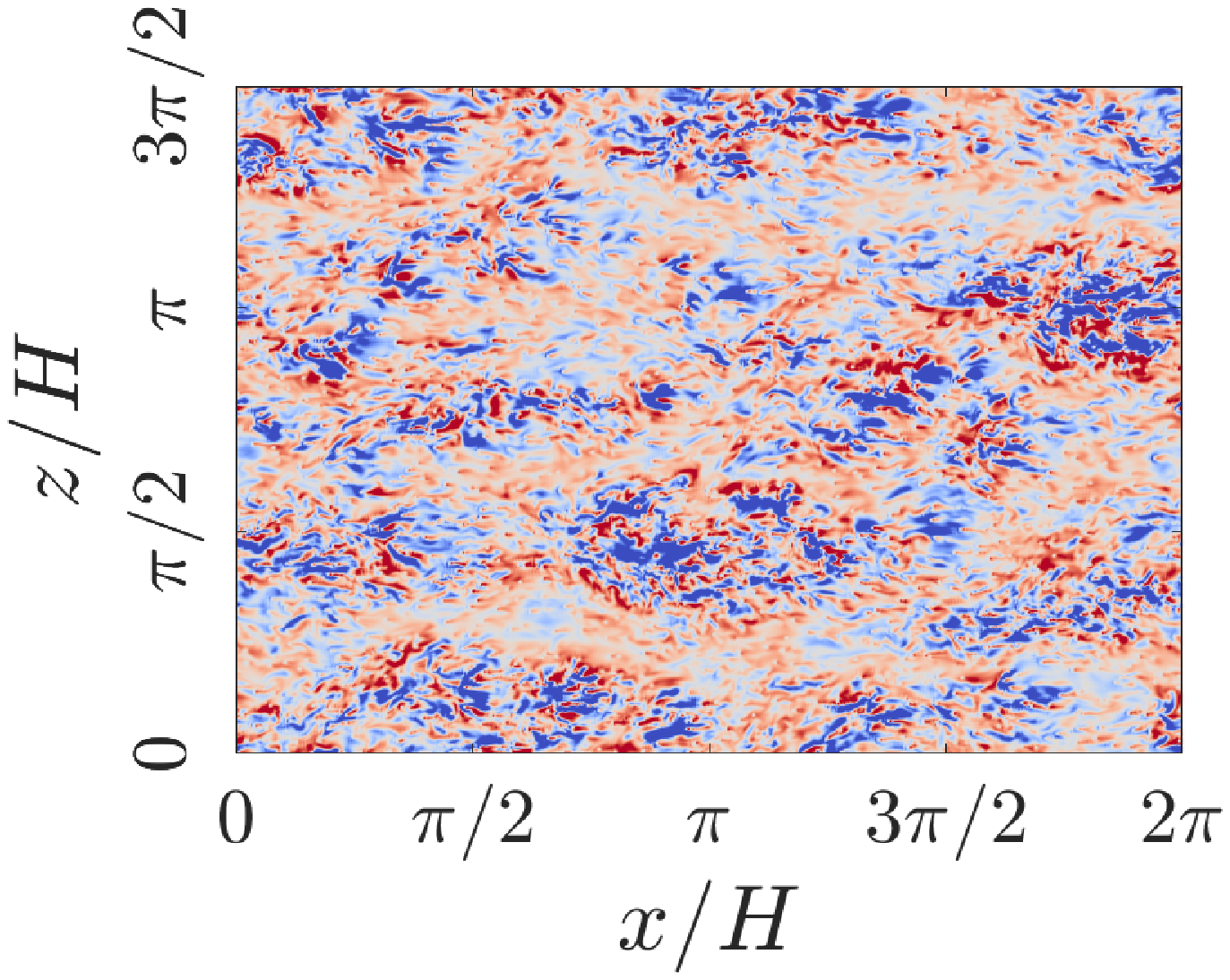}}
  \subfloat[]{\includegraphics[width=0.33\linewidth]{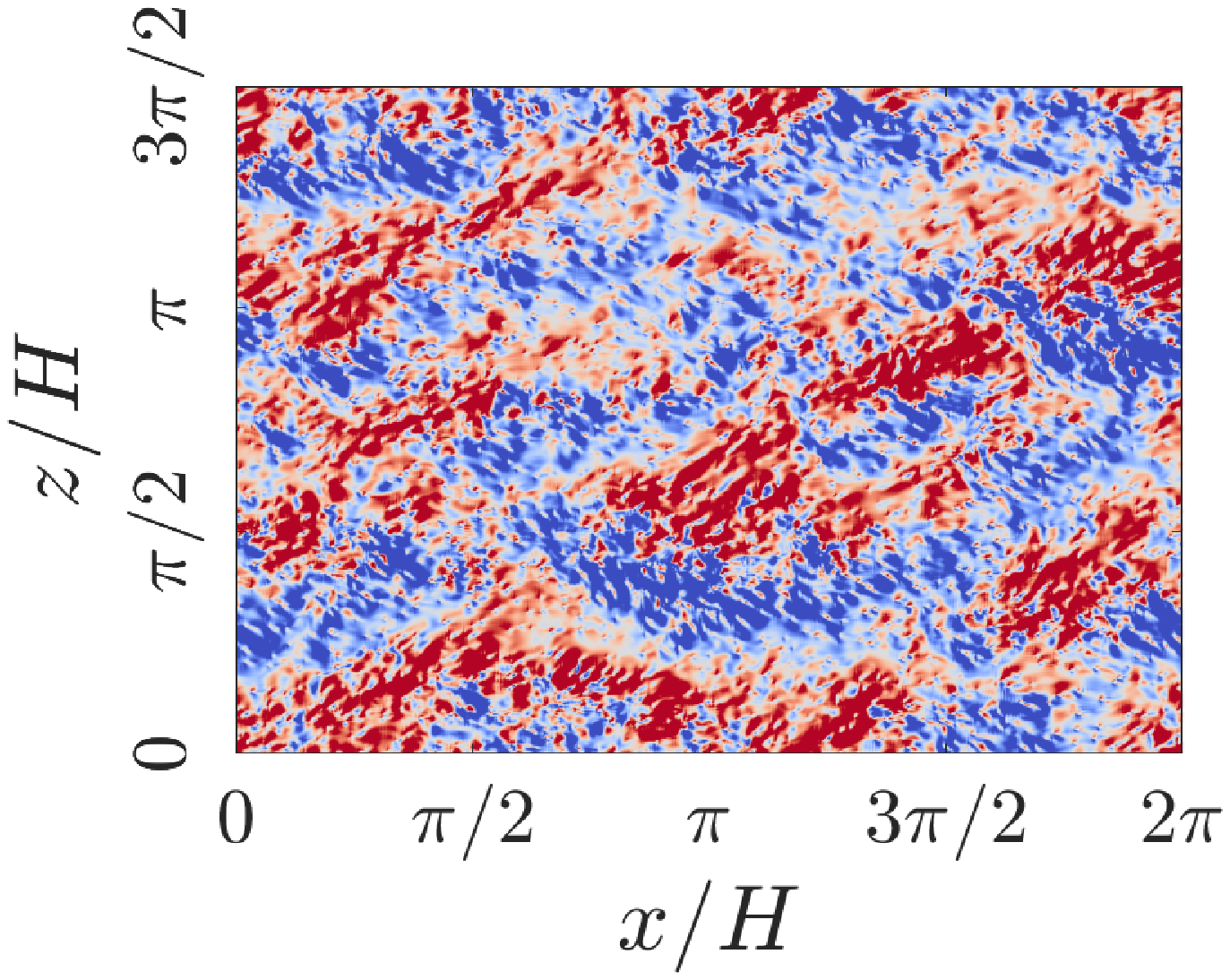}}\\
  \subfloat[]{\includegraphics[width=0.33\linewidth]{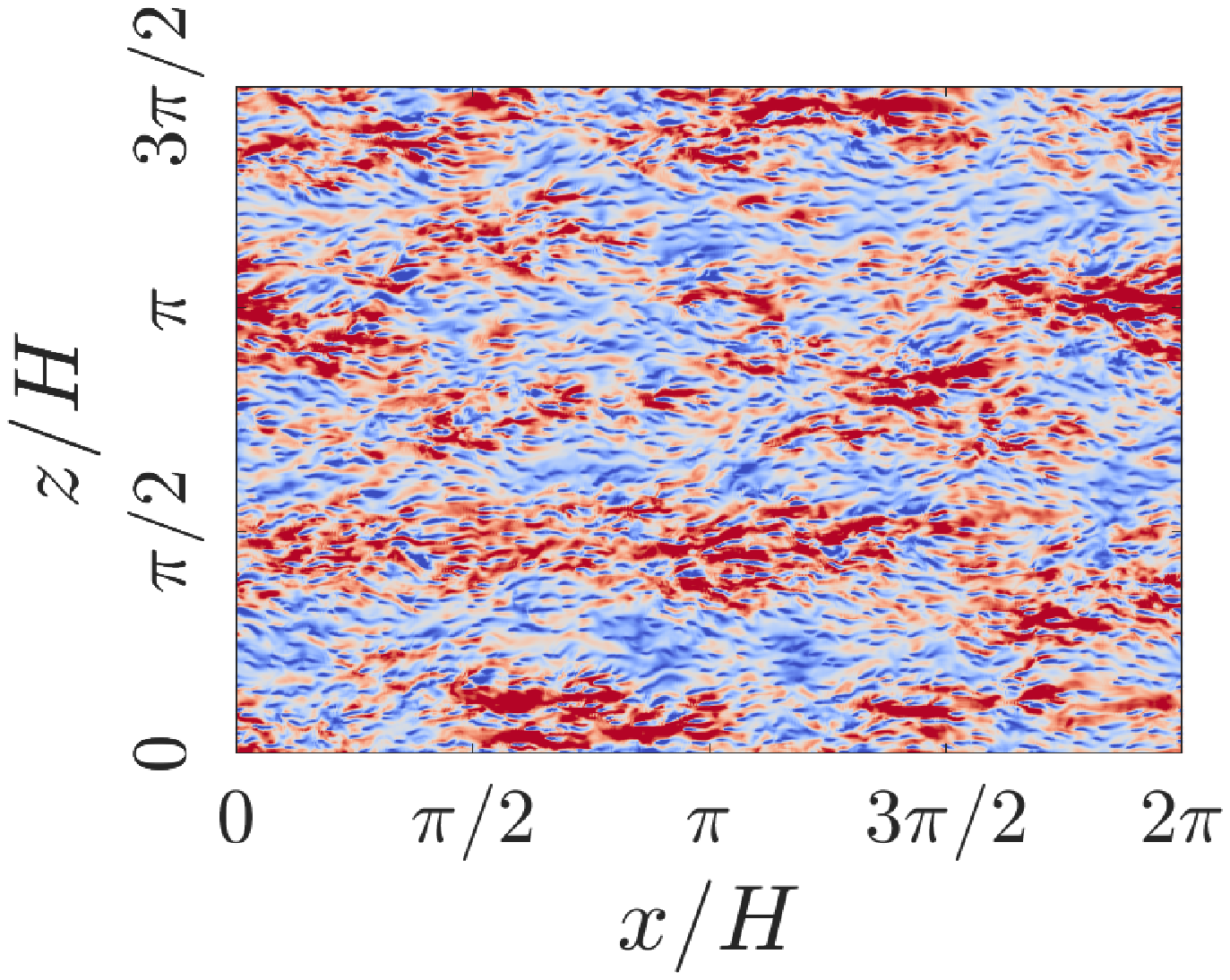}}
  \subfloat[]{\includegraphics[width=0.33\linewidth]{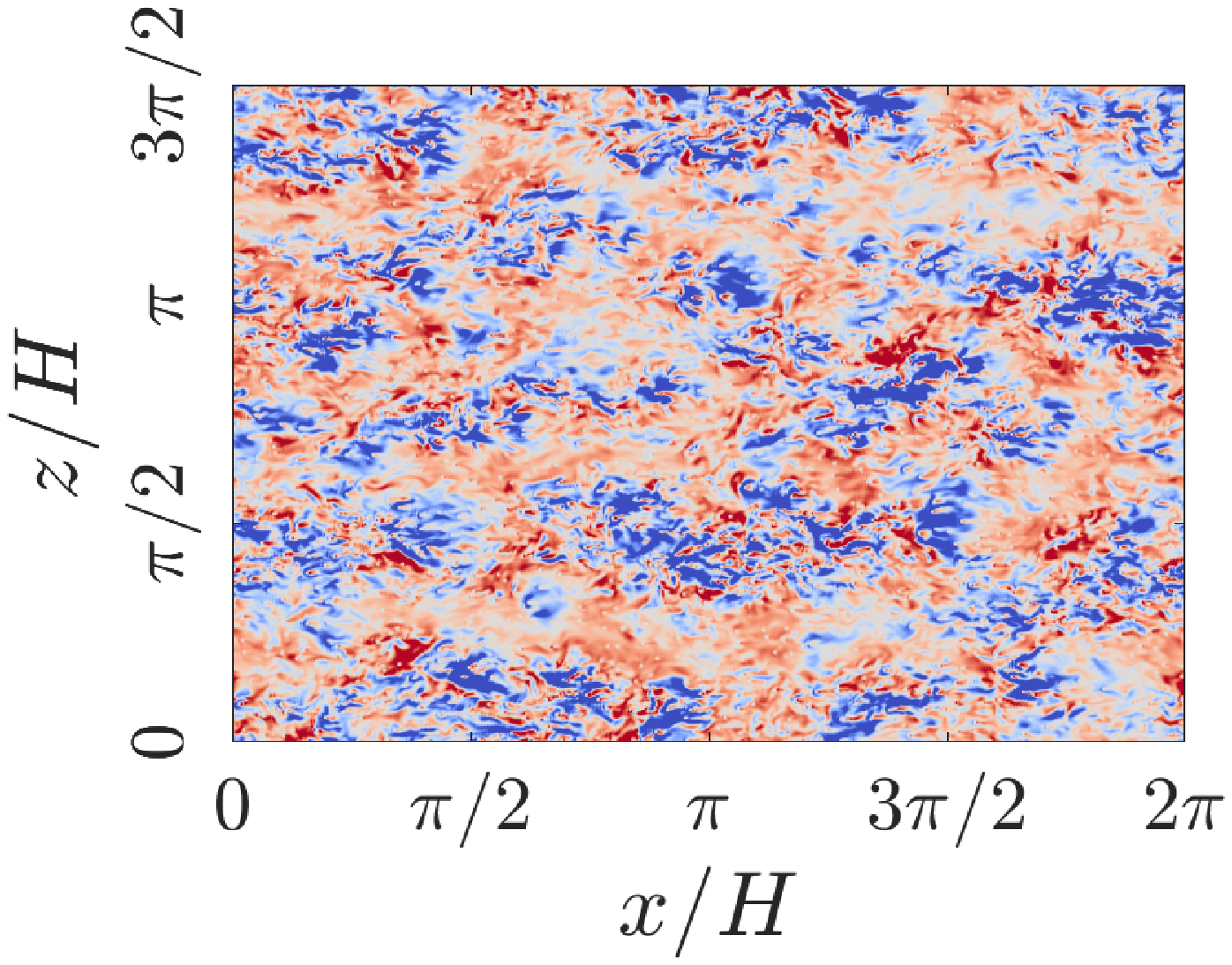}}
  \subfloat[]{\includegraphics[width=0.33\linewidth]{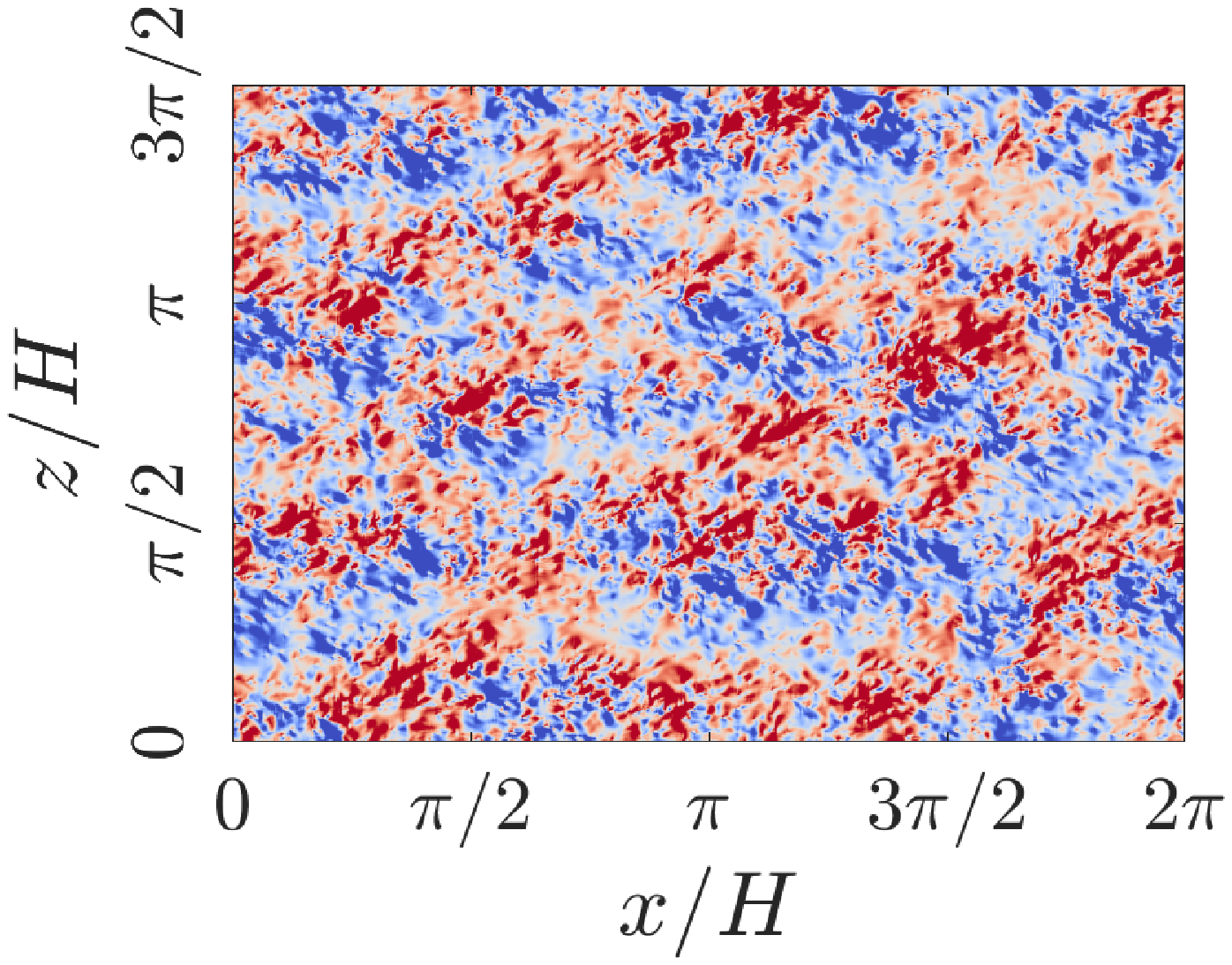}}\\
  \subfloat[]{\includegraphics[width=0.33\linewidth]{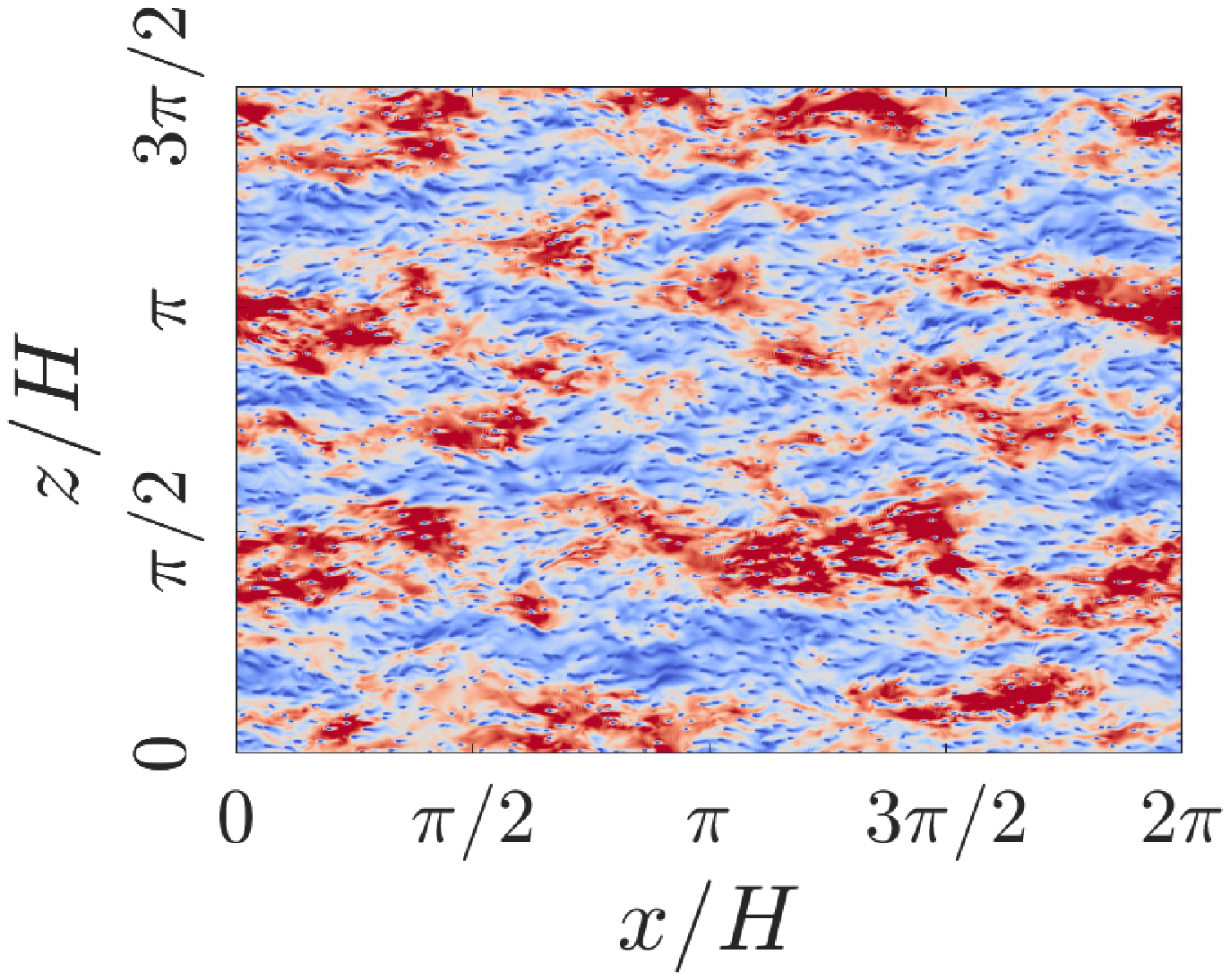}}
  \subfloat[]{\includegraphics[width=0.33\linewidth]{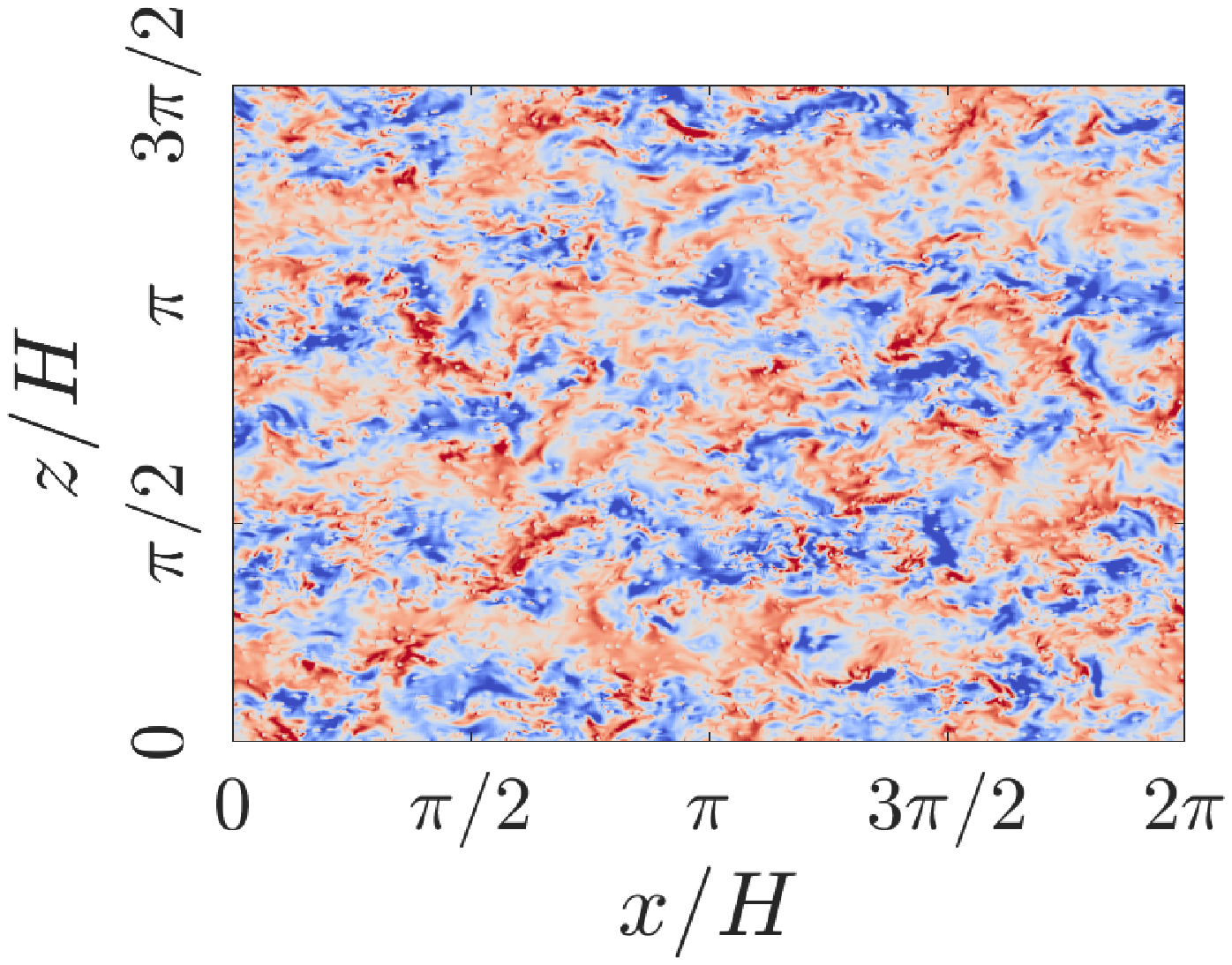}}
  \subfloat[]{\includegraphics[width=0.33\linewidth]{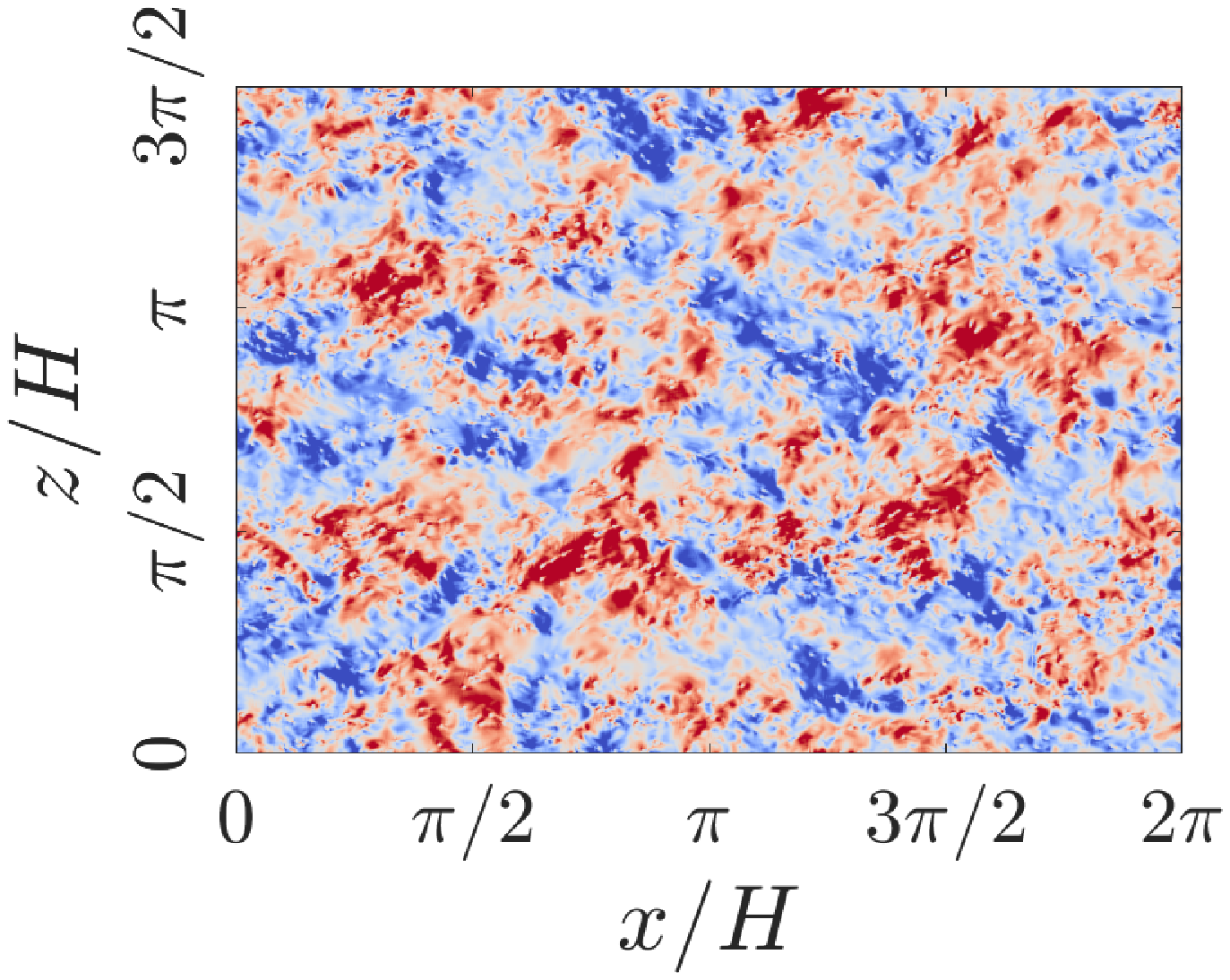}}\\
  \subfloat[]{\includegraphics[width=0.33\linewidth]{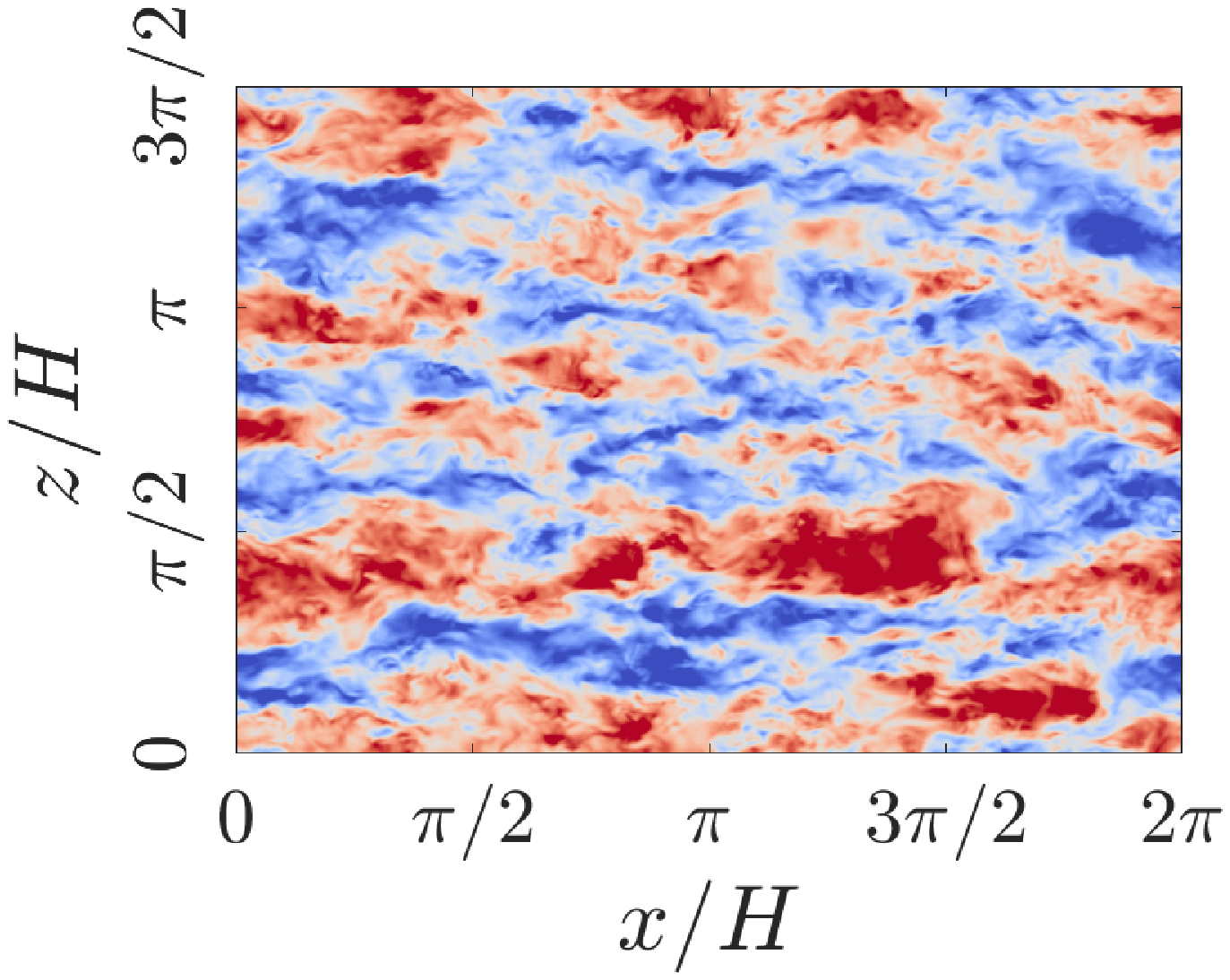}}
  \subfloat[]{\includegraphics[width=0.33\linewidth]{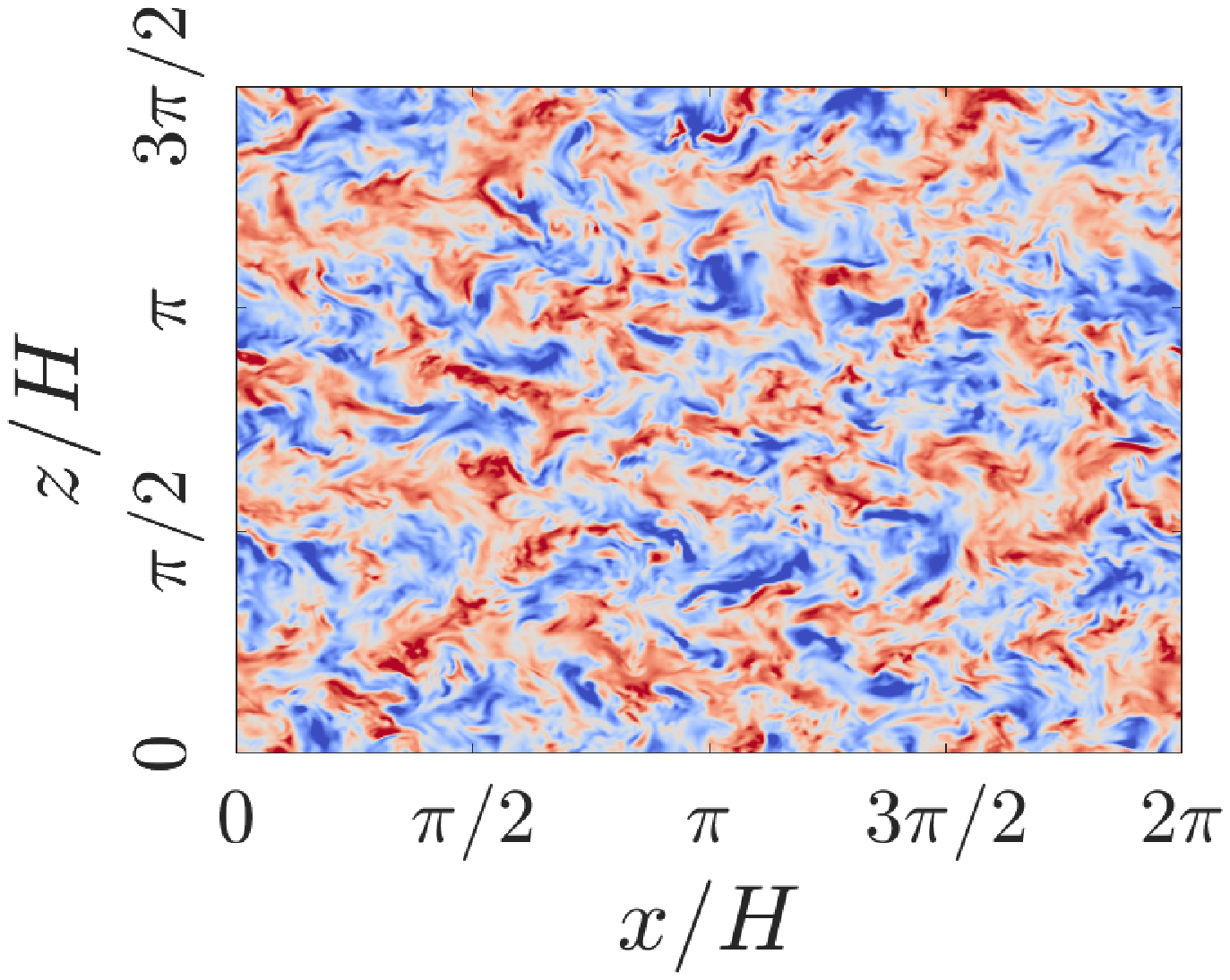}}
  \subfloat[]{\includegraphics[width=0.33\linewidth]{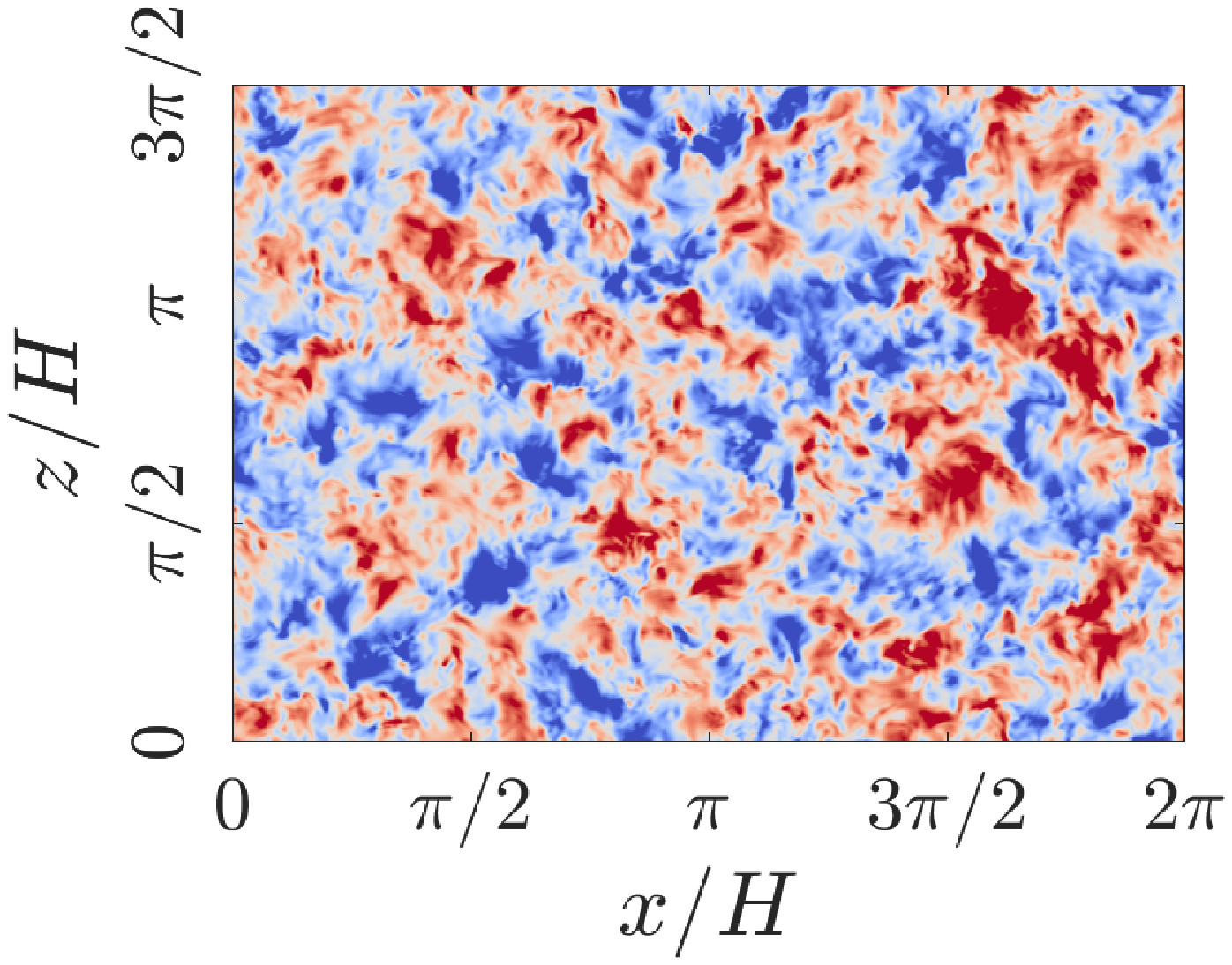}}\\
  \caption{Case $\theta=\ang{0}$. Contours of the instantaneous velocity fluctuations 
           on planes parallel to the wall.
           Panels (a), (d), (g) and (j): red $u'/u_{\tau,l}=3$, blue  $u'/u_{\tau,l}=-3$; 
           Panels (b), (e), (h) and (k): red $v'/u_{\tau,l}=2$, blue  $v'/u_{\tau,l}=-2$;
           Panels (c), (f), (i) and (l): red $w'/u_{\tau,l}=3$, blue  $w'/u_{\tau,l}=-3$. 
           The slides are located at: 
           $y=y_{lip}$, first row; $y=y_{vor}$, second row; $y=l_\perp$, third row;
           $y=0.35H$ (outer region), fourth row.}
  \label{fig:snapshotSliceV0d250}
  \end{figure}

Observing \cref{fig:premultipliedSpectraV0d250}, we note that large 
structures with wavelengths $\lambda_x=O(H)$ and $\lambda_z=O(H)$ populate
the outer region, that can be identified as very elongated coherent
structures and large spanwise rollers \citep{MONTI2020}. These structures are generated by the 
well-documented Kelvin-Helmholtz (KH) instability caused by the discontinuity of 
the drag offered by the finite-size canopy layer \citep{NEPF2012}. 
The KH instability triggers the formation of very large spanwise-coherent 
rollers that, trapped from the lower side by the canopy stems and transported 
by the high gradients of the mean flow that develop at the canopy tip, evolve
into large, very elongated structures in the streamwise direction. The footprints
of such elongated structures can be visualized in the panels (d), (g) and (j) of 
\cref{fig:snapshotSliceV0d250}, that show the contours of the
instantaneous fluctuations of the streamwise velocity component in a horizontal 
slice at the virtual origin, canopy tip and outside the canopy layer, 
respectively. The KH rollers, however, are not clearly visible
since their coherence is broken by the turbulence events at the canopy tip. 
A way to visualize them (not be shown here) consists in averaging the 
fluctuations along the spanwise directions and plot the resulting two-dimensional 
streamlines, see e.g. \citet{MONTI2020} and \citet{JIMENEZ2001}.

Before looking at the spectra of the inner layer, we consider the first
and the last rows of \cref{fig:snapshotSliceV0d250}, which represent 
slices of the instantaneous fluctuations of the velocity field at the location
of the inner inflection point (first row) and a region outside the canopy layer,
at $y=h + 0.1H$, above the canopy tip. From their comparison, it becomes
quite evident that the structures of the flow are uncorrelated, in the limit
of dense regimes.
With this in mind, we now consider again the spectra of velocity fluctuations,
shown in \cref{fig:premultipliedSpectraV0d250}, and we analyse the region 
within the canopy.
Particularly, in the inner layer, the premultiplied spectra of the streamwise
and the spanwise velocity fluctuations, panels (a), (c), (e) and (g) of
\cref{fig:premultipliedSpectraV0d250}, reveal two distinct peaks.
Specifically, the leftmost one refers to smaller structures and identifies
the high-momentum structures that form and meander between the canopy stems,
with lateral size $\lambda_z\approx \Delta S$ and streamwise size
$\lambda_x \approx 2$\texttt{\char`\~}$3 \Delta S$.
The rightmost peak, instead, detects structures with much larger size, 
suggesting the presence of other mechanisms that generate them. However,
it should be clear how, close to the wall, no particular length-scale, except
$\Delta S$ and the diameter of the stems $d$, is present and no relevant turbulent 
mechanisms can be activated, despite the presence of the wall and the (weak) shear 
flow developing from there.
These arise from the small high-momentum structures that, 
due to the limited disturbance coming from the outer flow 
in dense canopies, are free to organise along the entire horizontal domain
and coherently flow parallel to the wall, as the limit of the flow in 
emergent canopies at low Reynolds number suggests.  These large structures
are more easily recognisable in denser scenarios as the influence of the outer
flow decreases. 
Both kinds of structures living in the inner layer depicted above can be visualized
in panels (a) and (c) of \cref{fig:snapshotSliceV0d250}, where the
former shows the contours of the instantaneous fluctuations of the streamwise 
velocity component in a horizontal slice located in proximity of the inner inflection
point, and the latter the contours of the instantaneous fluctuations of the spanwise 
velocity component in a horizontal slice at the same location. The largest 
structures, however, do not cover the whole domain, especially along the streamwise
direction, since their coherence is broken by the vehement quasi-wall-normal jets 
that, due to the high permeability in the wall-normal direction \citep{rosti_brandt_2017a, rosti_brandt_pinelli_2018a},
penetrate from the outer layer through the canopy and reach the proximity of 
the bed \citep[similarly to the jets described by][]{BANYASSADY2015}.
These quasi-wall-normal jets are identified in the spectra of $v'$, 
panels (b) and (f) of \cref{fig:snapshotSliceV0d250}, with the large 
peak of size $\lambda_x\approx H$ that stretch from the outer layer 
beyond the virtual origin. In an instantaneous snapshot, they can be 
readily recognised, by looking at the contours of the wall-normal
component of the velocity fluctuations close to the canopy bed, as the large
spots of downward high-speed velocity shown in panel (b) of 
\cref{fig:snapshotSliceV0d250}. These jets, when approaching the wall, push
down the fluid, squeezing it against the wall, and break the horizontal 
coherence of the flow. In other terms, when such strong jets approach the 
solid wall, they slow down creating a region of high, negative gradients of 
the wall-normal velocity fluctuations $\partial_y v'$ which in turn, due 
to the incompressibility constraint, generate strong positive gradients 
of $\partial_x u'$ and $\partial_z w'$, which feed the structures close
to the wall, generating strong motions of positive and negative $u'$ and
$w'$, as they can be easily recognized in panel (a) and (c) of 
\cref{fig:snapshotSliceV0d250} \citep{MOSSA2017,MOSSA2021}.
Finally, the structures of $u'$ and $w'$ burst upwards losing momentum, when exiting the domain 
of influence of a jet and entering larger regions where the jets are 
upwardly reflected by the wall and the shear layer (vast regions of
weak, positive $v'$ in panel (b) of \cref{fig:snapshotSliceV0d250}). When these weakened horizontal
structures encounter another set of such structures caused by a consecutive 
wall-normal jet, they form region of negative $\partial_x u'$ and 
$\partial_z w'$ that give rise to positive, not-very-strong wall-normal 
jets $\partial_y v'$. This mechanism creates
coherent structures of $u'v'$ that generate a high-Reynolds 
shear stress region, reducing the contribution of the mean dissipation;
in other words, the turbulence mechanisms described are responsible
for setting the location of the lower inflection point. As a confirmation,
the magnitude of the cospectra of $u'v'$, panels (d) and (h) of 
\cref{fig:premultipliedSpectraV0d250} clearly show a peak located at the 
inner inflection point (yellow, horizontal dashed line). Moreover, the location
of the inner inflection point must saturate when reaching the dense regime,
eventually moving towards the wall (or the end of the inner shear layer in
a Darcy's flow fashion) when the layer of stems becomes denser. This behaviour can be
also seen in the cases analysed in this study, as shown by \cref{fig:remarkable_points}.

\begin{figure}
  \centering
  \subfloat[]{\includegraphics[width=0.25\linewidth]{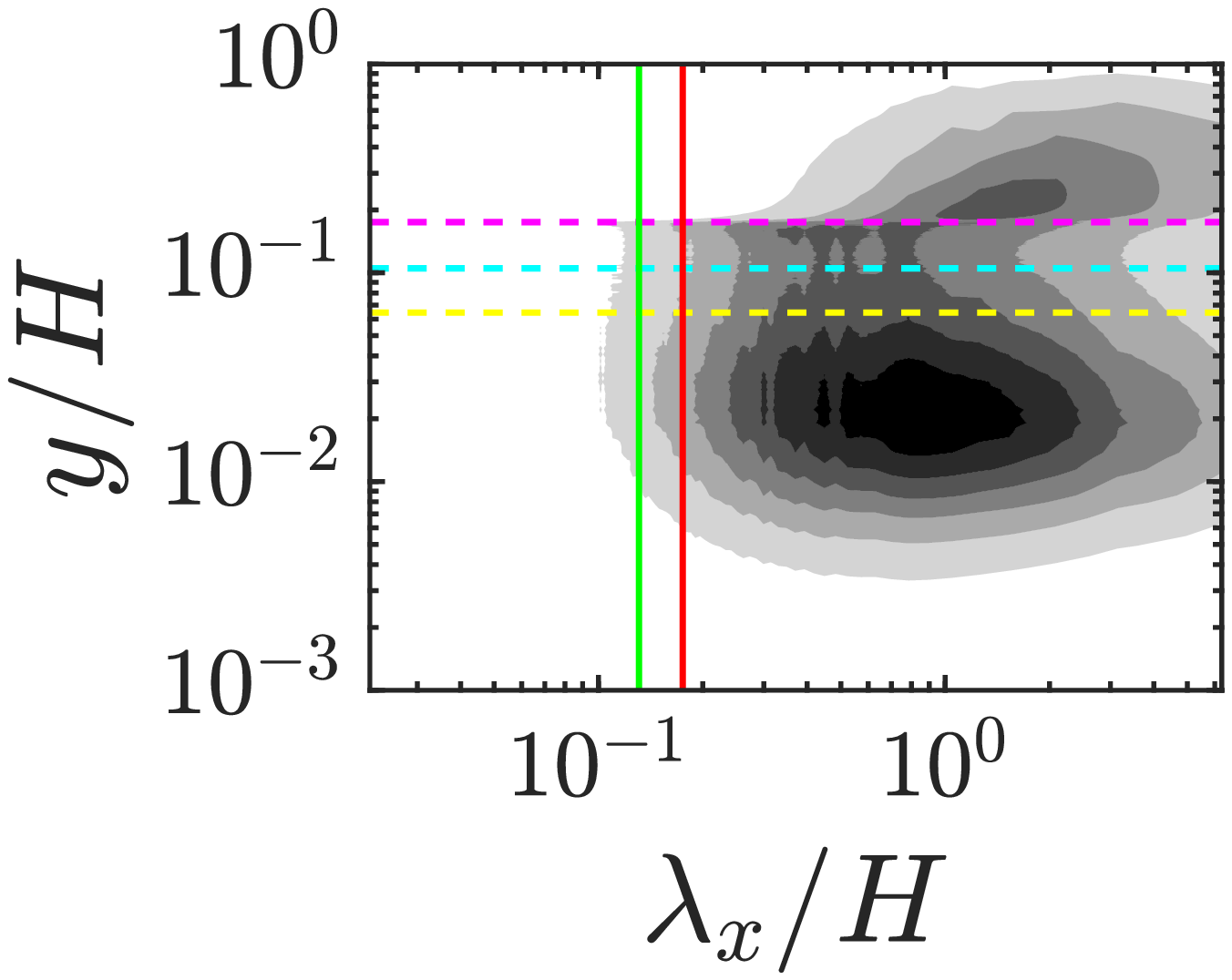}}
  \subfloat[]{\includegraphics[width=0.25\linewidth]{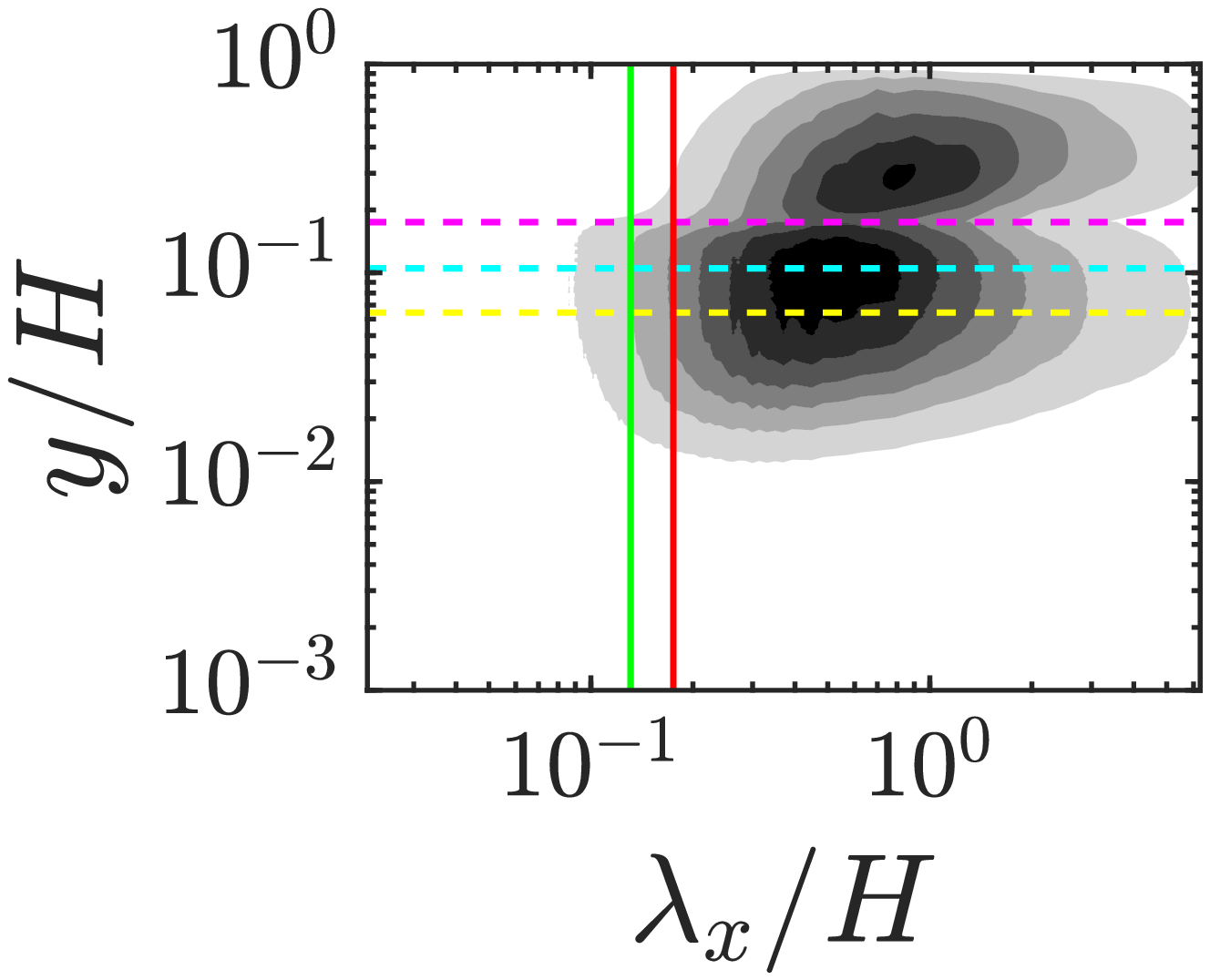}}
  \subfloat[]{\includegraphics[width=0.25\linewidth]{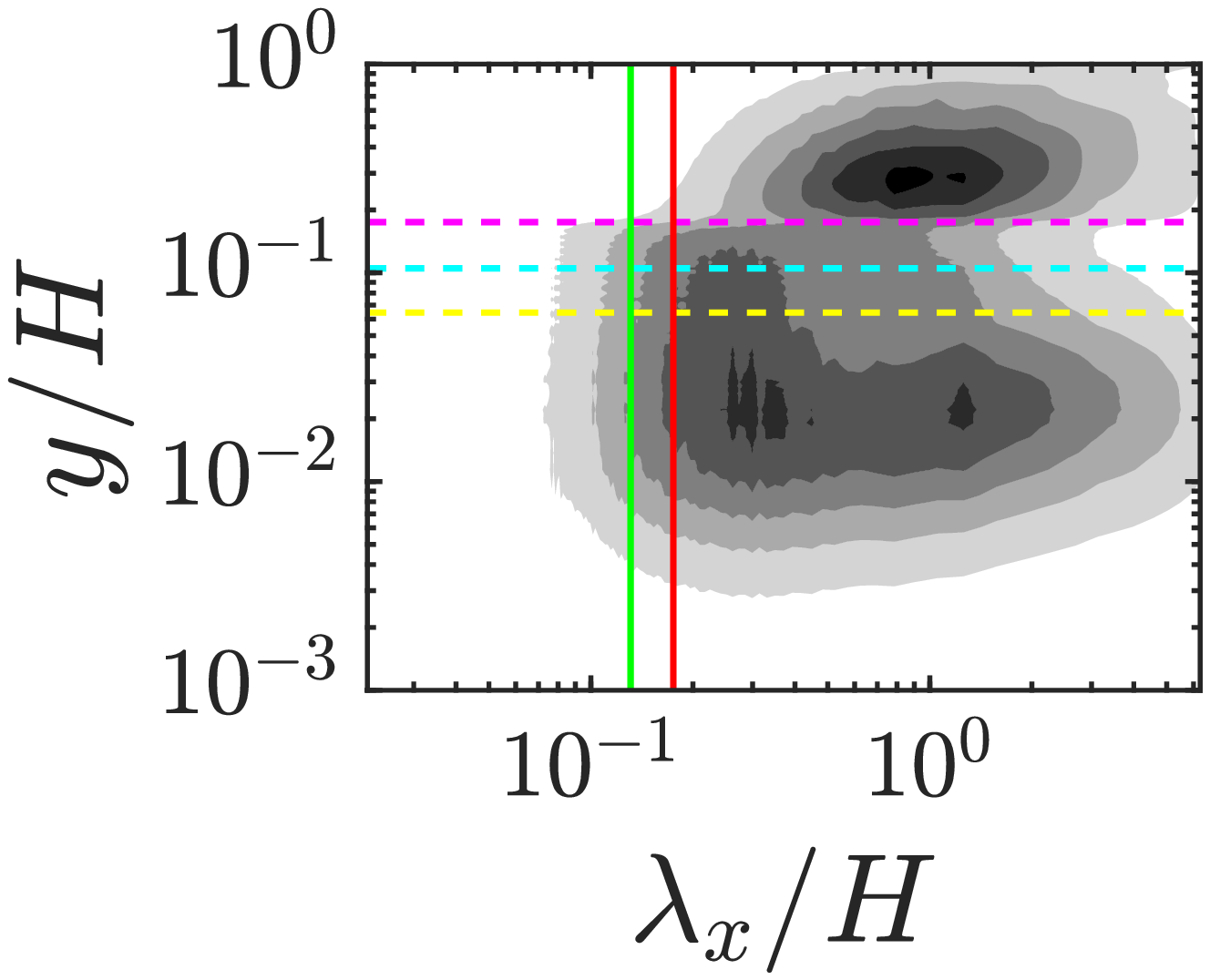}}
  \subfloat[]{\includegraphics[width=0.25\linewidth]{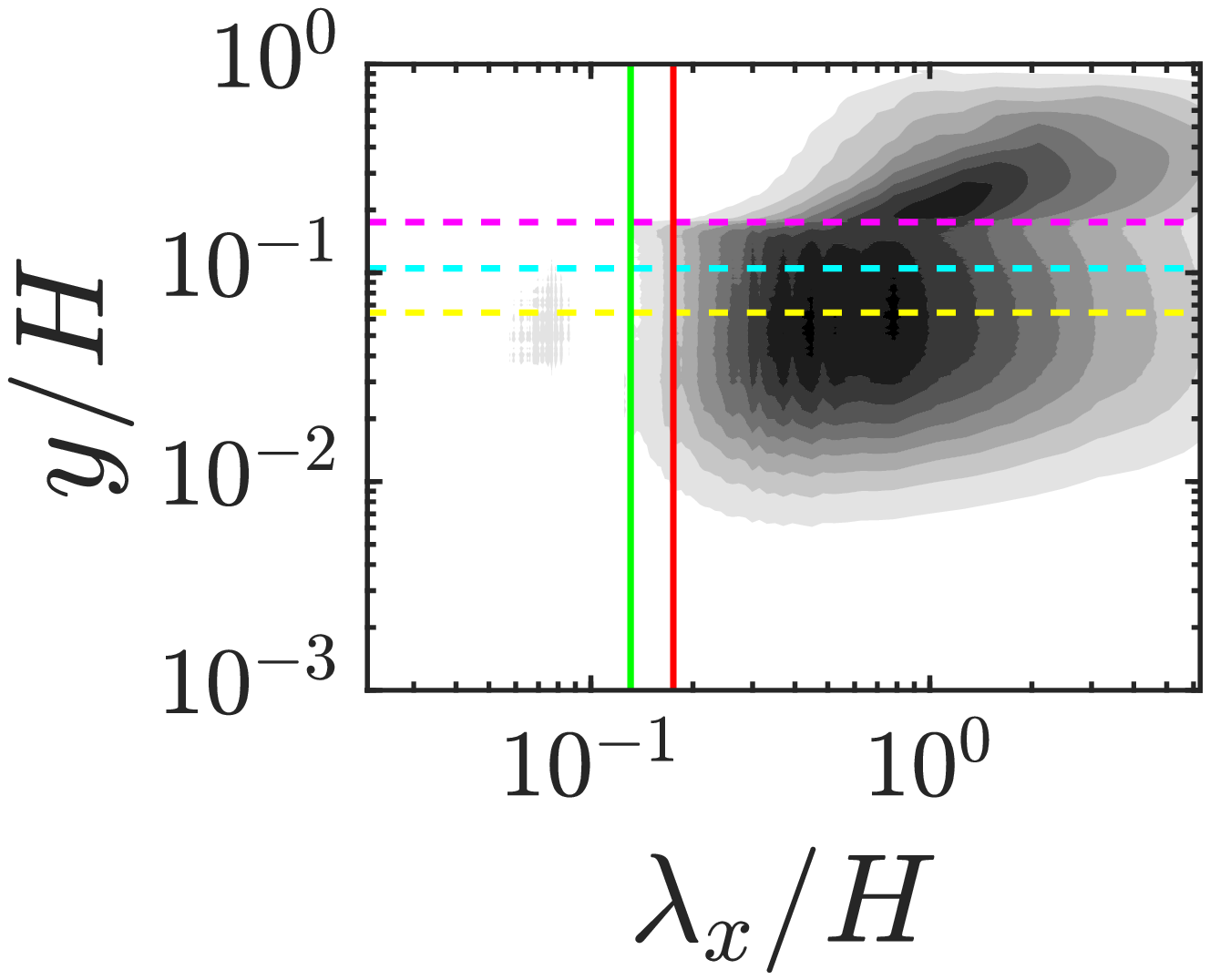}}\\
  \subfloat[]{\includegraphics[width=0.25\linewidth]{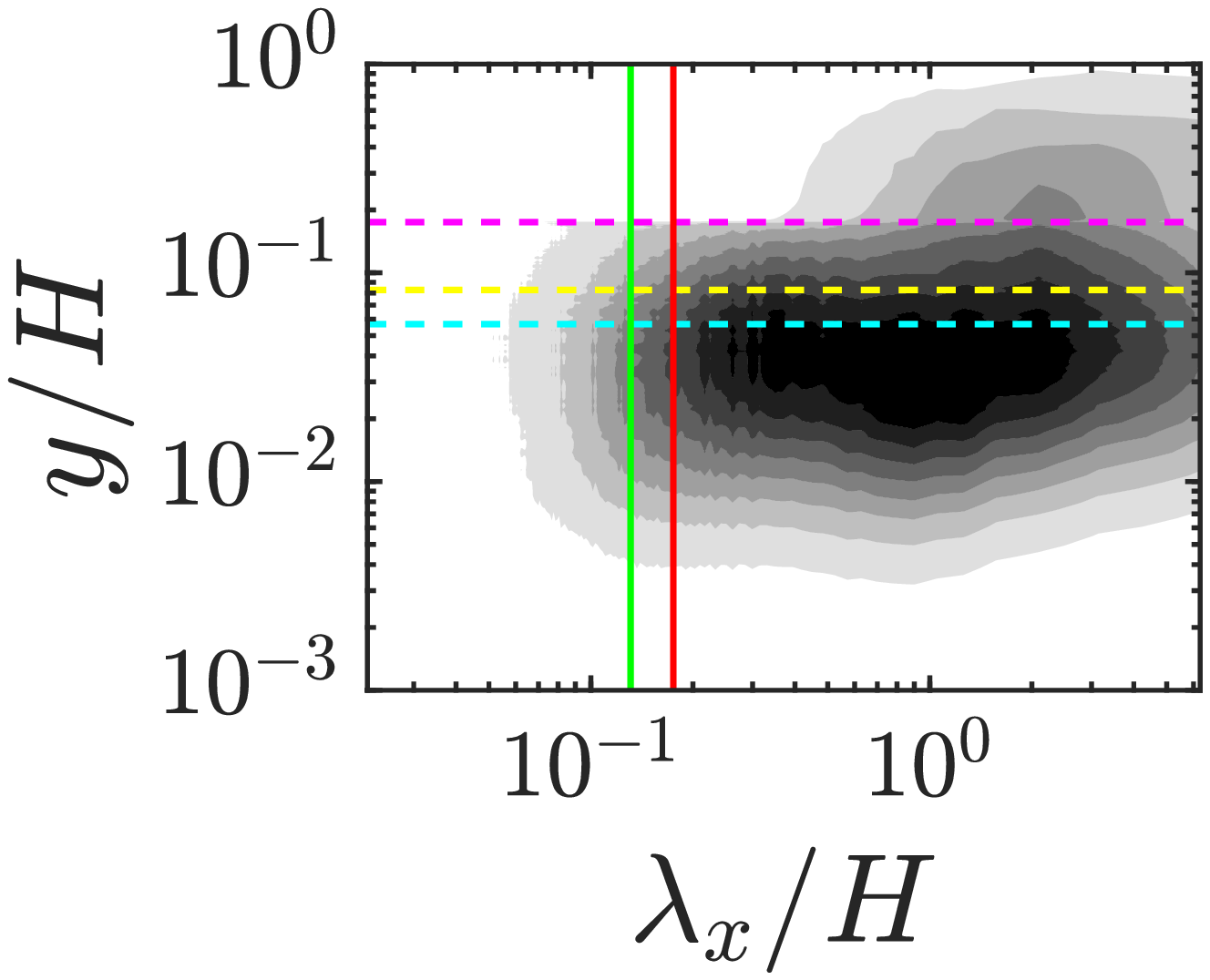}}
  \subfloat[]{\includegraphics[width=0.25\linewidth]{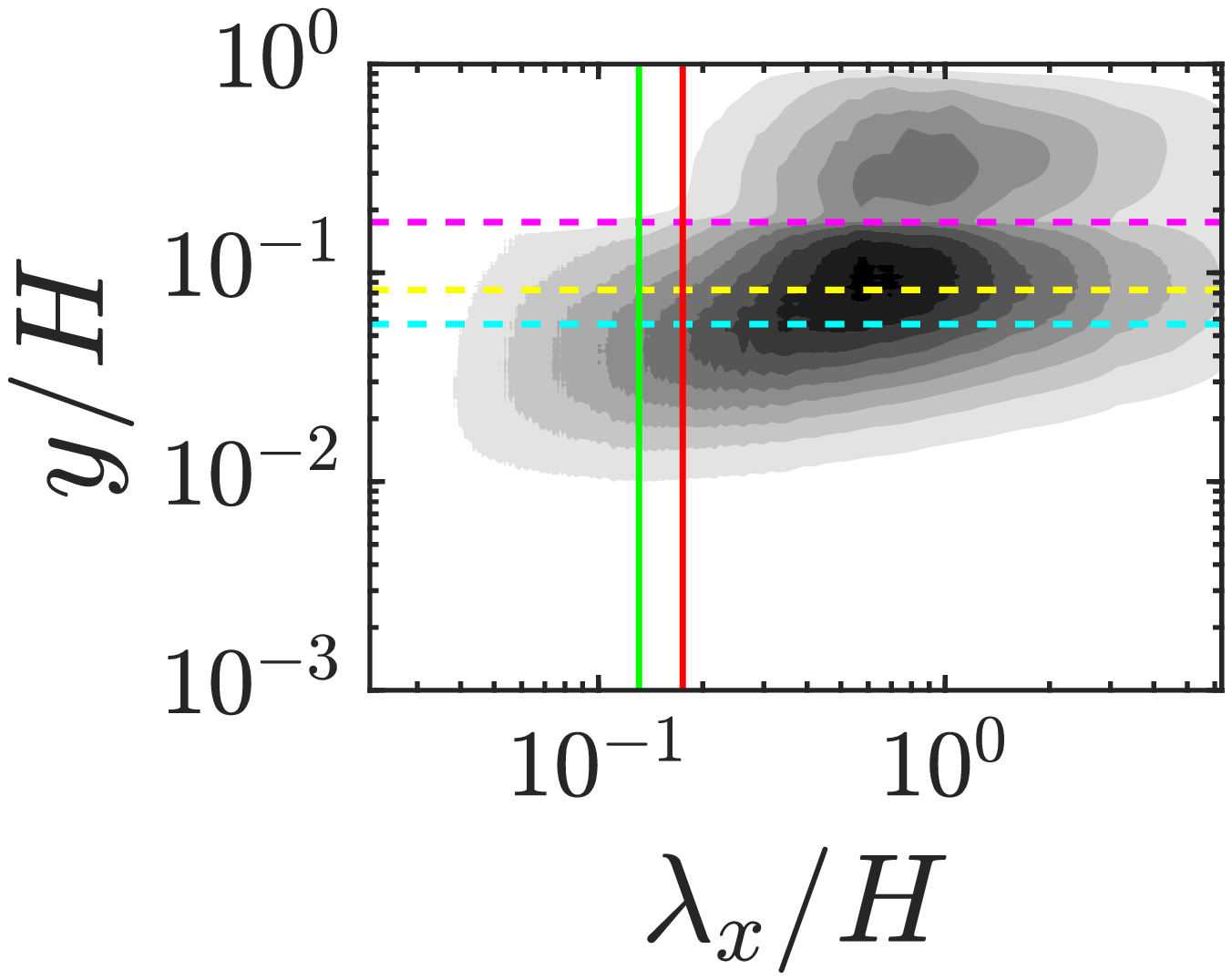}}
  \subfloat[]{\includegraphics[width=0.25\linewidth]{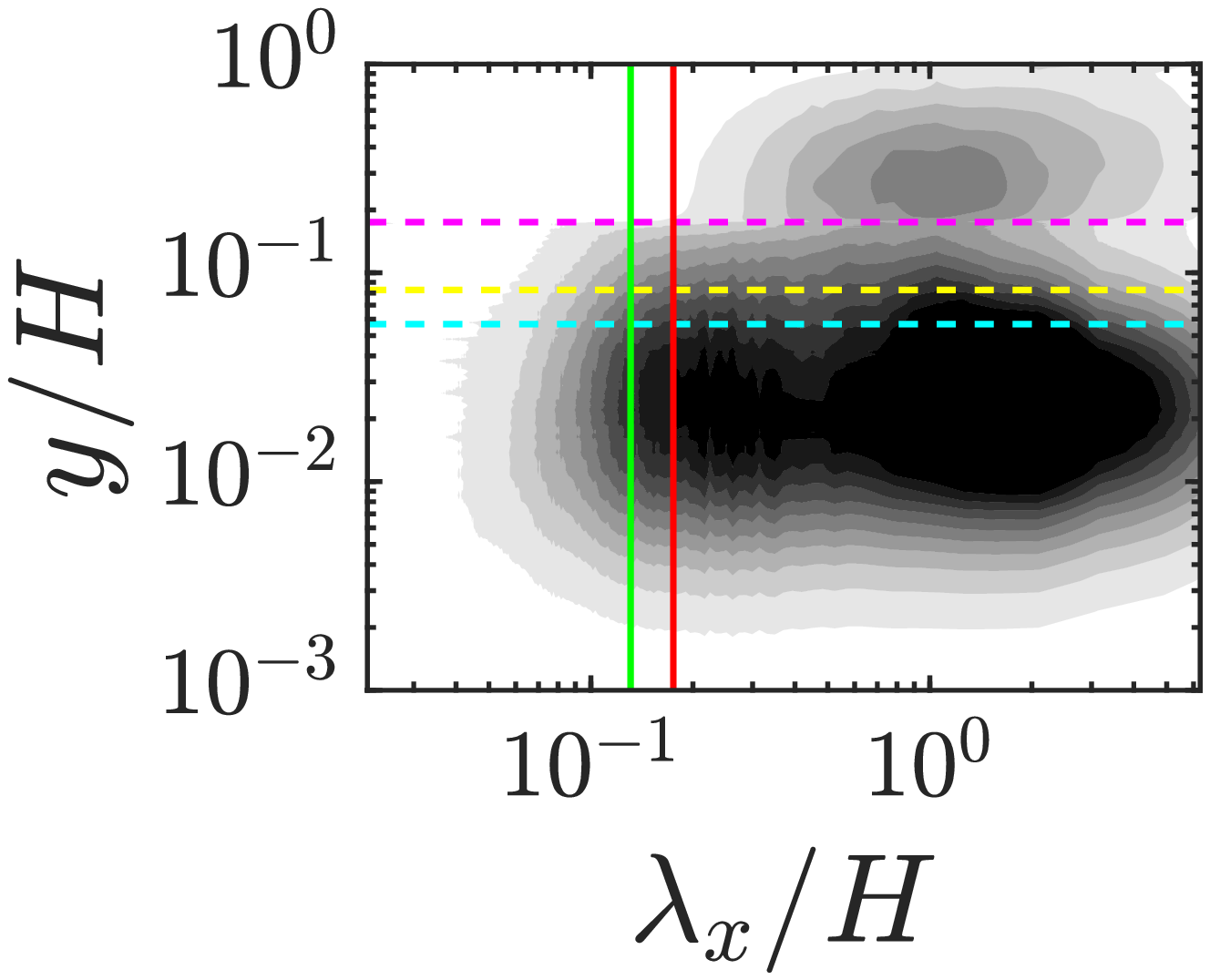}}
  \subfloat[]{\includegraphics[width=0.25\linewidth]{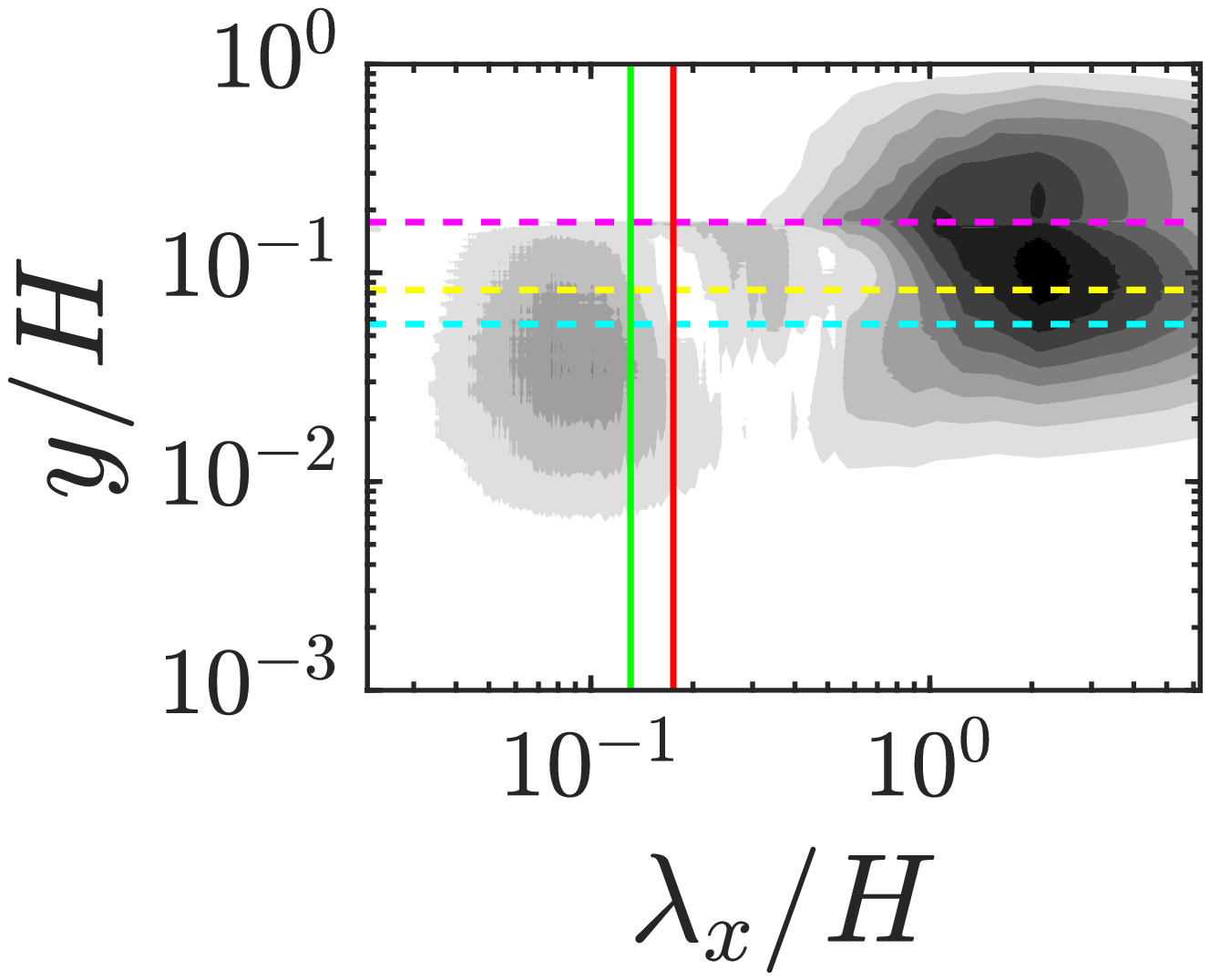}}\\
  \caption{Cases $\theta=\pm\ang{45}$. 
           Magnitude of the premultiplied spectra of the velocity components
           and cospectra of the Reynolds shear stress as a function of 
           the streamwise wavelength $\lambda_x/H$ and the wall-normal coordinates
           $y/H$. Top row: $\theta=\ang{45}$; bottom row $\theta=-\ang{45}$. 
           From left to right, the columns are:
           $\kappa_x\Phi_{u'u'}/u_{\tau,l}^2$ with 
           grey levels range in $[0,0.8]$ with a $0.1$ increment; 
           $\kappa_x\Phi_{v'v'}/u_{\tau,l}^2$ with grey levels range in 
           $[0,0.3]$ with a $0.03$ increment; 
           $\kappa_x\Phi_{w'w'}/u_{\tau,l}^2$ 
           with grey levels range in $[0,0.5]$ with a $0.05$ increment;
           $\kappa_x|\Phi_{u'v'}|/u_{\tau,l}^2$ 
           with grey levels range in $[0,0.4]$ with a $0.02$ increment.
           Colour lines have the same meaning as in 
           \cref{fig:premultipliedSpectraV0d250}.}
  \label{fig:premultipliedSpectraBFx0d175}
  \end{figure}
\begin{figure}
  \centering
  \subfloat[]{\includegraphics[width=0.25\linewidth]{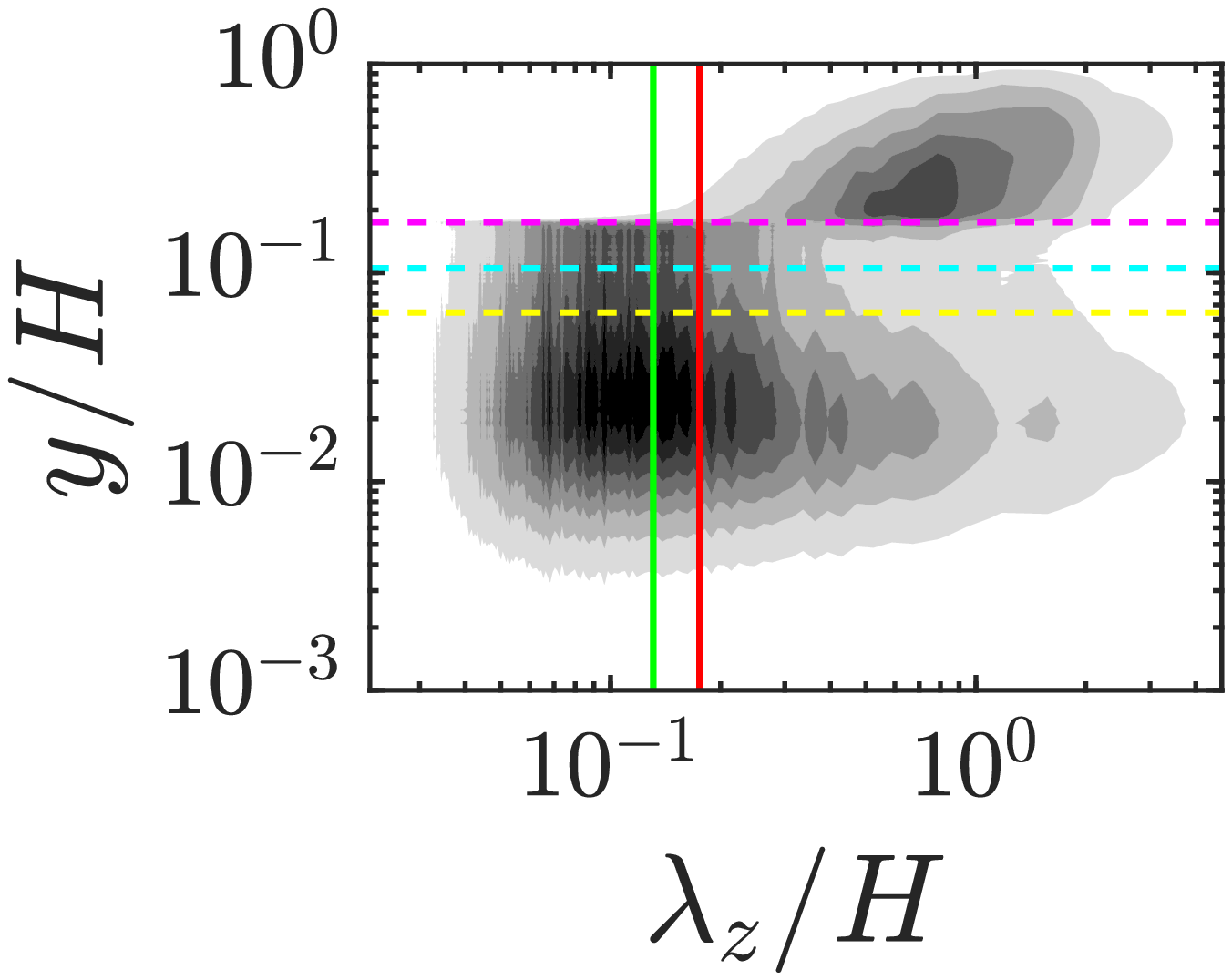}}
  \subfloat[]{\includegraphics[width=0.25\linewidth]{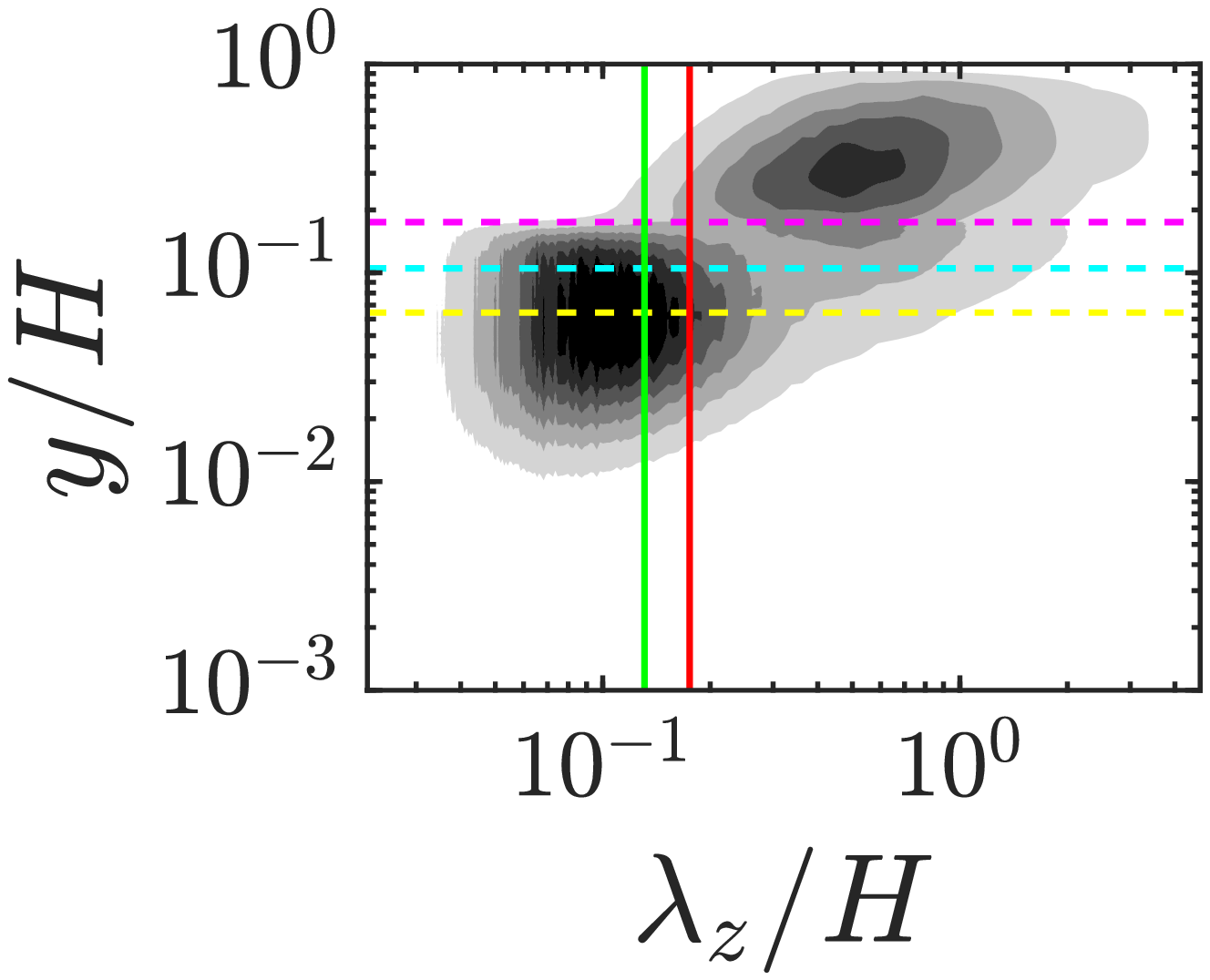}}
  \subfloat[]{\includegraphics[width=0.25\linewidth]{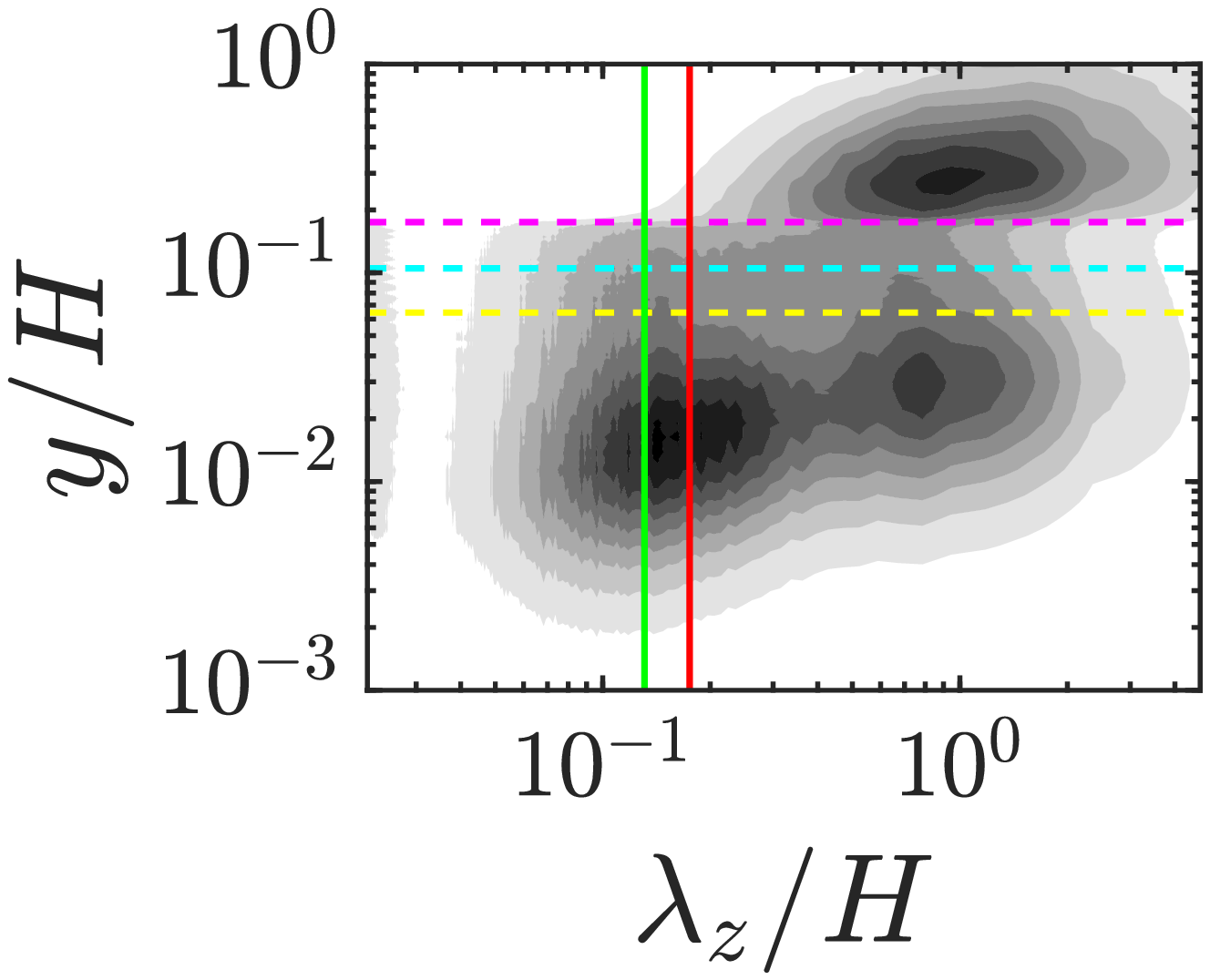}}
  \subfloat[]{\includegraphics[width=0.25\linewidth]{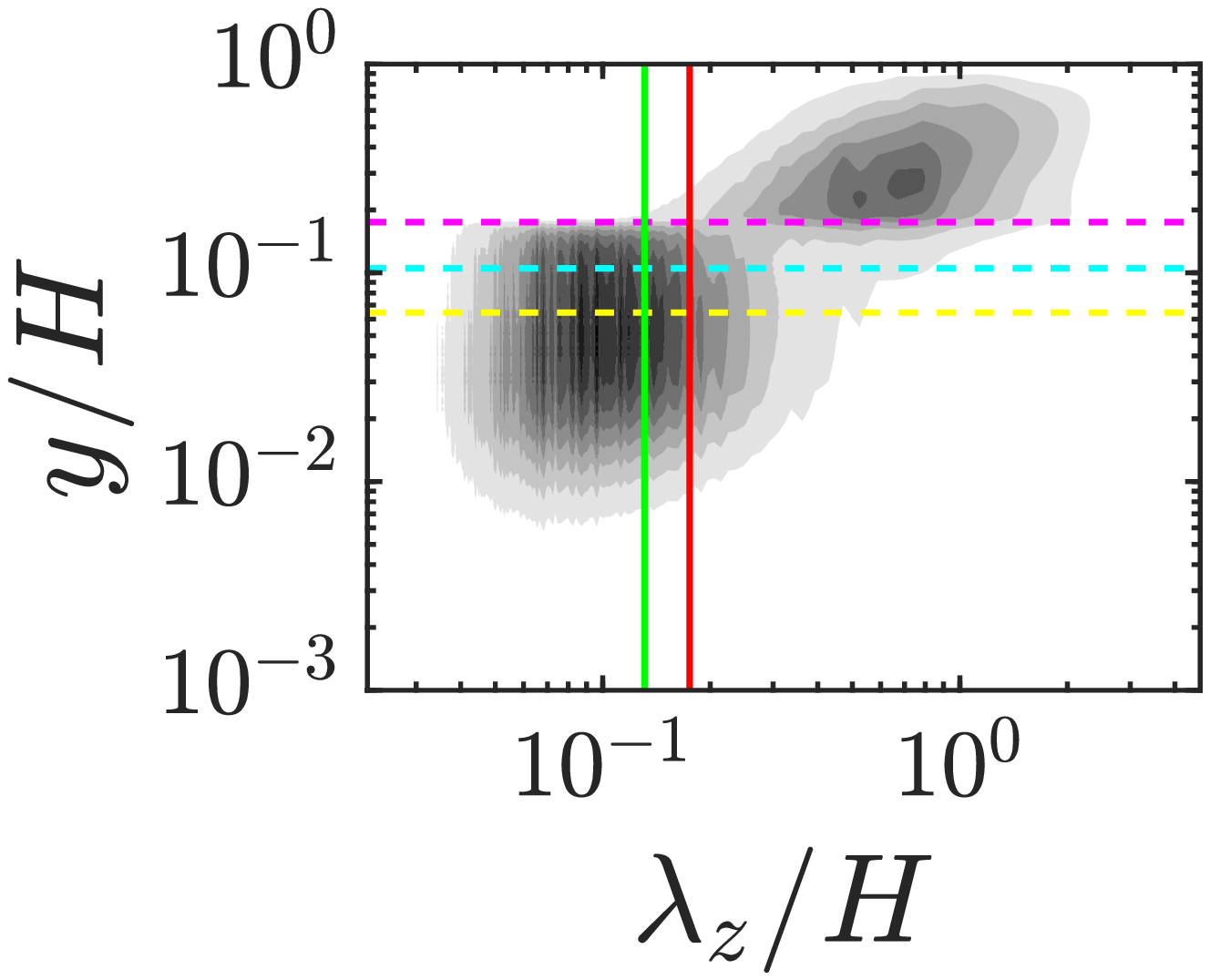}}\\
  \subfloat[]{\includegraphics[width=0.25\linewidth]{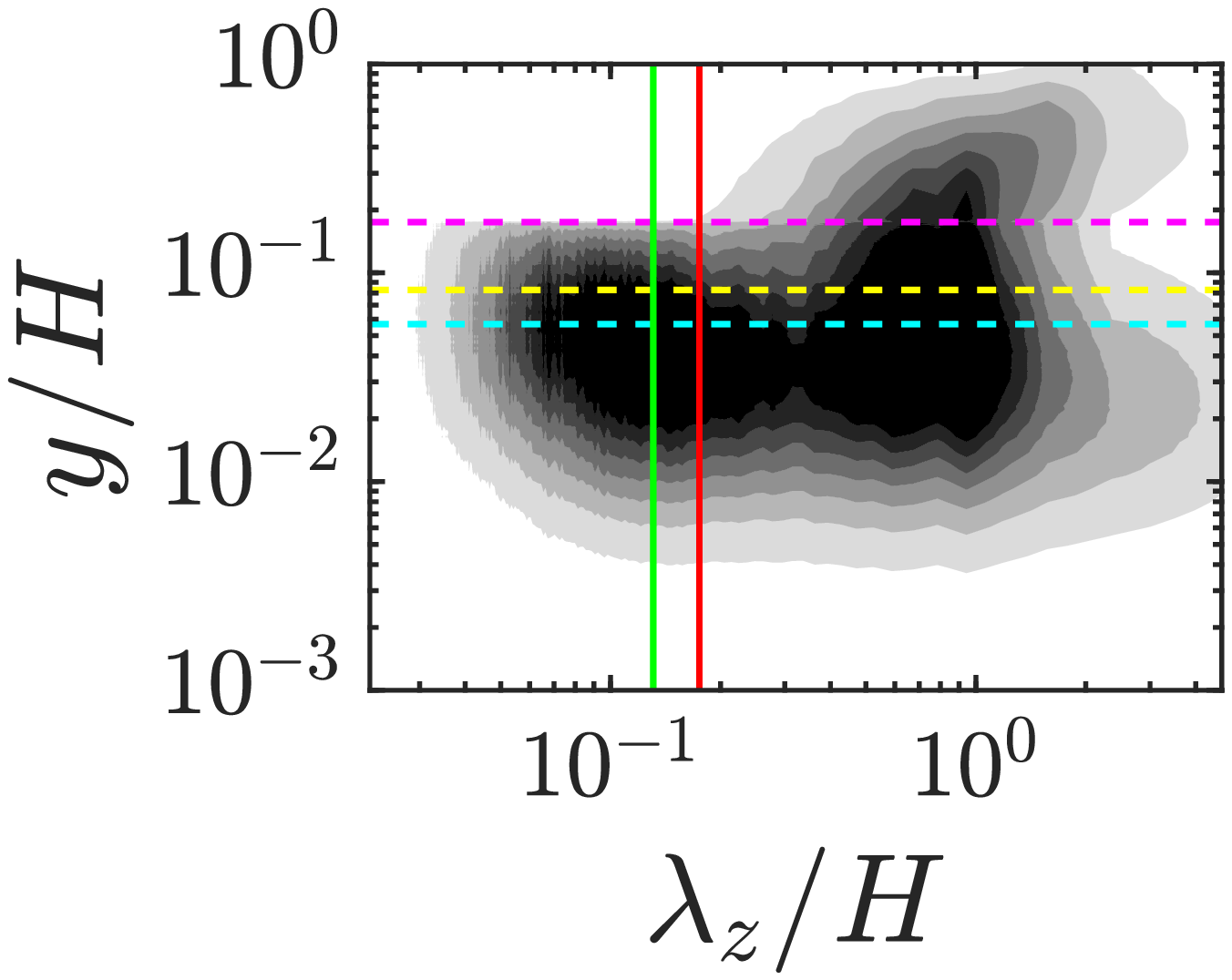}}
  \subfloat[]{\includegraphics[width=0.25\linewidth]{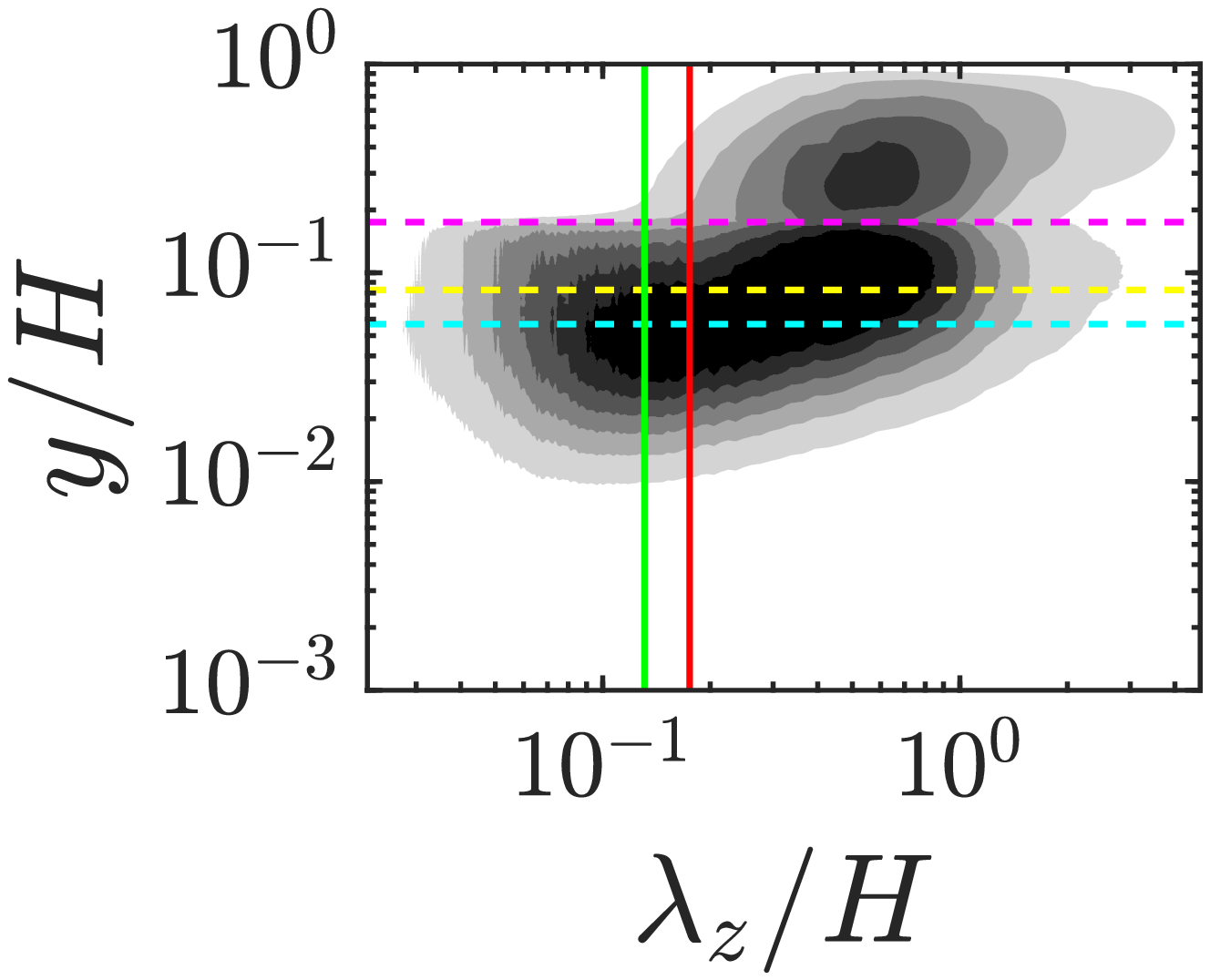}}
  \subfloat[]{\includegraphics[width=0.25\linewidth]{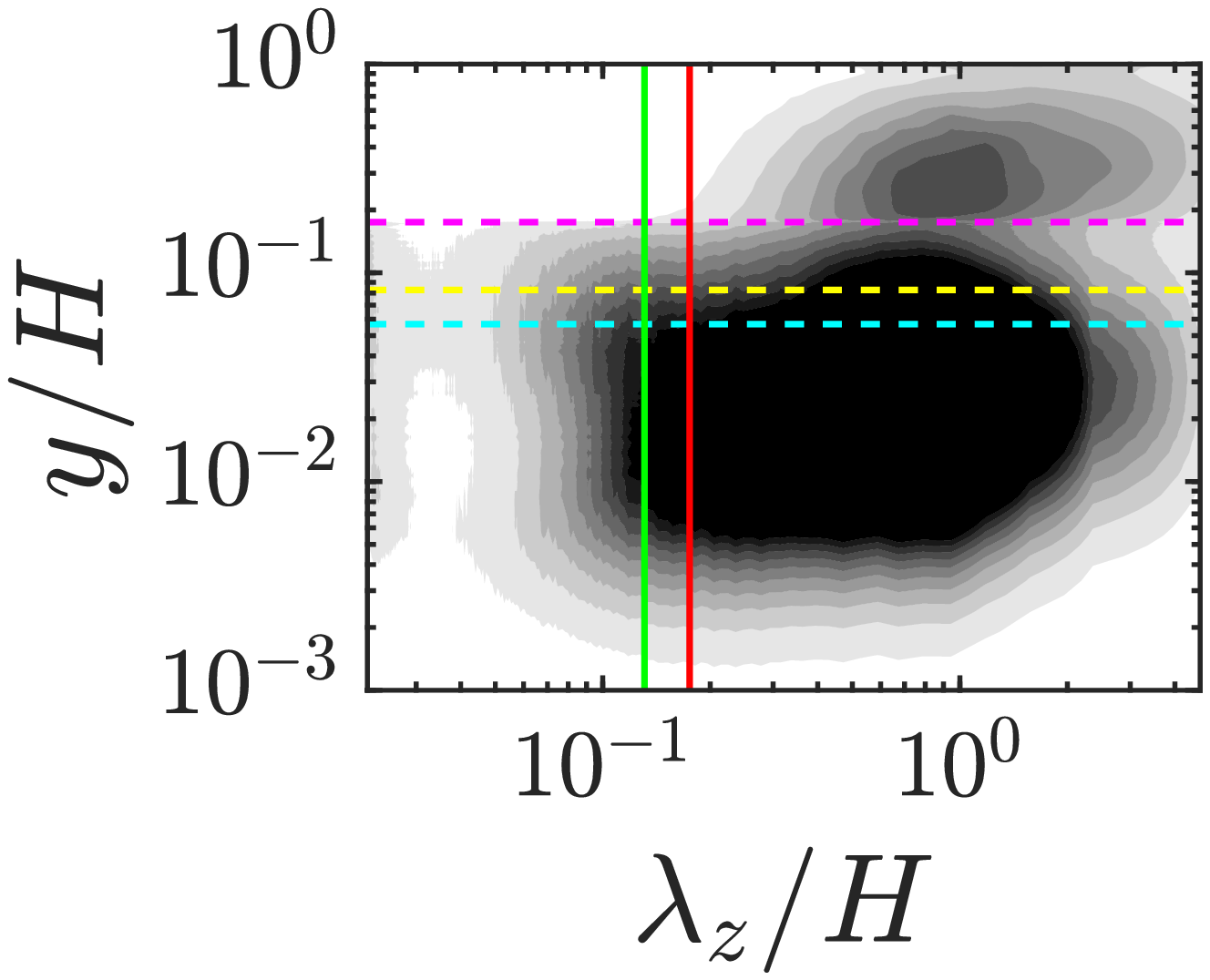}}
  \subfloat[]{\includegraphics[width=0.25\linewidth]{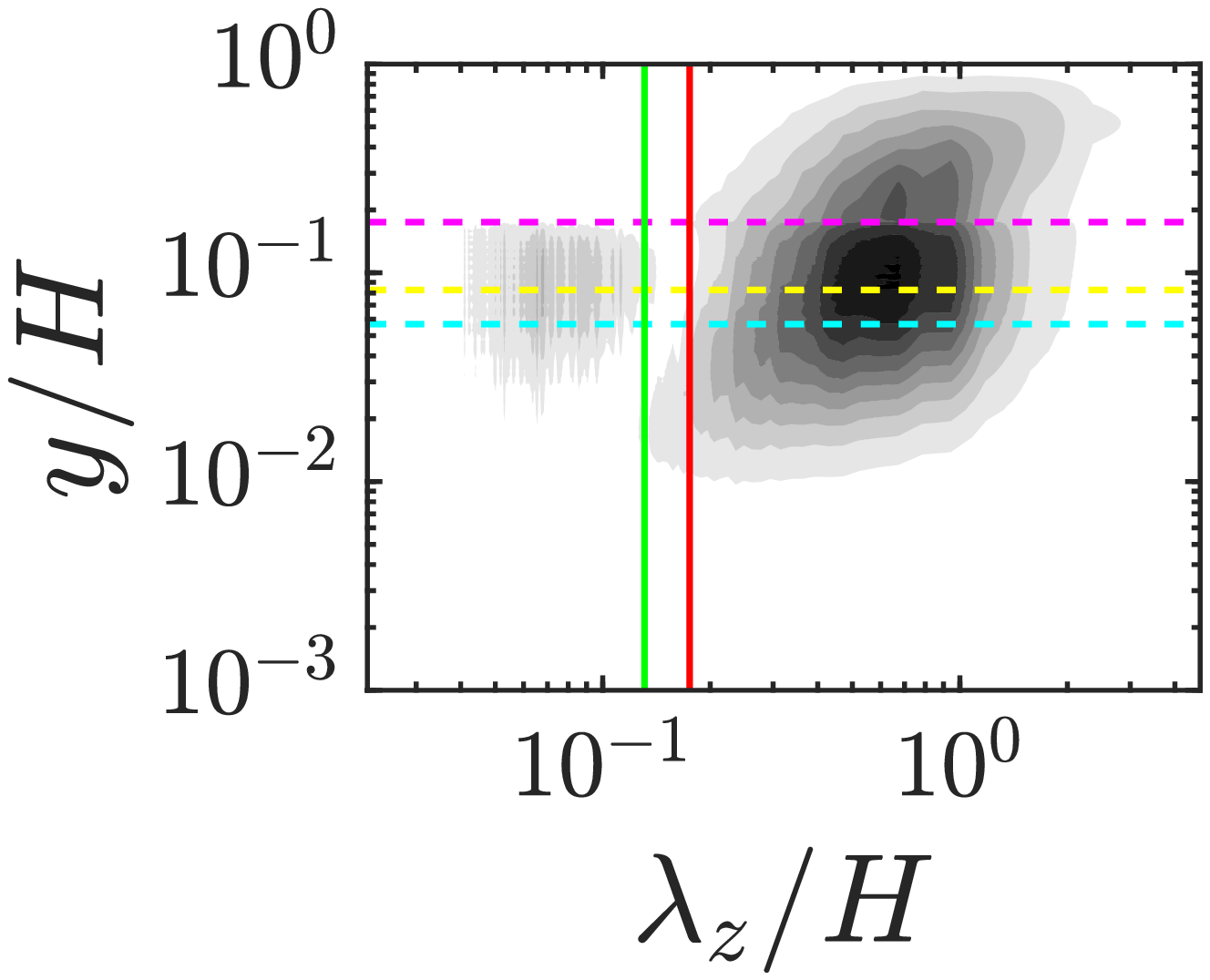}}\\
  \caption{Cases $\theta=\pm\ang{45}$. 
           Magnitude of the premultiplied spectra of the velocity components
           and cospectra of the Reynolds shear stress as a function of 
           the spanwise wavelength $\lambda_z/H$ and the wall-normal coordinates
           $y/H$. Top row: $\theta=\ang{45}$; bottom row $\theta=-\ang{45}$. 
           From left to right, the columns are:
           $\kappa_z\Phi_{u'u'}/u_{\tau,l}^2$ with 
           grey levels range in $[0,0.8]$ with a $0.1$ increment; 
           $\kappa_z\Phi_{v'v'}/u_{\tau,l}^2$ with grey levels range in 
           $[0,0.3]$ with a $0.03$ increment; 
           $\kappa_z\Phi_{w'w'}/u_{\tau,l}^2$ 
           with grey levels range in $[0,0.5]$ with a $0.05$ increment;
           $\kappa_z|\Phi_{u'v'}|/u_{\tau,l}^2$ 
           with grey levels range in $[0,0.4]$ with a $0.02$ increment.
           Colour lines have the same meaning as in 
           \cref{fig:premultipliedSpectraV0d250}.}
  \label{fig:premultipliedSpectraBFz0d175}
  \end{figure}
Introducing the angle of inclination, the preferential channel of 
communication described above between the outer and 
the inner layer, i.e the wall-normal jets, is highly affected. 
To prove this, we consider the two scenarios simulated in
an inclined positive and negative direction, i.e. $\theta=\pm \ang{45}$
which display a clear separation 
between the outer and the inner layer.
We examine the one-dimensional premultiplied spectra of
the velocity fluctuations and the magnitude of the one-dimensional 
premultiplied cospectra of the Reynolds shear stress, as a function of 
the distance from the wall. In particular, 
\cref{fig:premultipliedSpectraBFx0d175,fig:premultipliedSpectraBFz0d175} 
offer a direct comparison of 
the velocity structures of the two cases considered (top row $\theta=\ang{45}$
and bottom row $\theta=-\ang{45}$) as a function of the streamwise and spanwise
wavelengths, respectively. 
\Cref{fig:premultipliedSpectraBFx0d175,fig:premultipliedSpectraBFz0d175} confirm
the similarity of the spectral energy content of the horizontal velocity fluctuations
$u'$ and $w'$, with the presence of the three peaks, two of which within the canopy layer
(panels (a), (c), (e) and (g)). However, differences arise in the
wall-normal component of the velocity fluctuations, with the disappearance of the 
peak with large wavelengths within the canopy region for the case forwardly inclined (with the grain - panel (b)).
The latter consideration is the fundamental attribute added by the angle of inclination
$\theta$: when the canopy is forwardly inclined, the large jets that arise from the outer
region and tend to penetrate within the canopy layer are shielded and weakened by the
inclined stems, thus affecting the momentum transfer that takes place from the outer
to the inner layer; on the contrary, when the filaments are backwardly
inclined, the canopy configuration enhances the penetrations of the jets, 
largely influencing and strengthening the coherent turbulent structures living there. 
As a further confirmation, we consider panels (a), (c), (e) and (g) of
\cref{fig:premultipliedSpectraBFx0d175,fig:premultipliedSpectraBFz0d175},
i.e. the energy content of the streamwise and spanwise component of the 
velocity fluctuations. 
For the case $\theta=\ang{45}$, i.e. panels (a) and (c), the magnitude of the peaks within 
the inner layer is lower compared to panels (e) and (f), i.e. $\theta=-\ang{45}$, 
while in the outer region, the intensity of the peaks are higher in the first two
panels. This corroborates the role of the jets in transferring momentum from the outer
to inner layer and shows the impedance effect of the angle of inclination of the stems,
which is larger for positive angles.

\begin{figure}
  \centering
  \subfloat[]{\includegraphics[width=0.25\linewidth]{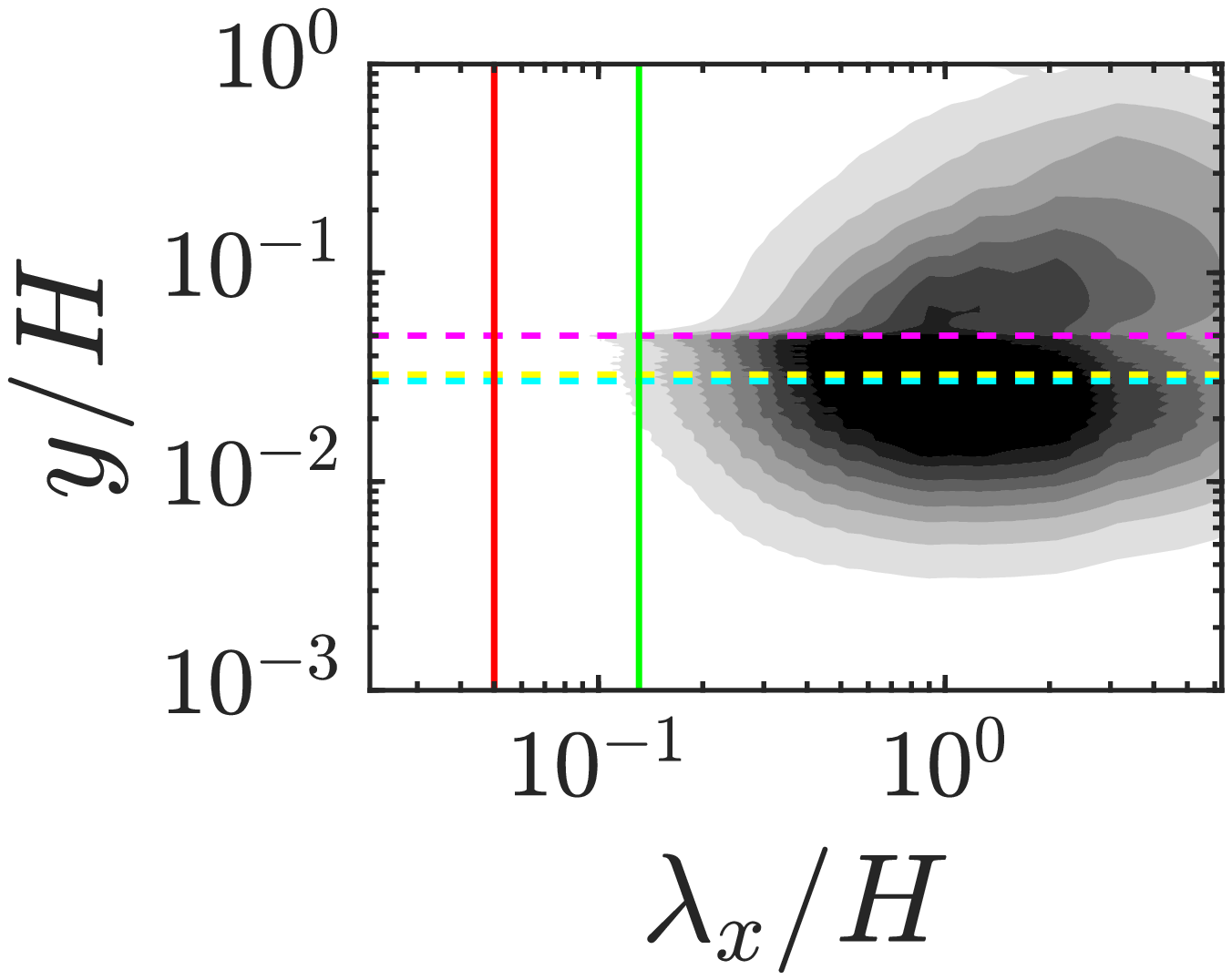}}
  \subfloat[]{\includegraphics[width=0.25\linewidth]{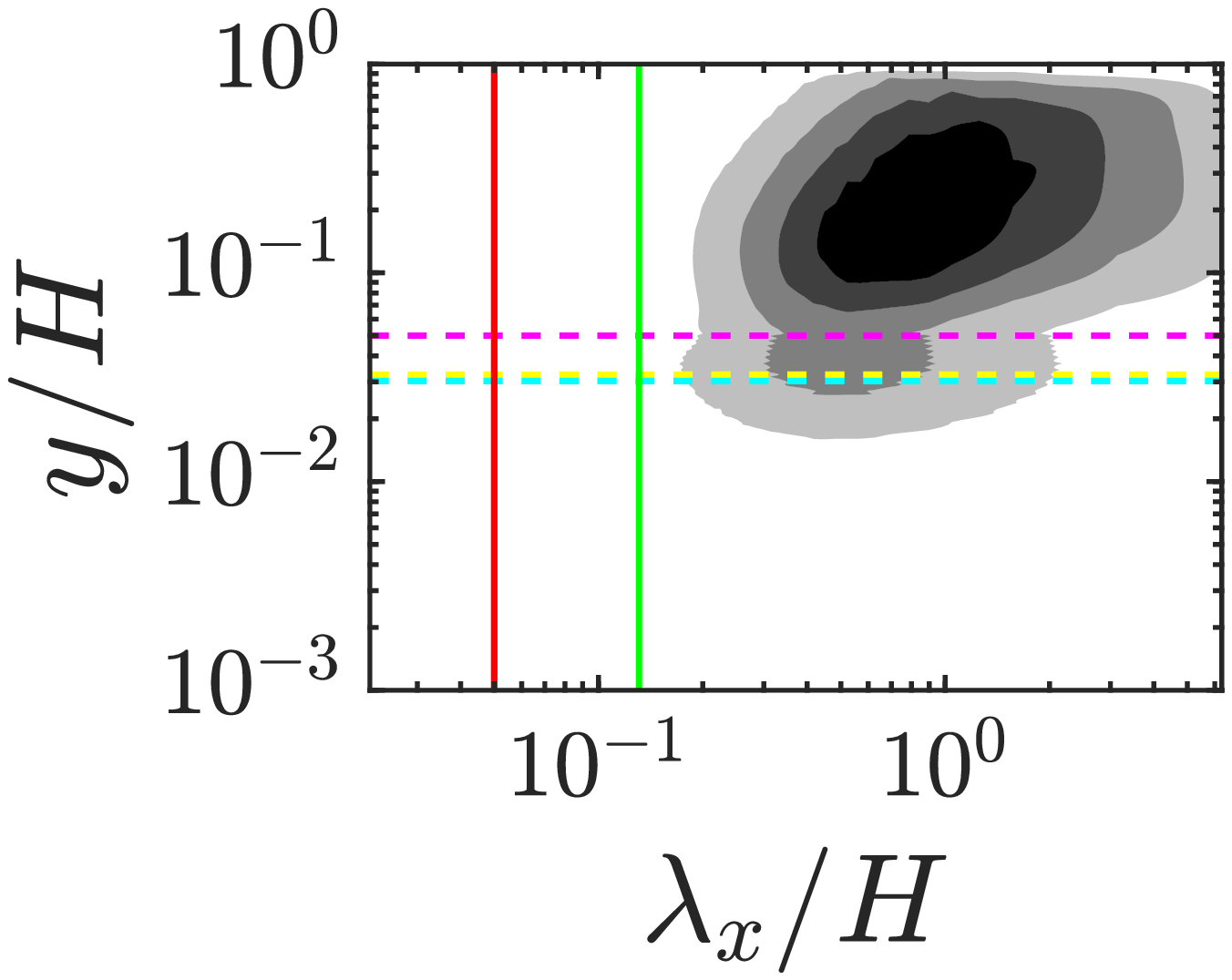}}
  \subfloat[]{\includegraphics[width=0.25\linewidth]{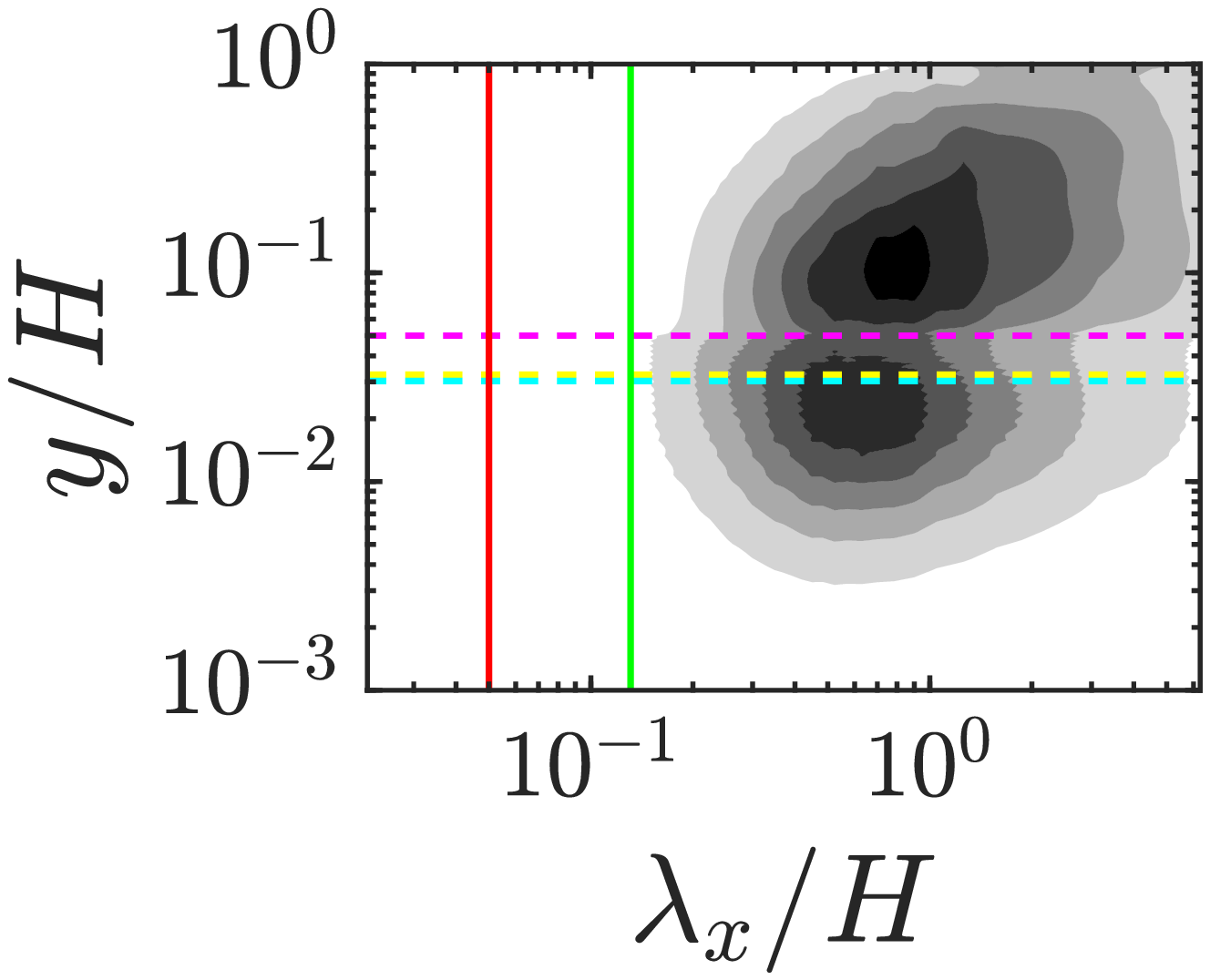}}
  \subfloat[]{\includegraphics[width=0.25\linewidth]{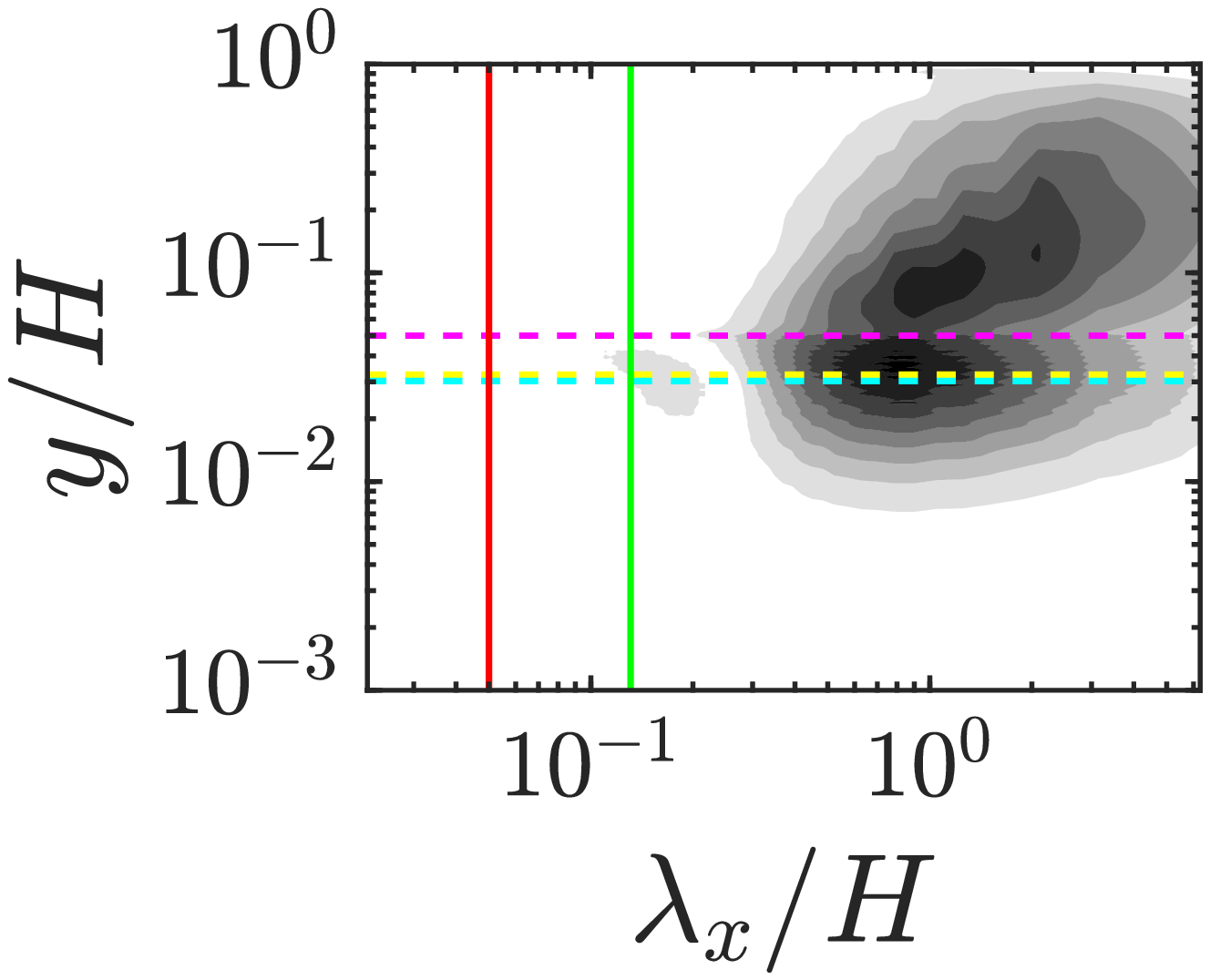}}\\
  \subfloat[]{\includegraphics[width=0.25\linewidth]{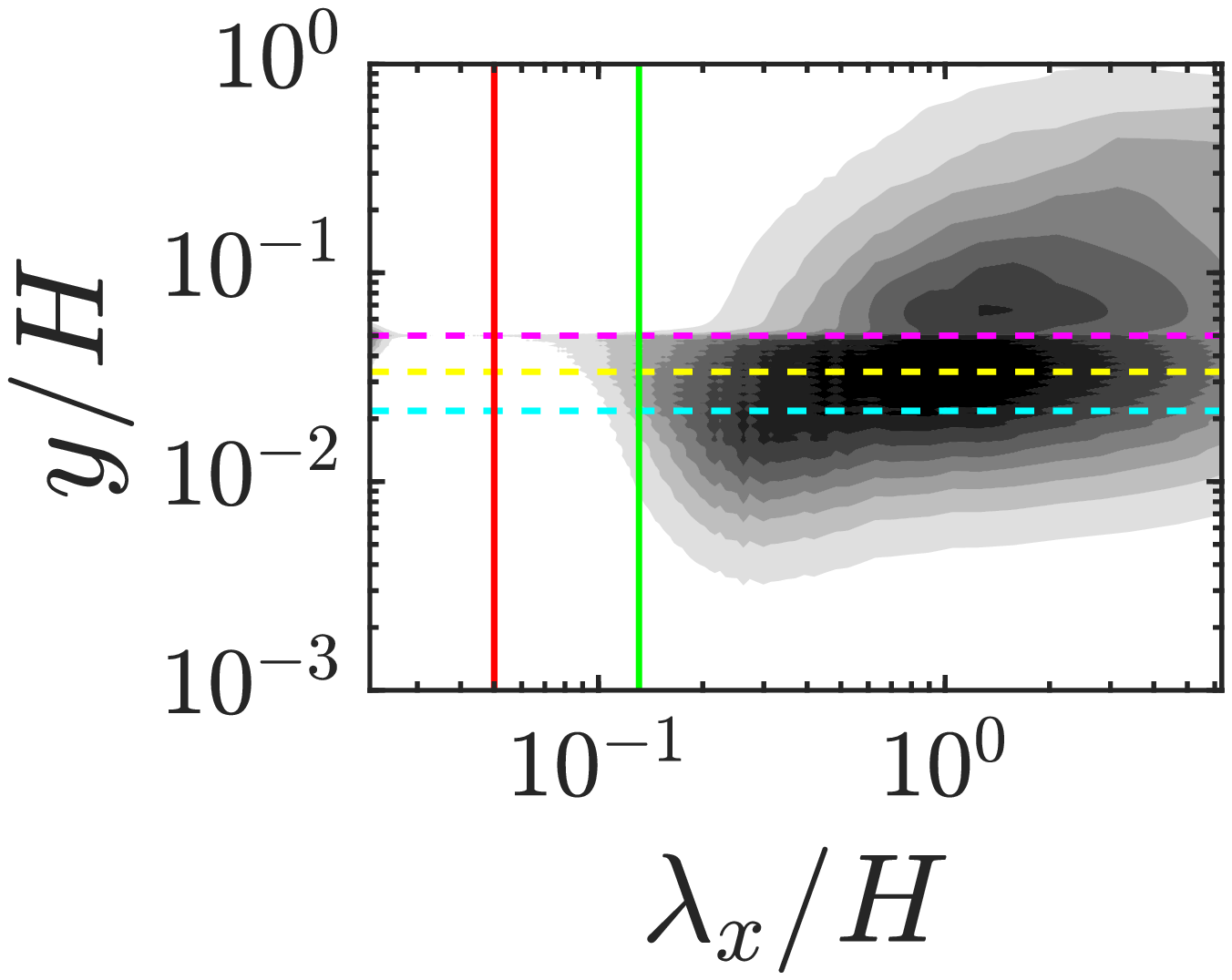}}
  \subfloat[]{\includegraphics[width=0.25\linewidth]{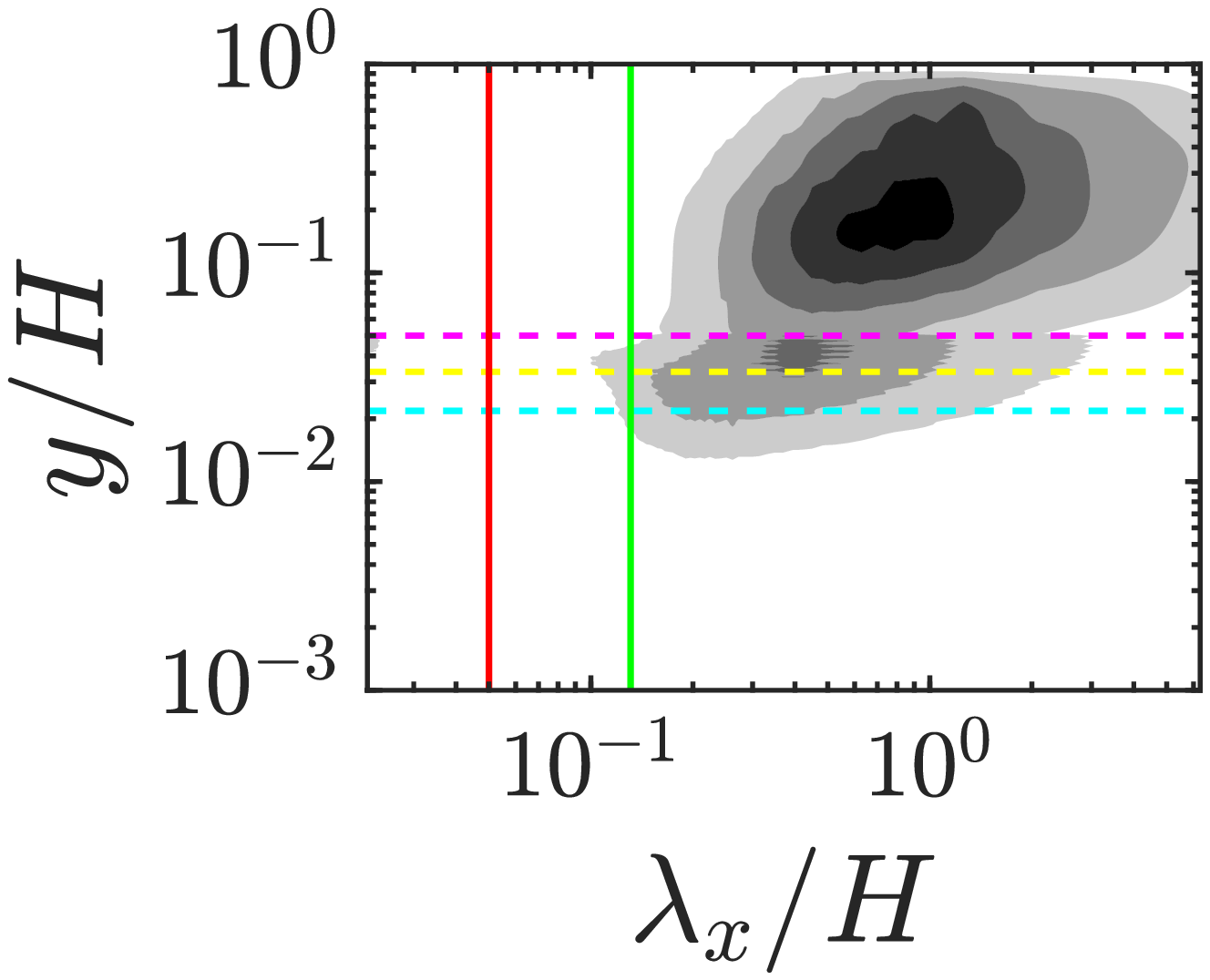}}
  \subfloat[]{\includegraphics[width=0.25\linewidth]{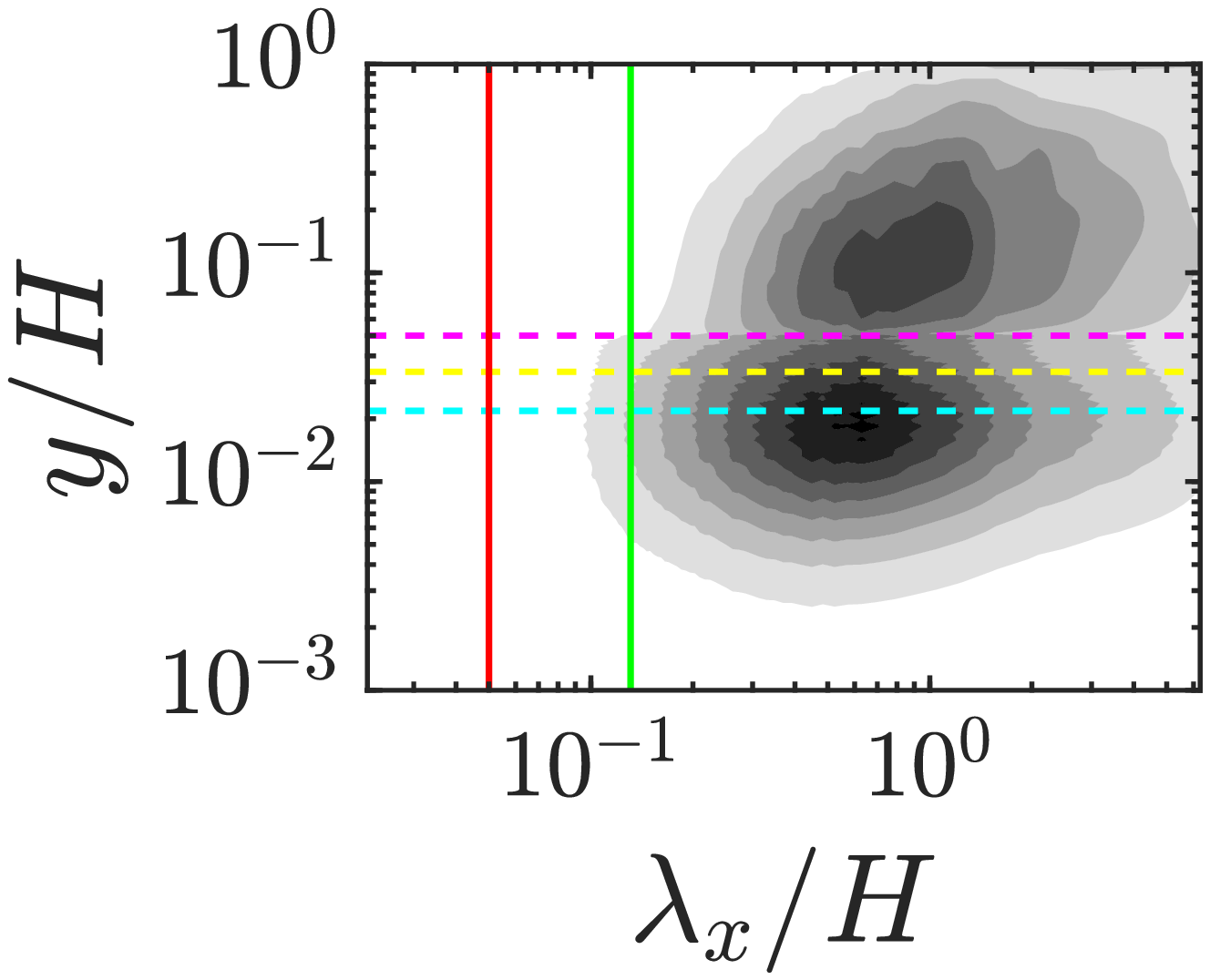}}
  \subfloat[]{\includegraphics[width=0.25\linewidth]{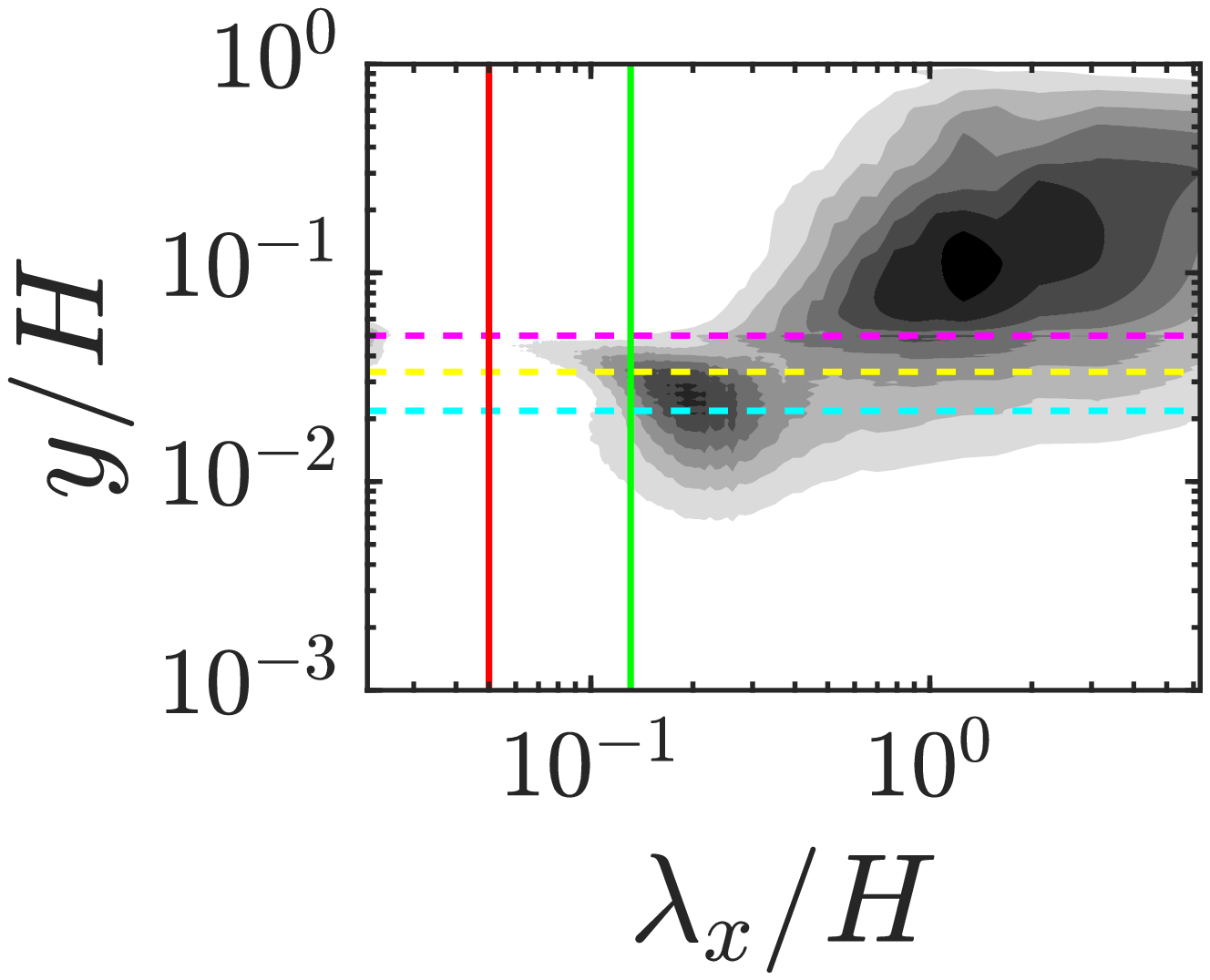}}\\
  \caption{Cases $\theta=\pm\ang{78.5}$. 
           Magnitude of the premultiplied spectra of the velocity components
           and cospectra of the Reynolds shear stress as a function of 
           the streamwise wavelength $\lambda_x/H$ and the wall-normal coordinates
           $y/H$. Top row: $\theta=\ang{78.5}$; bottom row $\theta=-\ang{78.5}$. 
           From left to right, the columns are:
           $\kappa_x\Phi_{u'u'}/u_{\tau,l}^2$ with 
           grey levels range in $[0,0.8]$ with a $0.1$ increment; 
           $\kappa_x\Phi_{v'v'}/u_{\tau,l}^2$ with grey levels range in 
           $[0,0.3]$ with a $0.03$ increment; 
           $\kappa_x\Phi_{w'w'}/u_{\tau,l}^2$ 
           with grey levels range in $[0,0.5]$ with a $0.05$ increment;
           $\kappa_x|\Phi_{u'v'}|/u_{\tau,l}^2$ 
           with grey levels range in $[0,0.4]$ with a $0.02$ increment.
           Colour lines have the same meaning as in 
           \cref{fig:premultipliedSpectraV0d250}.}
  \label{fig:premultipliedSpectraBFx0d050}
  \end{figure}
\begin{figure}
  \centering
  \subfloat[]{\includegraphics[width=0.25\linewidth]{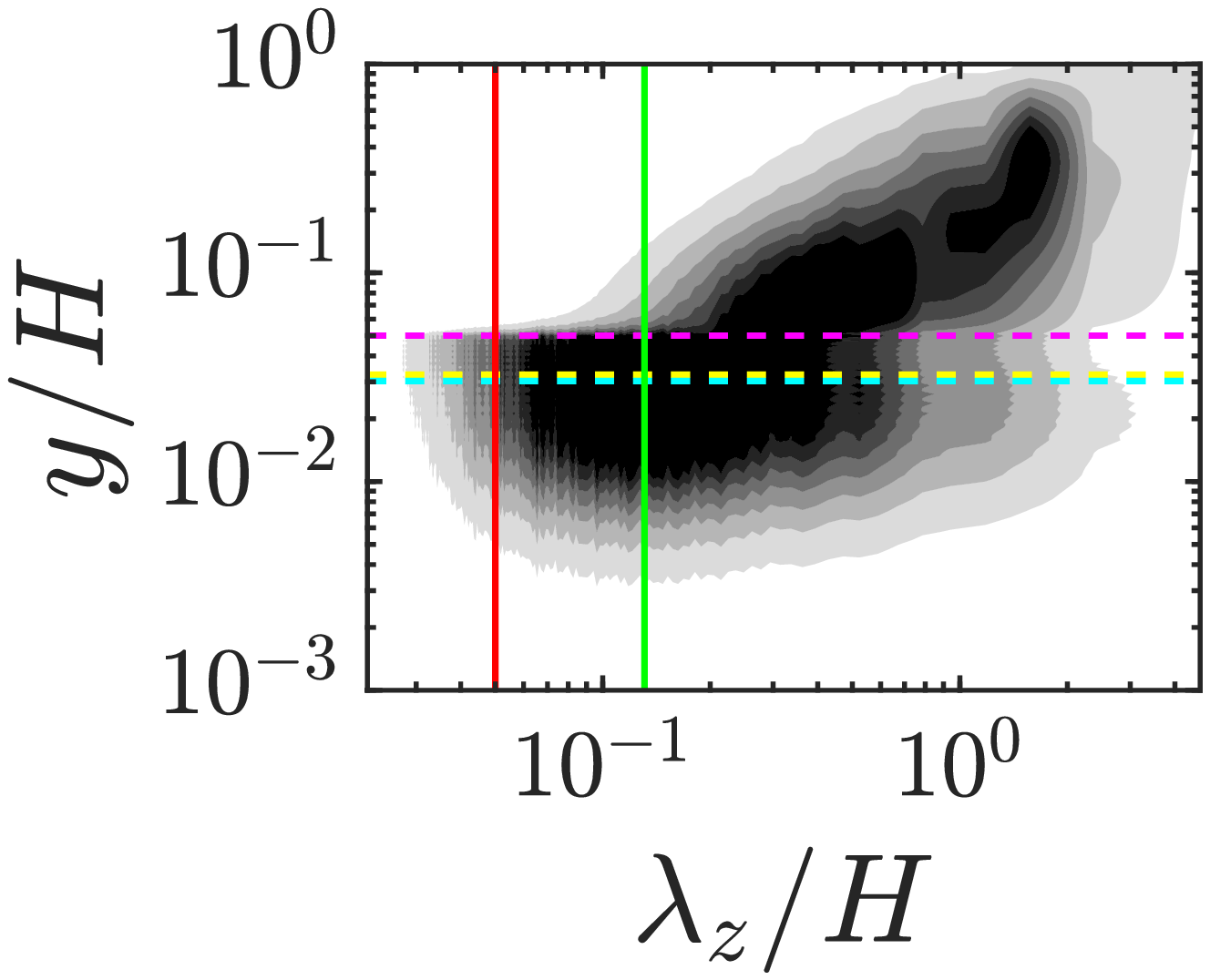}}
  \subfloat[]{\includegraphics[width=0.25\linewidth]{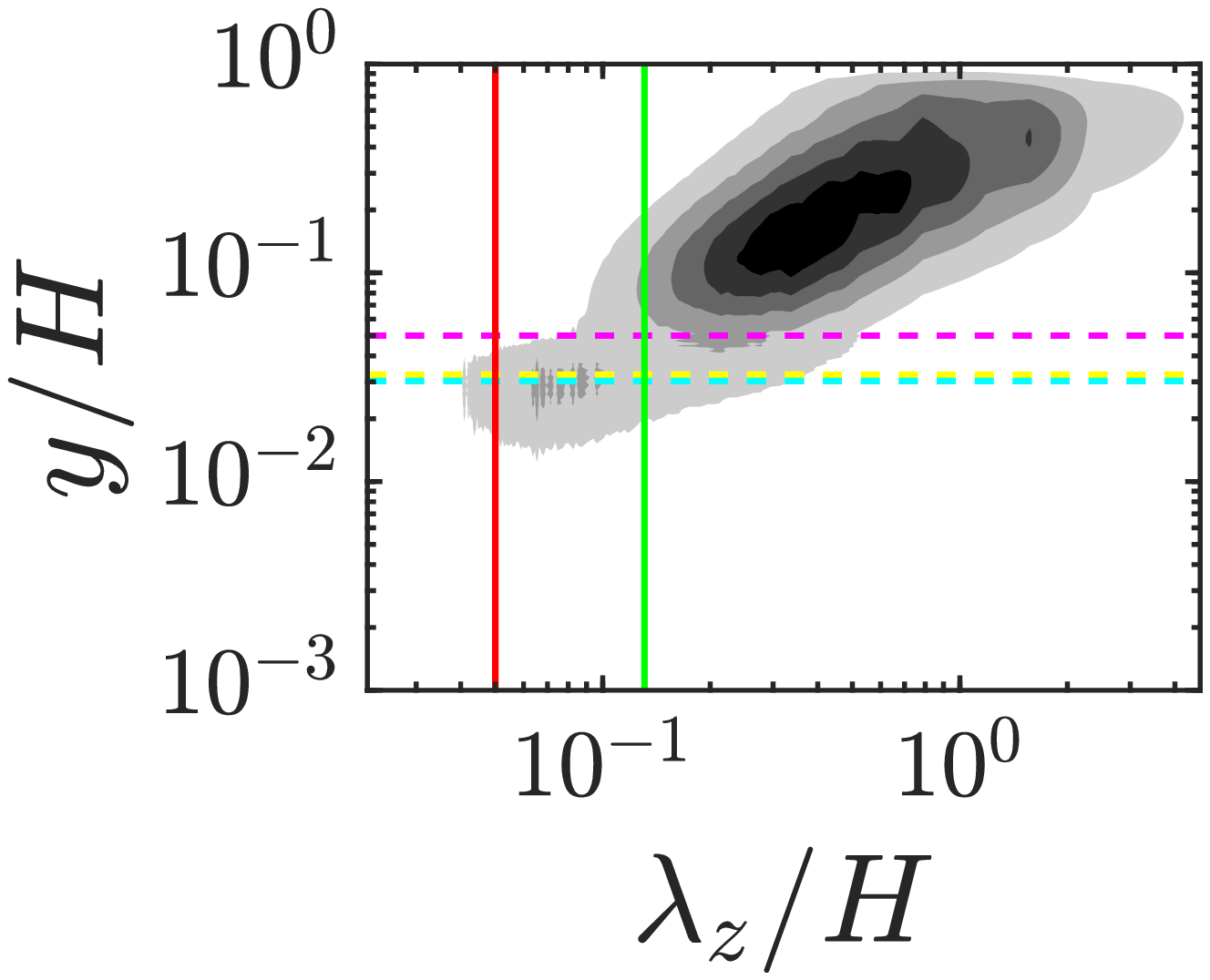}}
  \subfloat[]{\includegraphics[width=0.25\linewidth]{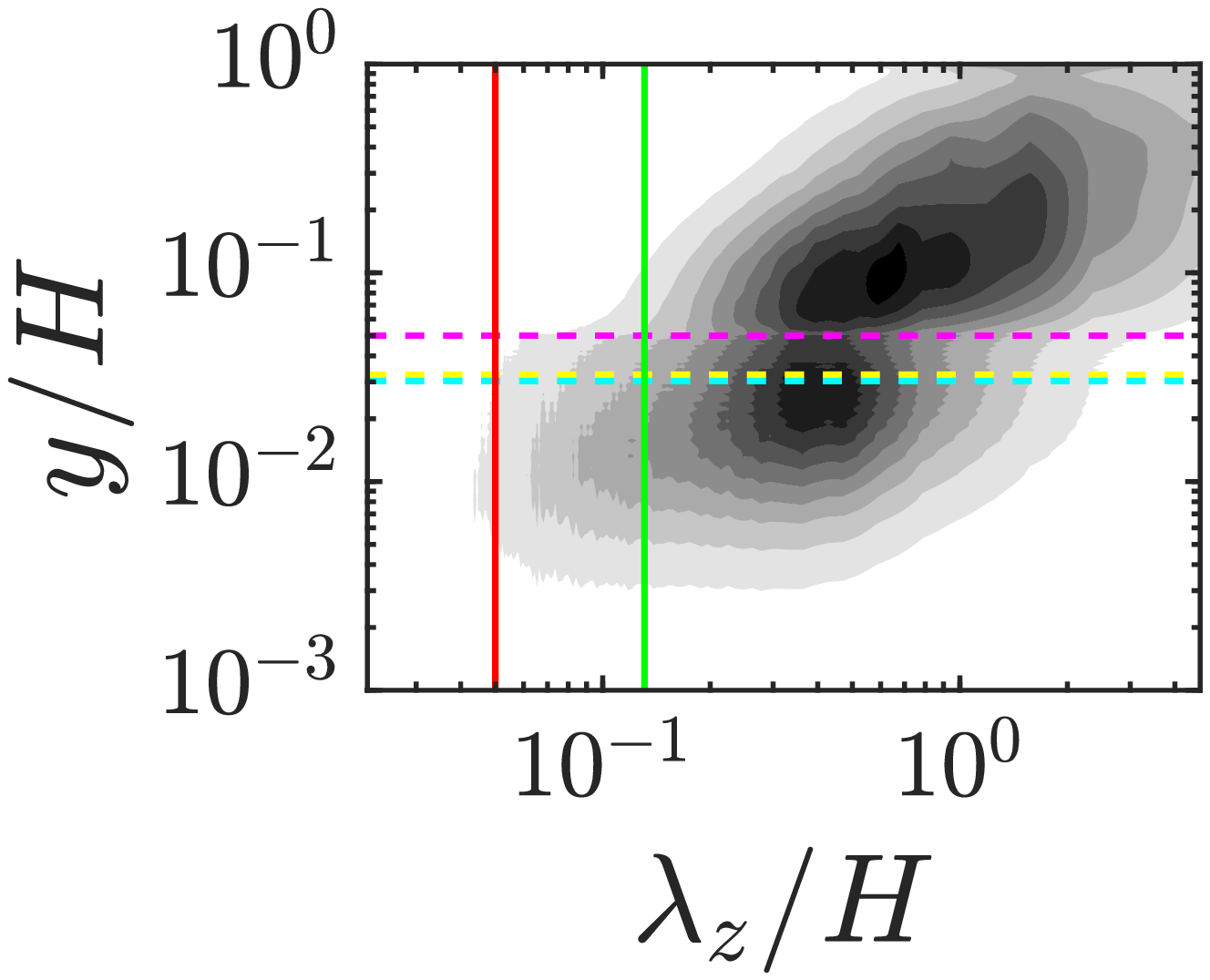}}
  \subfloat[]{\includegraphics[width=0.25\linewidth]{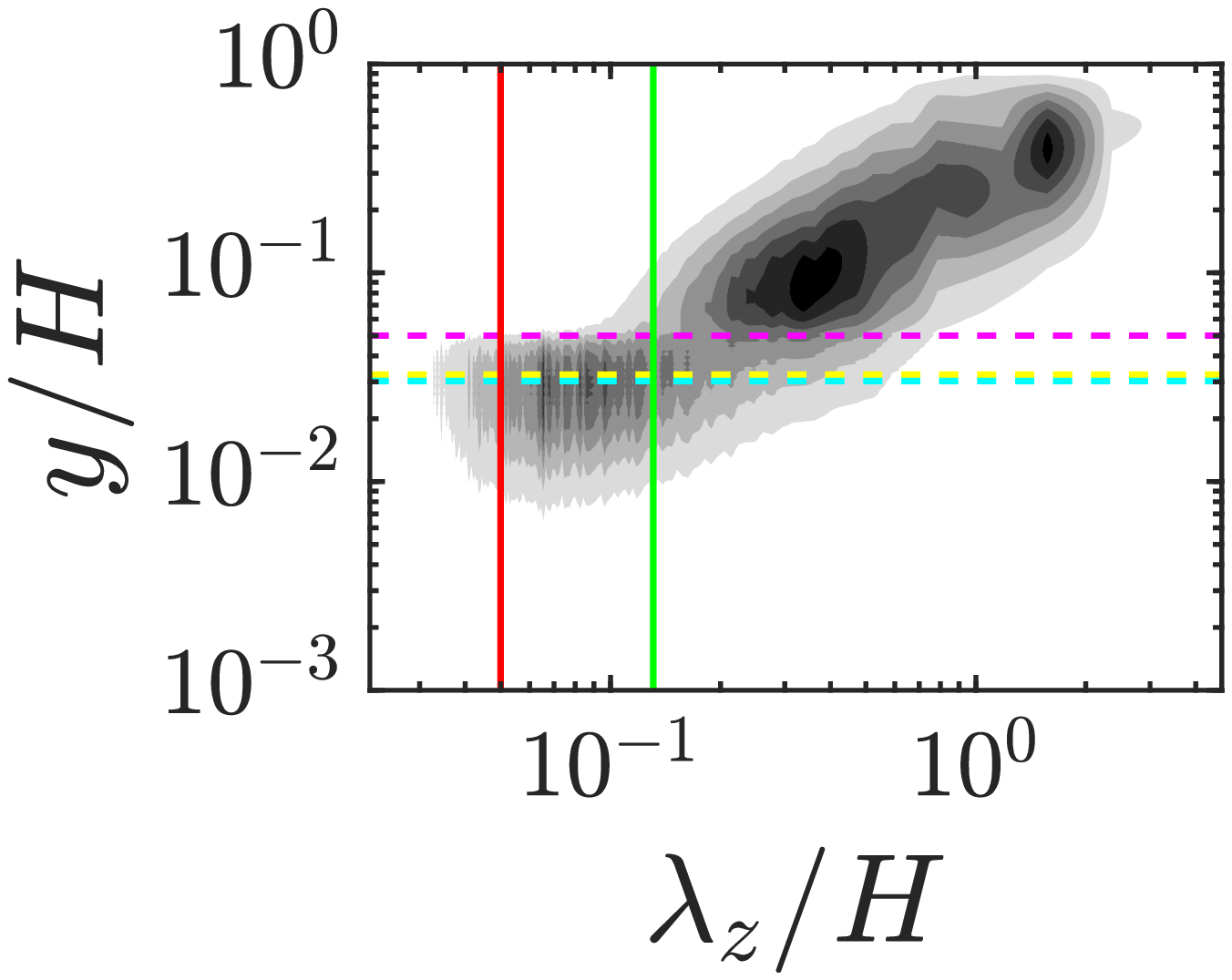}}\\
  \subfloat[]{\includegraphics[width=0.25\linewidth]{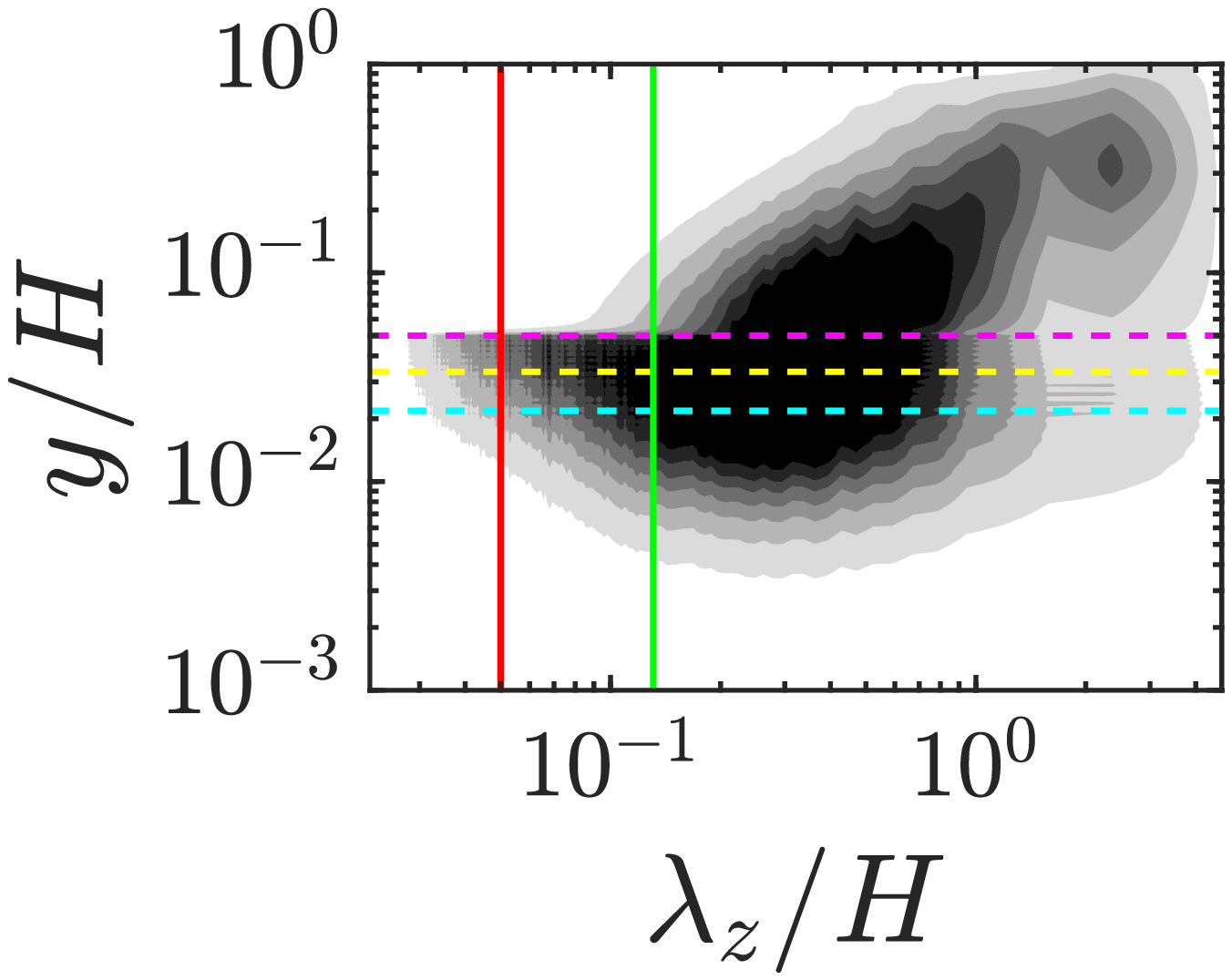}}
  \subfloat[]{\includegraphics[width=0.25\linewidth]{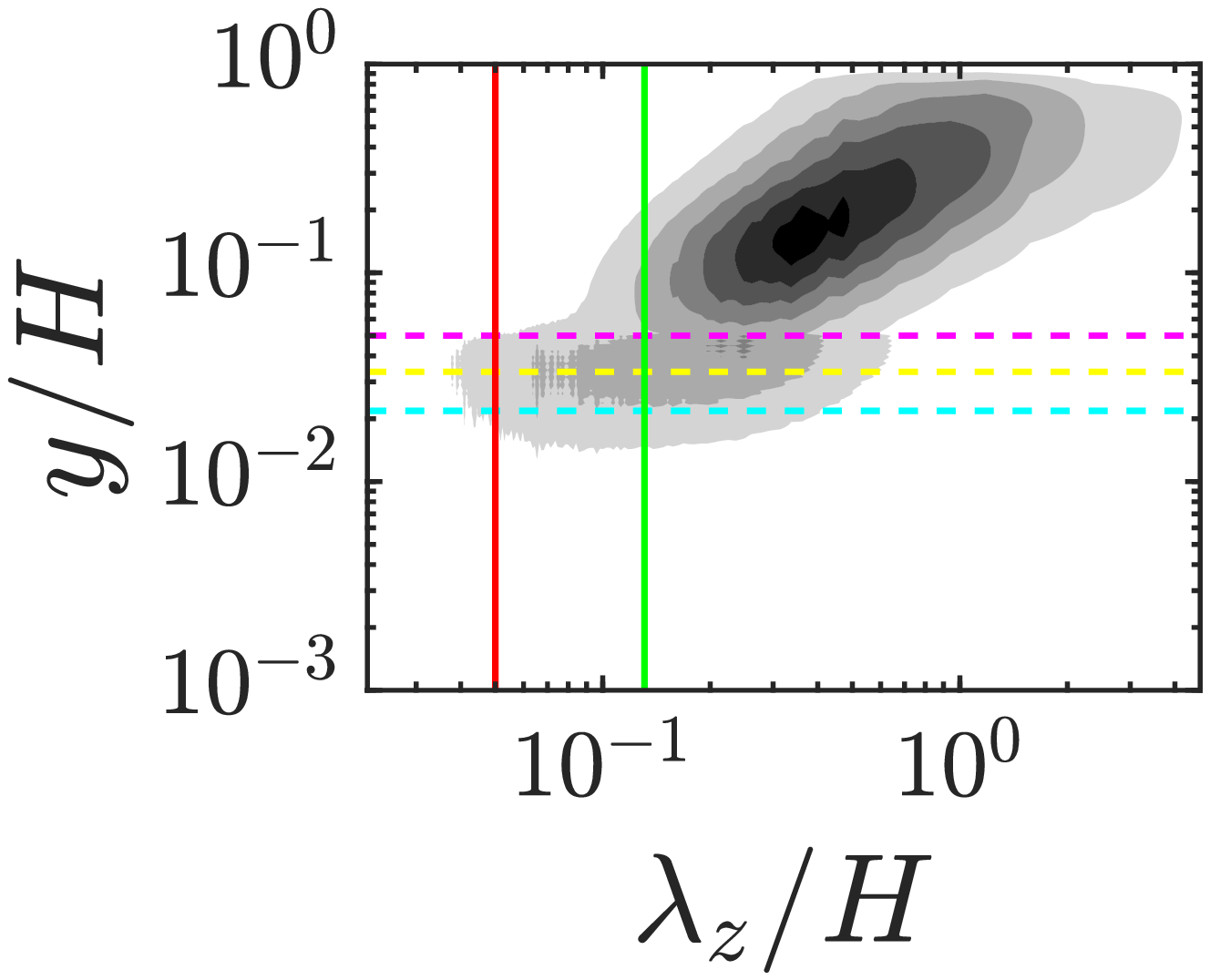}}
  \subfloat[]{\includegraphics[width=0.25\linewidth]{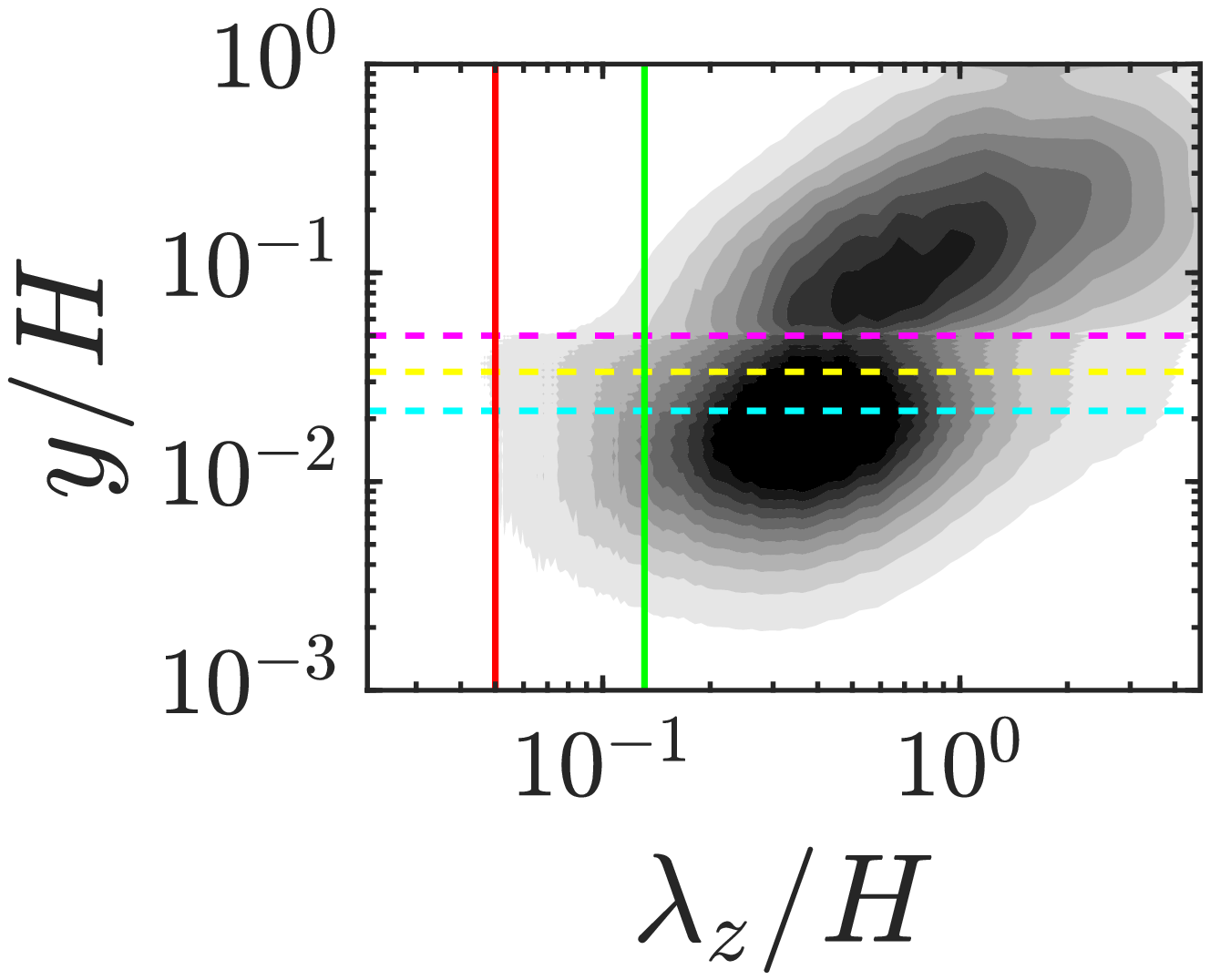}}
  \subfloat[]{\includegraphics[width=0.25\linewidth]{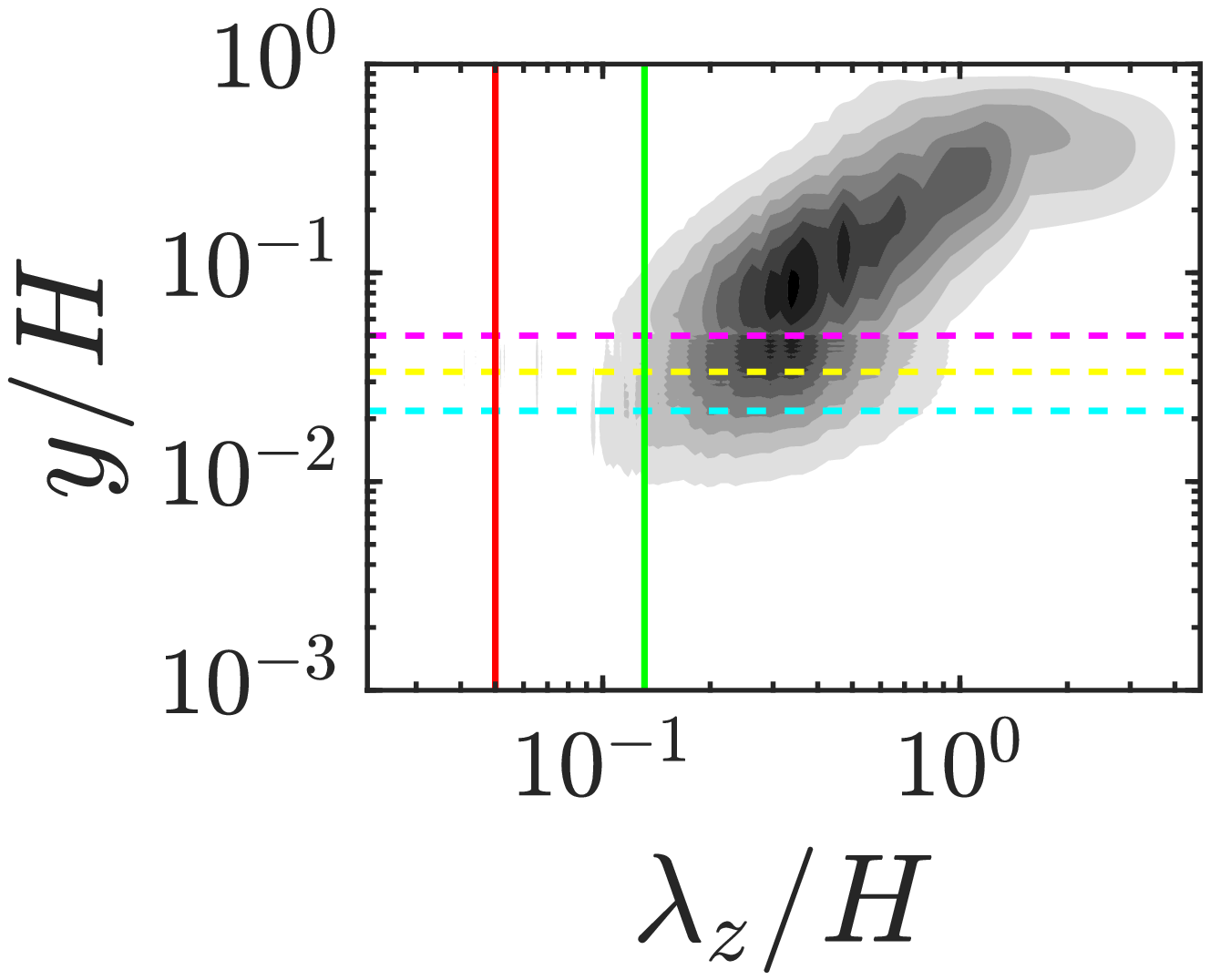}}\\
  \caption{Cases $\theta=\pm\ang{78.5}$. 
           Magnitude of the premultiplied spectra of the velocity components
           and cospectra of the Reynolds shear stress as a function of 
           the spanwise wavelength $\lambda_z/H$ and the wall-normal coordinates
           $y/H$. Top row: $\theta=\ang{78.5}$; bottom row $\theta=-\ang{78.5}$. 
           From left to right, the columns are:
           $\kappa_z\Phi_{u'u'}/u_{\tau,l}^2$ with 
           grey levels range in $[0,0.8]$ with a $0.1$ increment; 
           $\kappa_z\Phi_{v'v'}/u_{\tau,l}^2$ with grey levels range in 
           $[0,0.3]$ with a $0.03$ increment; 
           $\kappa_z\Phi_{w'w'}/u_{\tau,l}^2$ 
           with grey levels range in $[0,0.5]$ with a $0.05$ increment;
           $\kappa_z|\Phi_{u'v'}|/u_{\tau,l}^2$ 
           with grey levels range in $[0,0.4]$ with a $0.02$ increment.
           Colour lines have the same meaning as in 
           \cref{fig:premultipliedSpectraV0d250}.}
  \label{fig:premultipliedSpectraBFz0d050}
  \end{figure}
Finally, we consider the magnitude of the cospectra of the Reynolds shear stress
shown in panels (d) and (h). As we said, these cospectra are an indicator
of the large spanwise rollers that in these cases are triggered by the KH
instability at the canopy tip. Here, it is worth noticing how the large
structures deeply penetrate in the canopy layer when the stems are backwardly
inclined (panel (h)), moving the centre of the vortices well within the
canopy layer. This explains the drag increasing effect
of the negative inclination: the large vortices are pushed in the canopy 
layer that dissipates, through the drag, the high momentum carried by those 
structures, thus requiring more energy to move the flow.
The considerations made above are valid in the case of inclined canopies in the
limit of a transitional to dense regime. Further inclining the stems eventually
brings to a flow over a solid wall characterized by filamentous patterns of 
roughness, with size of the order of the diameter of the stems, $d$, with the 
direction of the inclination disappearing from the parameters.
Therefore, the structures of the two oppositely, very-inclined cases considered
in our study have a similar morphology between them and, approaching the sparsity
of the canopy layer \citep{NEPF2012,BRUNET2020}, behave very differently compared to the 
cases with $|\theta|=\ang{45}$. 
\Cref{fig:premultipliedSpectraBFx0d050,fig:premultipliedSpectraBFz0d050} show 
the velocity structures of the two cases considered (top row $\theta=\ang{78.5}$
and bottom row $\theta=-\ang{78.5}$) as a function of the streamwise and spanwise
wavelengths, respectively. The sets of figures show, as expected, very similar
coherency between the positively and negatively inclined canopies, with
slight differences within the canopy layer in the energy content of the streamwise 
component of the velocity fluctuations (and consequently of the Reynolds
shear stress) due to the different penetration of the outer layer imposed by 
the direction of the inclination.
Finally, concerning the morphology of the coherent structures, 
\cref{fig:premultipliedSpectraBFx0d050,fig:premultipliedSpectraBFz0d050} reveal
that the outer layer, as we expect from a canopy flow in a quasi-sparse regime,
is able to fully penetrate within the canopy layer, with the inner peaks that 
are simply an extension of the structures populating the outer
boundary layer, only shortened by the presence of the stems.

\begin{figure}
  \centering
  \subfloat[]{\includegraphics[width=0.33\linewidth]{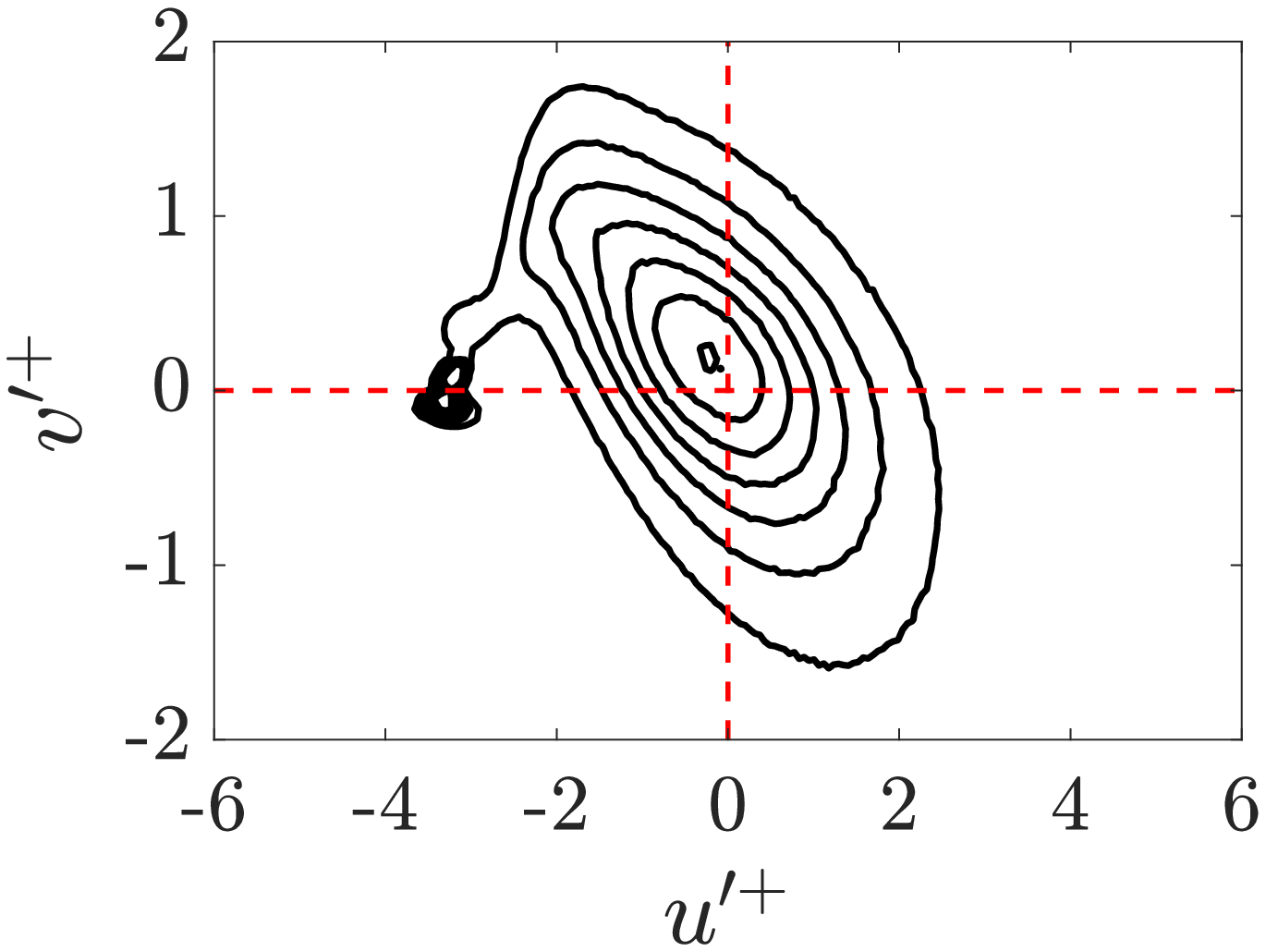}}%
  \subfloat[]{\includegraphics[width=0.33\linewidth]{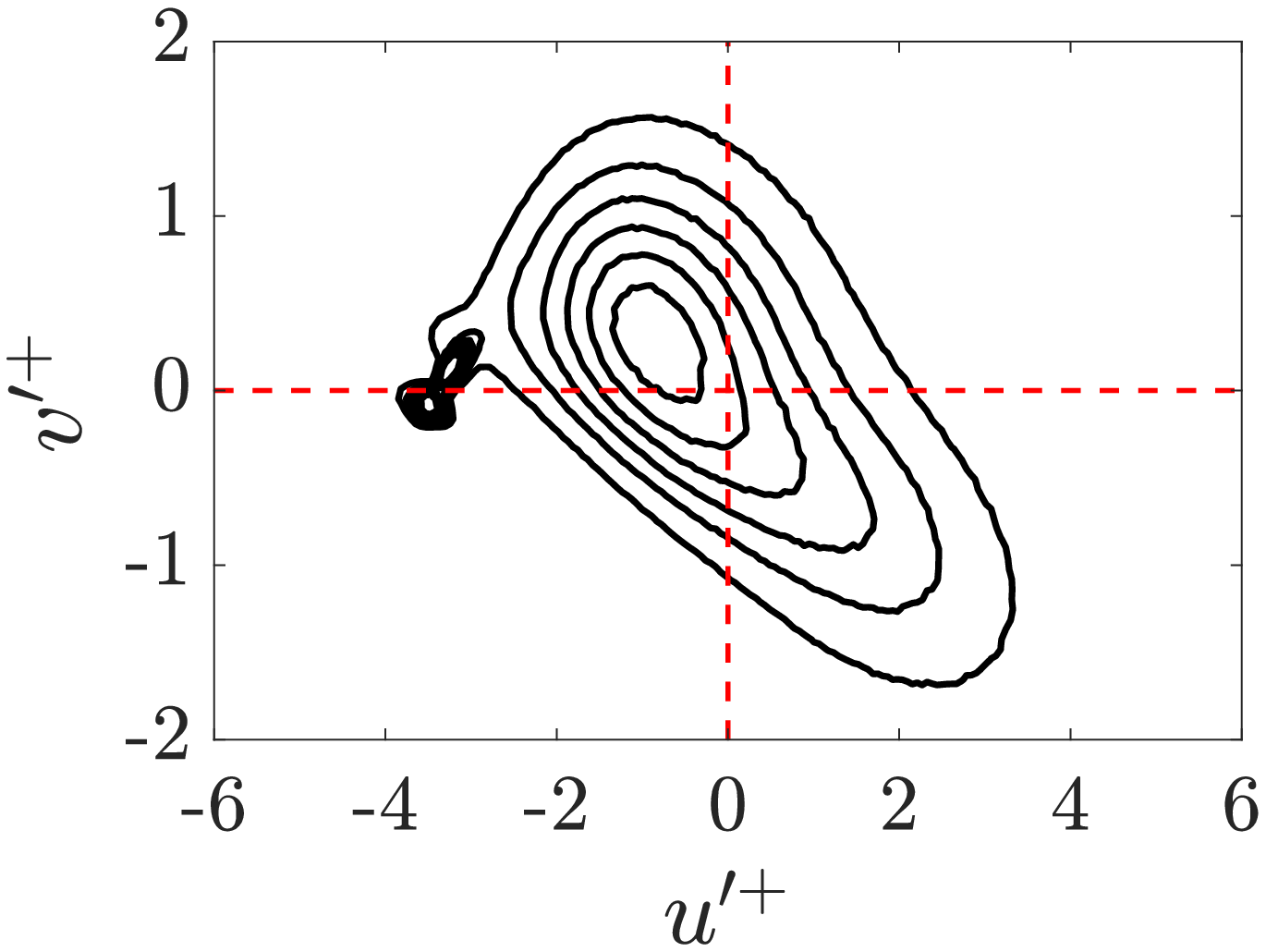}}%
  \subfloat[]{\includegraphics[width=0.33\linewidth]{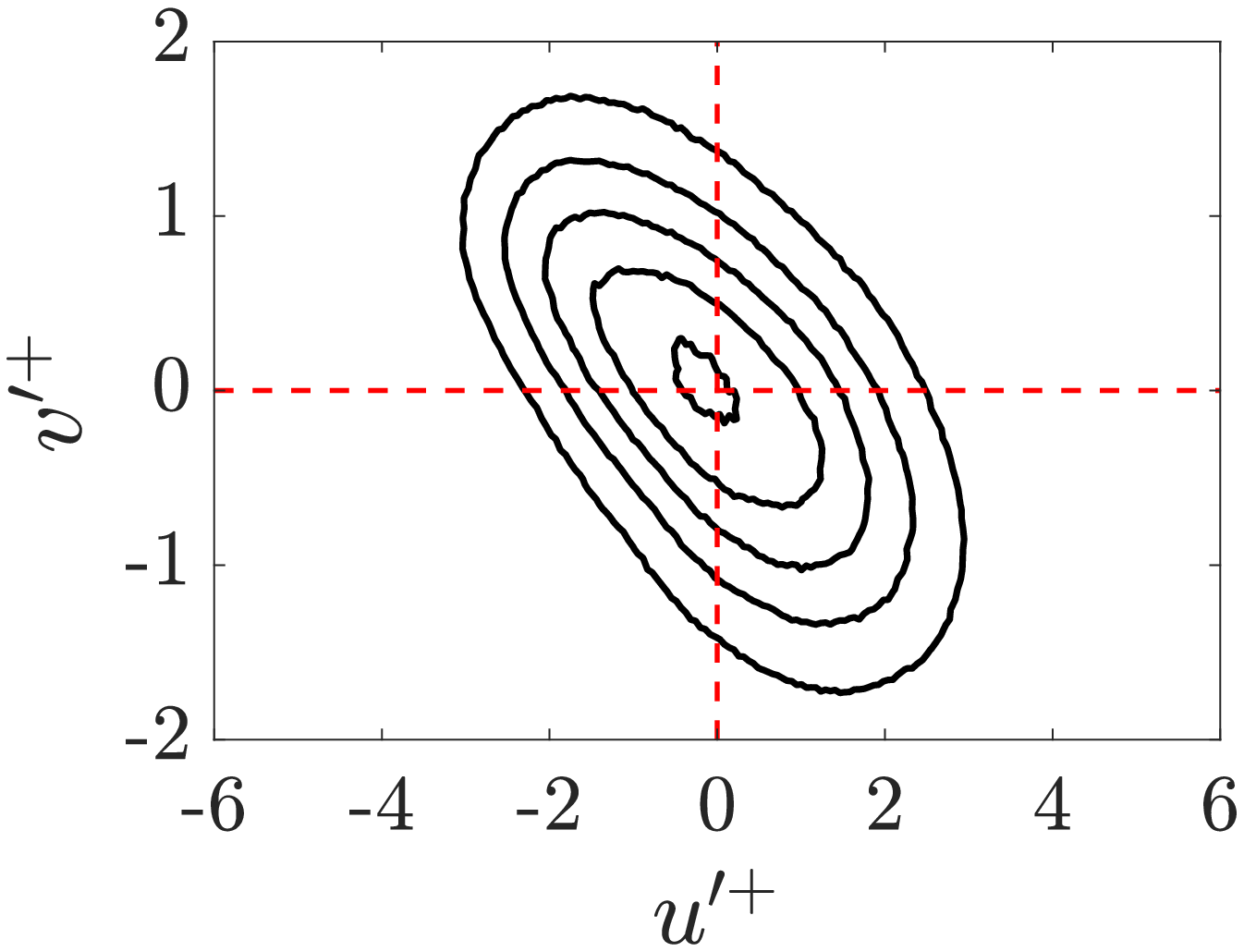}}\\%
  \subfloat[]{\includegraphics[width=0.33\linewidth]{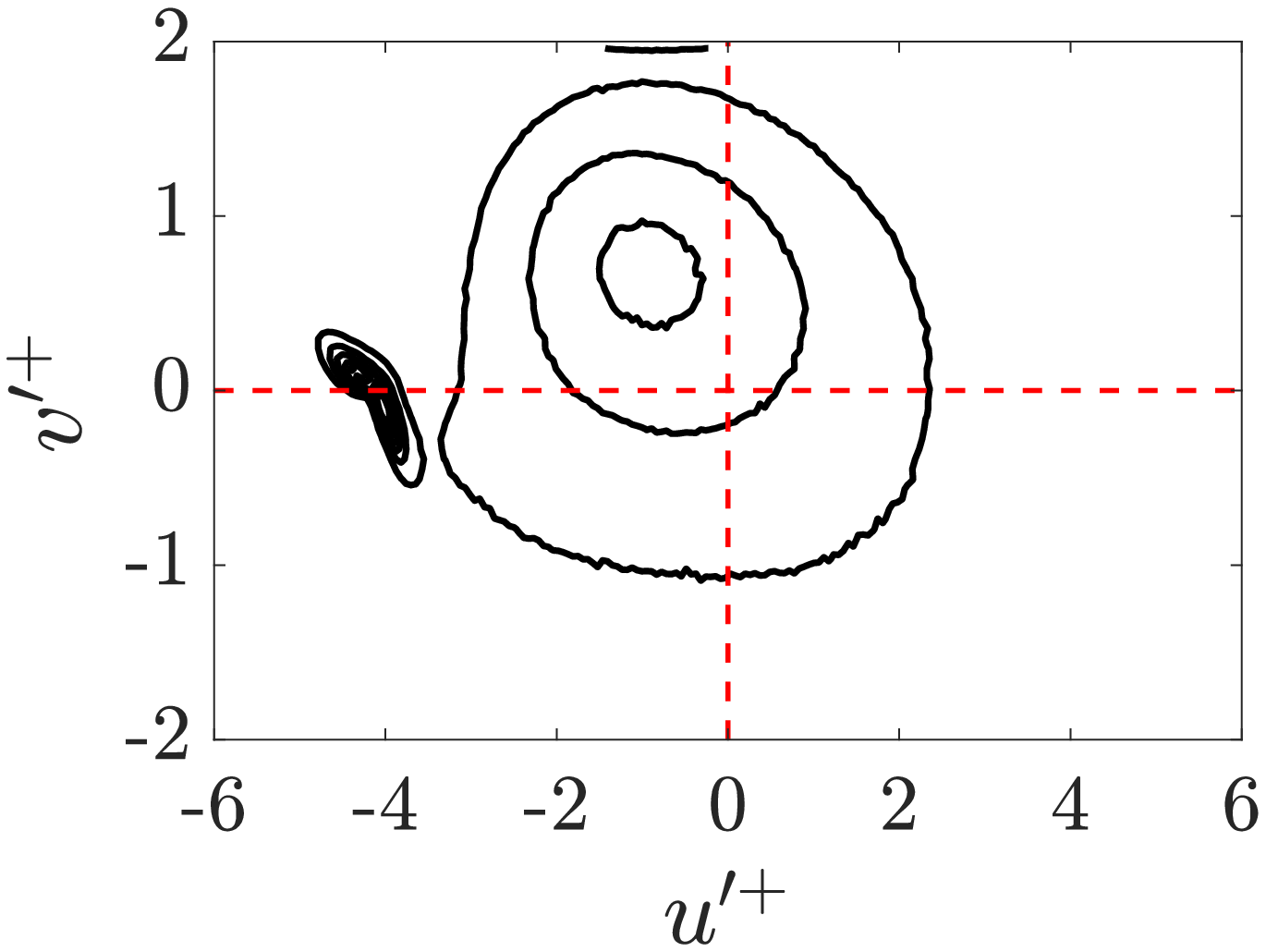}}%
  \subfloat[]{\includegraphics[width=0.33\linewidth]{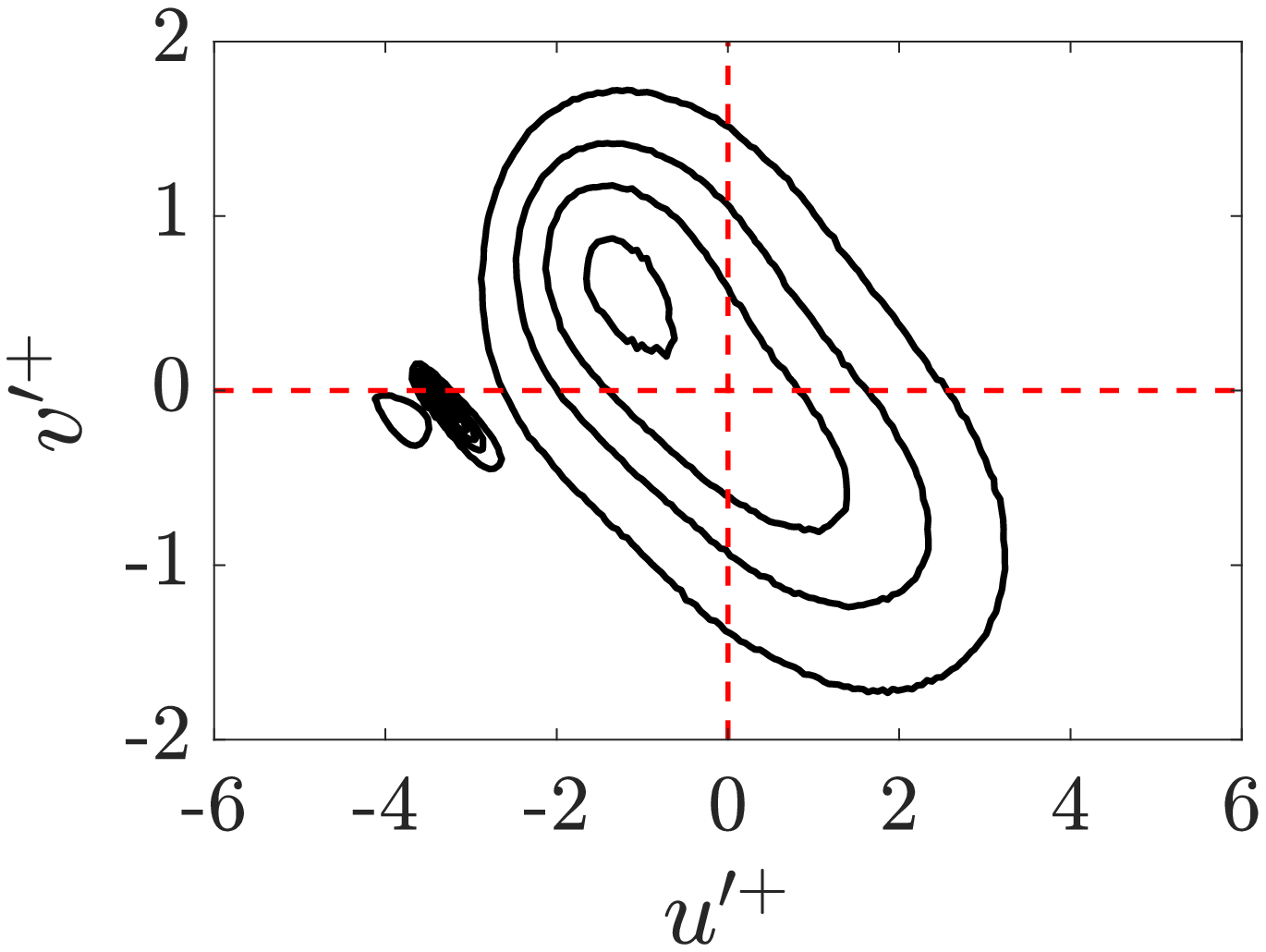}}%
  \subfloat[]{\includegraphics[width=0.33\linewidth]{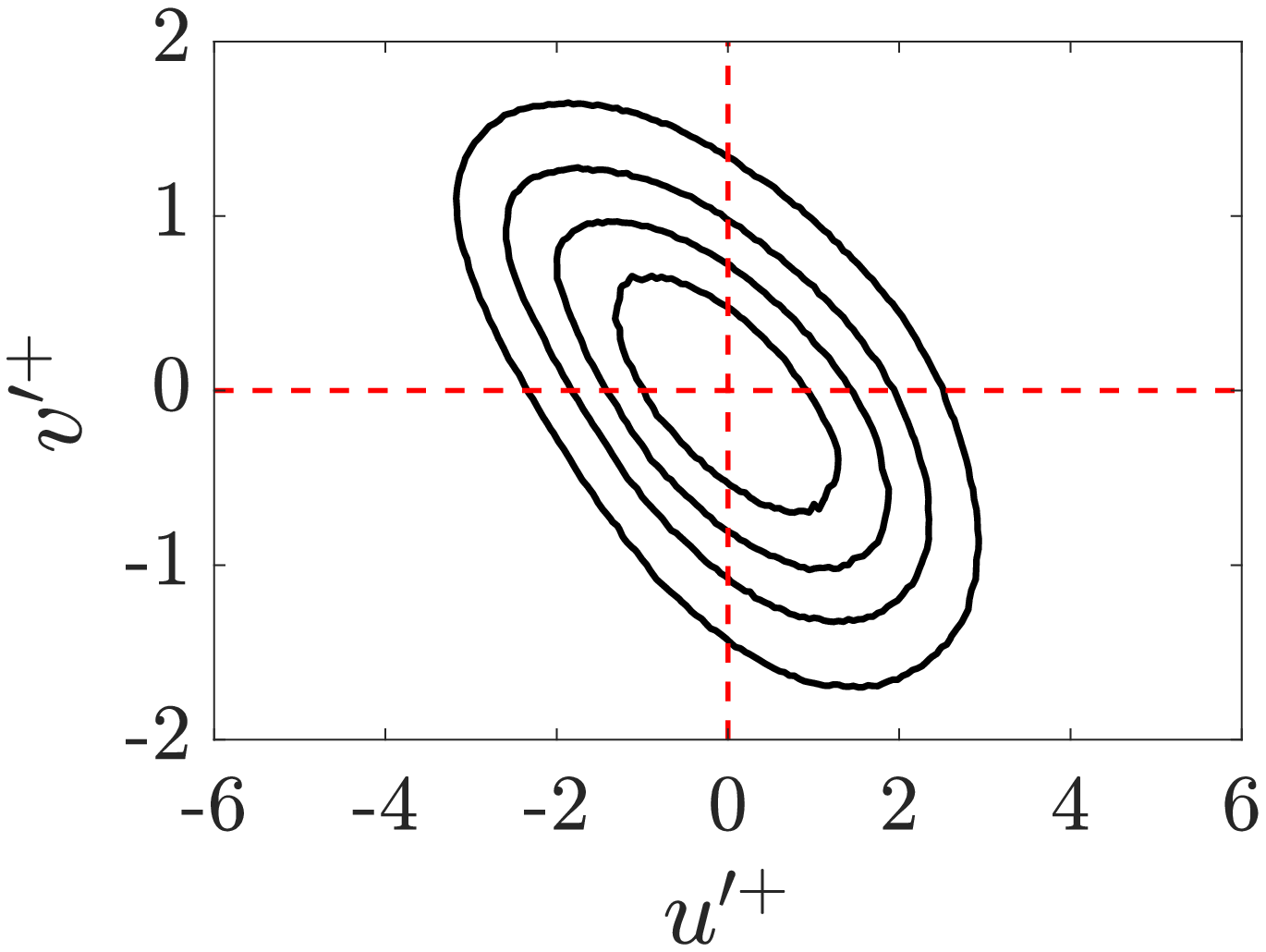}}\\%
  \caption{Cases $\theta=\pm\ang{45}$. 
           Contours of the joint probability function of the fluctuations of 
           the streamwise velocity component $u'^+$ and of the wall-normal 
           velocity component $v'^+$ normalised with the local friction 
           velocity $u_{\tau,l}$, on planes parallel to the wall.
           Top row: $\theta=\ang{45}$; bottom row $\theta=-\ang{45}$. 
           From left to right, the columns indicate the planes parallel
           to the wall:
           $y=0.10H<l_\perp$, first column; $y=l_\perp$, second column; 
           $y=0.25H>l_\perp$, third column.
           The levels of the contour lines start from $0.025$ (most external 
           line), and increase with increment $0.025$. The red dashed lines
           show the axis $u'=0$ and $v'=0$.}
  \label{fig:jpdfuv_he0d175H}
  \end{figure}
At last, a further proof of the discrepancy of the properties of the flow 
attributed to the sign of the angle of inclination of the filaments can be sought in the 
joint probability density functions (jPDF) of the fluctuations of the streamwise
velocity $u'$ and of the wall-normal velocity $v'$.
\Cref{fig:jpdfuv_he0d175H} shows the contour lines (spacing 0.025, starting from
the level 0.025 on the most external contour line) of the jPDF at three 
locations parallel to the wall for the cases $\theta=\pm\ang{45}$; 
such angle $|\theta|$ is chosen as a case that well represents the discrepancy 
between the positive and negative inclination of the stems.
The figure is organised in a $2\times 3$ matrix of panels, where 
the columns of the matrix, from left to right, shows the jPDF of 
$u'$-$v'$ at the location $y=0.10H<l_\perp$ (i.e.\ within the canopy layer), 
$y=l_\perp$ (the canopy edge) and $y=0.25H>l_\perp$ (i.e.\ above the canopy region)
for the cases $\theta=\ang{45}$ (top row) and $\theta=-\ang{45}$ (bottom row).
In the region above the canopy, the jPDF in panels (c) and (f) show a similar 
distribution of $u'$-$v'$ for the two canopies analysed, with prevalence 
of events almost equally distributed within the second and fourth quadrants 
(ejections and sweeps, respectively) as it has been observed in canopy 
flows \citep{BAILEY2016}. 
Moving toward the canopy region, the effect of the sign of the 
angle of inclination starts to appear. At first, we analyse the jPDF
at the edge of the canopy, shown in panels (b) and (e). We can promptly
notice that both the distributions show two peaks:
the smallest and very frequent one (higher number of levels) found on the left 
side of the figures with $u'<0$ and $v'\approx 0$ represents the 
recirculation forming behind each stem; the broadest one, instead, located within 
the second quadrant with $u'<0$ and $v'>0$ reports that ejections are 
the events that are more probable, especially for $\theta=-\ang{45}$.
The shape of the distributions for the two scenarios, however, is very similar,
with only higher probability of $u'$-$v'$ (more contour levels) for the case
forwardly inclined.
Within the canopy layer, panels (a) and (d), instead, the scenario is very different: 
while the canopy inclined with the grain (top panels) has a jPDF that keeps the shape
of the upper layers, with a prevalence of events within the second quadrant,
the canopy inclined against the grain (bottom panels) has a jPDF that shows the 
appearance of events within the third quadrant, evidencing the eased capability
of the flow to penetrate within the canopy. Finally, it is worth noticing the inclination
of the peak related to the wakes of the filaments: while in the case with 
$\theta=\ang{45}$ the wakes tend to have a positive $v'$, the opposite is true
for the case with $\theta=-\ang{45}$.

\section{Conclusion}\label{sec:conclusions}
We have carried out a set of high-fidelity numerical simulations of a turbulent
channel flow over inclined rigid canopies with the aim of \textit{i)} exploring the
effectiveness of the solidity $\lambda$ as selective parameter for determining
the behaviour of canopy flows and \textit{ii)} offering a complete description
of the interaction between the coherent structures populating the inner and
outer layers of the canopy flow.

At first, we have compared canopies inclined with opposite angles $\theta$,
finding that the stems inclined with the grain (i.e. $\theta > 0$) shield
the canopy layer from the outer coherent motions, thus increasing the separation
of the inner and outer layers while, on the contrary, filaments inclined against 
the grain (i.e. $\theta < 0$) promote the penetration of the high-momentum
large coherent motions of the outer flow within the canopy layer. Since these
large turbulent structures carry high-momentum that is dissipated by the stems, 
canopies against the grain introduce a higher drag. Finally, very inclined canopies
asymptotically behave as roughness, therefore losing their dependence from
the angle of inclination. These considerations conclude that the solidity may be a
misleading parameter for predicting the behaviour of canopy flows, and therefore
a need for a more complex model rises. Here, based on the scaling of the pressure
gradient, we proposed the outer quantities for the classification of the regimes.

In the second part of the manuscript, we have improved the description of mixing
between the outer boundary layer and the inner flow. In particular, we started 
treating the two layers as separated: the structures of the inner layer
can be compared to the high-speed and low-speed wakes generated passing through 
bidimensional cylinders, the former coming from the blockage effect and the latter
from the wakes of a bluff body; the outer layer, instead, is populated by large 
coherent motions triggered by the KH instability at the canopy tip. Following
the work by \citet{MONTI2020}, we suggest that the two layers communicate through 
a set of high-momentum large wall-normal jets \citep{BANYASSADY2015}
that detach from the outer coherent structures and penetrate 
within the canopy as a result of its high wall-normal permeability. At
the wall, these jets trigger the formation
of the very-large coherent motions of $u'$ and $w'$ observed in proximity of
the canopy bed. By reflection, outside the region of the jets, the flow 
is gently pushed away from the wall, creating regions of $u'v'$ vortices
similar to KH; this mechanism fixes the location of the inner inflection point.
Inclining the stems of the canopy, the mechanism of interaction is magnified 
or reduced depending on the direction of the inclination. In particular, when the
canopy is inclined against the grain ($\theta<0$), the outer large structures are
able to penetrate deeply within the canopy, with stronger wall-normal jets 
that energize the flow within the canopy layer; on the contrary, when the canopy is
inclined with the grain, the inclined stems shelter the inner region, impeding the
structures populating the boundary-layer above the canopy to interact with the
structures living close to the wall. However, this effect of the inclination 
tends to vanish when the height of the canopy layer becomes very small (very high 
$|\theta|$), since the latter tends to become roughness (sparse regime) and the
outer and inner structures become very well mixed.

The natural extension of the present research consists in considering
the flexibility of the stems of the canopy as additional  parameter. 
When the flexibility of the filaments is introduced, the complexity of flow 
considerably increases due to the coupling between the fluid and the structural elements, with the latter compliantly moving with the flow, thus having a variable and non-uniform inclination induced by the
presence of the strong coherent structures of the outer energetic layer.

\section*{Acknowledgments}
M.E.R and A.M. gratefully acknowledge the support of Okinawa Institute of Science and Technology Graduate University (OIST) with subsidy funding from the Cabinet Office, Government of Japan.
M.E.R. and A.M. acknowledge the computational time provided by HPCI on the Oakbridge-CX and Oakforest-PACS clusters, under the grant hp210025, and the computer time provided by the Scientific Computing section of Research Support Division at OIST. 

\section*{Declaration of Interests}
The authors report no conflict of interest.

\bibliographystyle{jfm}
\bibliography{biblio}

\begin{thebibliography}{50}
\expandafter\ifx\csname natexlab\endcsname\relax\def\natexlab#1{#1}\fi
\def\au#1{#1} \def\ed#1{#1} \def\yr#1{#1}\def\at#1{#1}\def\jt#1{\textit{#1}}
  \def\bt#1{#1}\def\bvol#1{\textbf{#1}} \def\vol#1{#1} \def\pg#1{#1}
  \def\publ#1{#1}\def\arxiv#1{#1}\def\org#1{#1}\def\st#1{\textit{#1}}

\bibitem[Alvarado {\em et~al.\/}(2017)Alvarado, Comtet, De~Langre \&
  Hosoi]{ALVARADO2017}
{\sc \au{Alvarado, J.}, \au{Comtet, J.}, \au{De~Langre, E.} \& \au{Hosoi,
  A.E.}} \yr{2017}  \at{Nonlinear flow response of soft hair beds}.  \jt{Nature
  Physics}  \bvol{13}~(10),  \pg{1014--1019}.

\bibitem[Bailey \& Stoll(2013)]{BAILEY2013}
{\sc \au{Bailey, B.N.} \& \au{Stoll, R.}} \yr{2013}  \at{Turbulence in sparse,
  organized vegetative canopies: a large-eddy simulation study}.
  \jt{Boundary-Layer Meteorology}  \bvol{147}~(3),  \pg{369--400}.

\bibitem[Bailey \& Stoll(2016)]{BAILEY2016}
{\sc \au{Bailey, B.N.} \& \au{Stoll, R.}} \yr{2016}  \at{The creation and
  evolution of coherent structures in plant canopy flows and their role in
  turbulent transport}.  \jt{Journal of Fluid Mechanics}  \bvol{789},
  \pg{425--460}.

\bibitem[Banyassady \& Piomelli(2015)]{BANYASSADY2015}
{\sc \au{Banyassady, R.} \& \au{Piomelli, U.}} \yr{2015}  \at{Interaction of
  inner and outer layers in plane and radial wall jets}.  \jt{Journal of
  Turbulence}  \bvol{16}~(5),  \pg{460--483}.

\bibitem[Belcher {\em et~al.\/}(2003)Belcher, Jerram \& Hunt]{BELCHER2003}
{\sc \au{Belcher, S.E.}, \au{Jerram, N.} \& \au{Hunt, J.C.R.}} \yr{2003}
  \at{Adjustment of a turbulent boundary layer to a canopy of roughness
  elements}.  \jt{Journal of Fluid Mechanics}  \bvol{488},  \pg{369--398}.

\bibitem[Brizzolara {\em et~al.\/}(2021)Brizzolara, Rosti, Olivieri, Brandt,
  Holzner \& Mazzino]{BRIZZOLARA2020}
{\sc \au{Brizzolara, S.}, \au{Rosti, M.E.}, \au{Olivieri, S.}, \au{Brandt, L.},
  \au{Holzner, M.} \& \au{Mazzino, A.}} \yr{2021}  \at{Fiber tracking
  velocimetry for two-point statistics of turbulence}.  \jt{Physical review X}
  \bvol{11}~(3),  \pg{031060}.

\bibitem[Br\"{u}cker \& Weidner(2014)]{BRUECKER2014}
{\sc \au{Br\"{u}cker, C.} \& \au{Weidner, C.}} \yr{2014}  \at{Influence of
  self-adaptive hairy flaps on the stall delay of an airfoil in ramp-up
  motion}.  \jt{Journal of Fluids and Structures}  \bvol{47},  \pg{31--40}.

\bibitem[Brunet(2020)]{BRUNET2020}
{\sc \au{Brunet, Y.}} \yr{2020}  \at{Turbulent flow in plant canopies:
  historical perspective and overview}.  \jt{Boundary-Layer Meteorology}
  \bvol{177}~(2),  \pg{315--364}.

\bibitem[Fadlun {\em et~al.\/}(2000)Fadlun, Verzicco, Orlandi \&
  Mohd-Yusof]{FADLUN2000}
{\sc \au{Fadlun, E.A.}, \au{Verzicco, R.}, \au{Orlandi, P.} \& \au{Mohd-Yusof,
  J.}} \yr{2000}  \at{Combined immersed-boundary finite-difference methods for
  three-dimensional complex flow simulations}.  \jt{Journal of Computational
  Physics}  \bvol{161}~(1),  \pg{35--60}.

\bibitem[Finnigan(2000)]{FINNIGAN2000}
{\sc \au{Finnigan, J.J.}} \yr{2000}  \at{Turbulence in plant canopies}.
  \jt{Annual Review of Fluid Mechanics}  \bvol{32}~(1),  \pg{519--571}.

\bibitem[Ghisalberti \& Nepf(2002)]{GHISALBERTI2002}
{\sc \au{Ghisalberti, M.} \& \au{Nepf, H.M.}} \yr{2002}  \at{Mixing layers and
  coherent structures in vegetated aquatic flows}.  \jt{Journal of Geophysical
  Research: Oceans}  \bvol{107}~(C2),  \pg{3--1}.

\bibitem[Ghisalberti \& Nepf(2004)]{GHISALBERTI2004}
{\sc \au{Ghisalberti, M.} \& \au{Nepf, H.M.}} \yr{2004}  \at{The limited growth
  of vegetated shear layers}.  \jt{Water Resources Research}  \bvol{40}~(7).

\bibitem[Itoh {\em et~al.\/}(2006)Itoh, Iguchi, Yokota, Akino, R. \&
  S.]{ITOH2006}
{\sc \au{Itoh, M.}, \au{Iguchi, R.}, \au{Yokota, K.}, \au{Akino, N.}, \au{R.,
  Hino} \& \au{S., Kubo}} \yr{2006}  \at{Turbulent drag reduction by the seal
  fur surface}.  \jt{Physics of Fluids}  \bvol{18 - 065102}.

\bibitem[Jim{\'e}nez(2004)]{JIMENEZ2004}
{\sc \au{Jim{\'e}nez, J.}} \yr{2004}  \at{Turbulent flows over rough walls}.
  \jt{Annual Review of Fluid Mechanics}  \bvol{36},  \pg{173--196}.

\bibitem[Jim{\'e}nez \& Pinelli(1999)]{JIMENEZ1999}
{\sc \au{Jim{\'e}nez, J.} \& \au{Pinelli, A.}} \yr{1999}  \at{The autonomous
  cycle of near-wall turbulence}.  \jt{Journal of Fluid Mechanics}  \bvol{389},
   \pg{335--359}.

\bibitem[Jim{\'e}nez {\em et~al.\/}(2001)Jim{\'e}nez, Uhlmann, Pinelli \&
  Kawahara]{JIMENEZ2001}
{\sc \au{Jim{\'e}nez, J.}, \au{Uhlmann, M.}, \au{Pinelli, A.} \& \au{Kawahara,
  G.}} \yr{2001}  \at{Turbulent shear flow over active and passive porous
  surfaces}.  \jt{Journal of Fluid Mechanics}  \bvol{442},  \pg{89--117}.

\bibitem[Kim \& Moin(1985)]{KIM1985}
{\sc \au{Kim, J.} \& \au{Moin, P.}} \yr{1985}  \at{Application of a
  fractional-step method to incompressible {N}avier-{S}tokes equations}.
  \jt{Journal of Computational Physics}  \bvol{59}~(2),  \pg{308--323}.

\bibitem[Kim {\em et~al.\/}(1987)Kim, Moin \& Moser]{KIM1987}
{\sc \au{Kim, J.}, \au{Moin, P.} \& \au{Moser, R.}} \yr{1987}  \at{Turbulence
  statistics in fully developed channel flow at low {R}eynolds number}.
  \jt{Journal of Fluid Mechanics}  \bvol{177},  \pg{133--166}.

\bibitem[Lauga \& Powers(2009)]{LAUGA2009}
{\sc \au{Lauga, E.} \& \au{Powers, T.R.}} \yr{2009}  \at{The hydrodynamics of
  swimming microorganisms}.  \jt{Reports on Progress in Physics}
  \bvol{72}~(9),  \pg{096601}.

\bibitem[Lee {\em et~al.\/}(1974)Lee, Vaseleski \& Metzner]{LEE1974}
{\sc \au{Lee, W.K.}, \au{Vaseleski, R.C.} \& \au{Metzner, A.B.}} \yr{1974}
  \at{Turbulent drag reduction in polymeric solutions containing suspended
  fibers}.  \jt{AIChE Journal}  \bvol{20}~(1),  \pg{128--133}.

\bibitem[Leonard(1975)]{LEONARD1975}
{\sc \au{Leonard, A.}} \yr{1975}  \at{Energy cascade in large-eddy simulations
  of turbulent fluid flows}.  \jt{Advances in Geophysics}  \bvol{18},
  \pg{237--248}.

\bibitem[Lodish {\em et~al.\/}(2007)Lodish, Berk \& Kaiser]{LODISH2007}
{\sc \au{Lodish, H.}, \au{Berk, A.} \& \au{Kaiser, C.A.}} \yr{2007} {\em
  Molecular cell biology\/}.  \publ{W.H. Freeman \& Co. Ltd}.

\bibitem[Luhar {\em et~al.\/}(2008)Luhar, Rominger \& Nepf]{LUHAR2008}
{\sc \au{Luhar, M.}, \au{Rominger, J.} \& \au{Nepf, H.M.}} \yr{2008}
  \at{Interaction between flow, transport and vegetation spatial structure}.
  \jt{Environmental Fluid Mechanics}  \bvol{8}~(5-6),  \pg{423}.

\bibitem[Mars {\em et~al.\/}(1999)Mars, Mathew \& Ho]{MARS1999}
{\sc \au{Mars, R.}, \au{Mathew, K.} \& \au{Ho, G.}} \yr{1999}  \at{The role of
  the submergent macrophyte {\em triglochin huegelii} in domestic greywater
  treatment}.  \jt{Ecological Engineering}  \bvol{12}~(1),  \pg{57--66}.

\bibitem[Monti {\em et~al.\/}(2019)Monti, Omidyeganeh \& A.]{MONTI2019}
{\sc \au{Monti, A.}, \au{Omidyeganeh, M.} \& \au{A., Pinelli}} \yr{2019}
  \at{Large eddy simulation of of an open-channel flow bounded by a semi-dense
  rigid filamentous canopy: scaling and flow structure}.  \jt{Physics of
  Fluids}  \bvol{31}~(065108).

\bibitem[Monti {\em et~al.\/}(2020)Monti, Omidyeganeh, Eckhardt \&
  Pinelli]{MONTI2020}
{\sc \au{Monti, A.}, \au{Omidyeganeh, M.}, \au{Eckhardt, B.} \& \au{Pinelli,
  A.}} \yr{2020}  \at{On the genesis of different regimes in canopy flows: a
  numerical investigation}.  \jt{Journal of Fluid Mechanics}  \bvol{891}.

\bibitem[Mossa {\em et~al.\/}(2017)Mossa, Ben~Meftah, De~Serio \&
  Nepf]{MOSSA2017}
{\sc \au{Mossa, Michele}, \au{Ben~Meftah, Mouldi}, \au{De~Serio, Francesca} \&
  \au{Nepf, Heidi~M}} \yr{2017}  \at{How vegetation in flows modifies the
  turbulent mixing and spreading of jets}.  \jt{Scientific reports}
  \bvol{7}~(1),  \pg{1--14}.

\bibitem[Mossa {\em et~al.\/}(2021)Mossa, Goldshmid, Liberzon, Negretti,
  Sommeria, Termini \& De~Serio]{MOSSA2021}
{\sc \au{Mossa, Michele}, \au{Goldshmid, Roni~H}, \au{Liberzon, Dan},
  \au{Negretti, M~Eletta}, \au{Sommeria, Joel}, \au{Termini, Donatella} \&
  \au{De~Serio, Francesca}} \yr{2021}  \at{Quasi-geostrophic jet-like flow with
  obstructions}.  \jt{Journal of Fluid Mechanics}  \bvol{921}.

\bibitem[Nepf(2012)]{NEPF2012}
{\sc \au{Nepf, H.M.}} \yr{2012}  \at{Flow and transport in regions with aquatic
  vegetation}.  \jt{Annual Review of Fluid Mechanics}  \bvol{44},
  \pg{123--142}.

\bibitem[Nepf \& Vivoni(2000)]{NEPF2000}
{\sc \au{Nepf, H.M.} \& \au{Vivoni, E.R.}} \yr{2000}  \at{Flow structure in
  depth-limited, vegetated flow}.  \jt{Journal of Geophysical Research: Oceans}
   \bvol{105}~(C12),  \pg{28547--28557}.

\bibitem[Nezu \& Sanjou(2008)]{NEZU2008}
{\sc \au{Nezu, I.} \& \au{Sanjou, M.}} \yr{2008}  \at{Turburence structure and
  coherent motion in vegetated canopy open-channel flows}.  \jt{Journal of
  Hydro-Environment Research}  \bvol{2}~(2),  \pg{62--90}.

\bibitem[Olivieri {\em et~al.\/}(2021)Olivieri, Mazzino \&
  Rosti]{olivieri_mazzino_rosti_2021d}
{\sc \au{Olivieri, S.}, \au{Mazzino, A.} \& \au{Rosti, M.E.}} \yr{2021}
  \at{Universal flapping states of elastic fibers in modulated turbulence}.
  \jt{Physics of Fluids}  \bvol{33},  \pg{071704}.

\bibitem[Omidyeganeh \& Piomelli(2013)]{OMID2013a}
{\sc \au{Omidyeganeh, M.} \& \au{Piomelli, U.}} \yr{2013}  \at{Large-eddy
  simulation of three-dimensional dunes in a steady, unidirectional flow.
  {P}art 1. turbulence statistics}.  \jt{Journal of Fluid Mechanics}
  \bvol{721},  \pg{454--483}.

\bibitem[Paschkewitz {\em et~al.\/}(2004)Paschkewitz, Dubief, Dimitropoulos,
  Shaqfeh \& Moin]{PASCHKEWITZ2004}
{\sc \au{Paschkewitz, J.S.}, \au{Dubief, Y.V.E.S.}, \au{Dimitropoulos, C.D.},
  \au{Shaqfeh, E.S.G.} \& \au{Moin, P.}} \yr{2004}  \at{Numerical simulation of
  turbulent drag reduction using rigid fibres}.  \jt{Journal of Fluid
  Mechanics}  \bvol{518},  \pg{281--317}.

\bibitem[Pinelli {\em et~al.\/}(2010)Pinelli, Naqavi, Piomelli \&
  Favier]{PINELLI2010}
{\sc \au{Pinelli, A.}, \au{Naqavi, I.Z.}, \au{Piomelli, U.} \& \au{Favier, J.}}
  \yr{2010}  \at{Immersed-boundary methods for general finite-difference and
  finite-volume {N}avier--{S}tokes solvers}.  \jt{Journal of Computational
  Physics}  \bvol{229}~(24),  \pg{9073--9091}.

\bibitem[Piomelli {\em et~al.\/}(2015)Piomelli, Rouhi \& Geurts]{PIOMELLI2015}
{\sc \au{Piomelli, U.}, \au{Rouhi, A.} \& \au{Geurts, B.J.}} \yr{2015}  \at{A
  grid-independent length scale for large-eddy simulations}.  \jt{Journal of
  Fluid Mechanics}  \bvol{766},  \pg{499--527}.

\bibitem[Poggi {\em et~al.\/}(2004)Poggi, Porporato, Ridolfi, Albertson \&
  Katul]{POGGI2004a}
{\sc \au{Poggi, D.}, \au{Porporato, A.}, \au{Ridolfi, L.}, \au{Albertson, J.D.}
  \& \au{Katul, G.G.}} \yr{2004}  \at{The effect of vegetation density on
  canopy sub-layer turbulence}.  \jt{Boundary-Layer Meteorology}
  \bvol{111}~(3),  \pg{565--587}.

\bibitem[Raupach {\em et~al.\/}(1996)Raupach, Finnigan \& Brunet]{RAUPACH1996}
{\sc \au{Raupach, M.R.}, \au{Finnigan, J.J.} \& \au{Brunet, Y.}} \yr{1996}
  \at{Coherent eddies and turbulence in vegetation canopies: the mixing-layer
  analogy}.  \jt{Boundary-Layer Meteorology}  \bvol{78}~(3-4),  \pg{351--382}.

\bibitem[Raupach \& Thom(1981)]{RAUPACH1981}
{\sc \au{Raupach, M.R.} \& \au{Thom, A.S.}} \yr{1981}  \at{Turbulence in and
  above plant canopies}.  \jt{Annual Review of Fluid Mechanics}  \bvol{13}~(1),
   \pg{97--129}.

\bibitem[Rhie \& Chow(1983)]{RHIE1983}
{\sc \au{Rhie, C.M.} \& \au{Chow, W.L.}} \yr{1983}  \at{Numerical study of the
  turbulent flow past an airfoil with trailing edge separation}.  \jt{AIAA
  Journal}  \bvol{21}~(11),  \pg{1525--1532}.

\bibitem[Rosti {\em et~al.\/}(2018{\natexlab{{\em a\/}}})Rosti, Banaei, Brandt
  \& Mazzino]{ROSTI2018b}
{\sc \au{Rosti, M.E.}, \au{Banaei, A.A.}, \au{Brandt, L.} \& \au{Mazzino, A.}}
  \yr{2018{\natexlab{{\em a\/}}}}  \at{Flexible fiber reveals the two-point
  statistical properties of turbulence}.  \jt{Physical review letters}
  \bvol{121}~(4),  \pg{044501}.

\bibitem[Rosti \& Brandt(2017)]{rosti_brandt_2017a}
{\sc \au{Rosti, M.E.} \& \au{Brandt, L.}} \yr{2017}  \at{Numerical simulation
  of turbulent channel flow over a viscous hyper-elastic wall}.  \jt{Journal of
  Fluid Mechanics}  \bvol{830},  \pg{708--735}.

\bibitem[Rosti {\em et~al.\/}(2018{\natexlab{{\em b\/}}})Rosti, Brandt \&
  Pinelli]{rosti_brandt_pinelli_2018a}
{\sc \au{Rosti, M.E.}, \au{Brandt, L.} \& \au{Pinelli, A.}}
  \yr{2018{\natexlab{{\em b\/}}}}  \at{Turbulent channel flow over an
  anisotropic porous wall -- drag increase and reduction}.  \jt{Journal of
  Fluid Mechanics}  \bvol{842},  \pg{381--394}.

\bibitem[Rosti {\em et~al.\/}(2020)Rosti, Olivieri, Banaei, Brandt \&
  Mazzino]{ROSTI2020}
{\sc \au{Rosti, M.E.}, \au{Olivieri, S.}, \au{Banaei, A.A.}, \au{Brandt, L.} \&
  \au{Mazzino, A.}} \yr{2020}  \at{Flowing fibers as a proxy of turbulence
  statistics}.  \jt{Meccanica}  \bvol{55}~(2),  \pg{357--370}.

\bibitem[Rouhi {\em et~al.\/}(2016)Rouhi, Piomelli \& Geurts]{ROUHI2016}
{\sc \au{Rouhi, A.}, \au{Piomelli, U.} \& \au{Geurts, B.J.}} \yr{2016}
  \at{Dynamic subfilter-scale stress model for large-eddy simulations}.
  \jt{Physical Review Fluids}  \bvol{1}~(4),  \pg{044401}.

\bibitem[Sharma \& Garc{\'\i}a-Mayoral(2018)]{SHARMA2018}
{\sc \au{Sharma, A.} \& \au{Garc{\'\i}a-Mayoral, R.}} \yr{2018} Turbulent flows
  over sparse canopies.  \bt{In {\em Journal of Physics: Conference Series\/}},
  ,  \vol{vol. 1001},  \pg{p. 012012}. IOP Publishing.

\bibitem[Shimizu {\em et~al.\/}(1991)Shimizu, Tsujimoto, Nakagawa \&
  Kitamura]{SHIMIZU1991}
{\sc \au{Shimizu, Y.}, \au{Tsujimoto, T.}, \au{Nakagawa, H.} \& \au{Kitamura,
  T.}} \yr{1991}  \at{Experimental study on flow over rigid vegetation
  simulated by cylinders with equi-spacing}.  \jt{Doboku Gakkai Ronbunshu}
  \bvol{1991}~(438),  \pg{31--40}.

\bibitem[Tschisgale {\em et~al.\/}(2021)Tschisgale, L{\"o}hrer, Meller \&
  Fr{\"o}hlich]{TSCHISGALE2021}
{\sc \au{Tschisgale, S.}, \au{L{\"o}hrer, B.}, \au{Meller, R.} \&
  \au{Fr{\"o}hlich, J.}} \yr{2021}  \at{Large eddy simulation of the
  fluid--structure interaction in an abstracted aquatic canopy consisting of
  flexible blades}.  \jt{Journal of Fluid Mechanics}  \bvol{916}.

\bibitem[Wilcock {\em et~al.\/}(1999)Wilcock, Champion, Nagels \&
  Croker]{WILCOCK1999}
{\sc \au{Wilcock, R.J.}, \au{Champion, P.D.}, \au{Nagels, J.W.} \& \au{Croker,
  G.F.}} \yr{1999}  \at{The influence of aquatic macrophytes on the hydraulic
  and physico-chemical properties of a {N}ew {Z}ealand lowland stream}.
  \jt{Hydrobiologia}  \bvol{416},  \pg{203--214}.

\bibitem[Yang {\em et~al.\/}(2002)]{YANG2002}
{\sc \au{Yang, U.M.} \& \au{others}} \yr{2002}  \at{Boomeramg: a parallel
  algebraic multigrid solver and preconditioner}.  \jt{Applied Numerical
  Mathematics}  \bvol{41}~(1),  \pg{155--177}.

\end{thebibliography}

\end{document}